\documentclass[twocolumn,prd,aps,nofootinbib,showpacs,superscriptaddress]{revtex4-1}
\usepackage{longtable}
\usepackage{multirow}
\usepackage{graphicx}
\usepackage[dvips]{color}
\usepackage{amssymb,amsmath}
\usepackage{supertabular}
\allowdisplaybreaks[4]
\begin{document}

\title{$I=\frac{3}{2}$ $\pi K$ $s$-wave scattering length from lattice QCD}

\author{Ziwen Fu}
\email{Corresponding Author: fuziwen@scu.edu.cn}
\affiliation{
Key Laboratory of Radiation Physics and Technology of Education Ministry;
Institute of Nuclear Science and Technology, Sichuan University, Chengdu 610064, People's Republic of China
}

\author{Qu-Zhi Li}
\email{Corresponding Author: liquzhi@scu.edu.cn}
\affiliation{
College of Physics, Sichuan University, Chengdu, Sichuan 610064, People's Republic of China
}

\author{Wang  Jun}
\email{Corresponding Author: wangjun@scu.edu.cn\\
\\
}

\affiliation{
Key Laboratory of Radiation Physics and Technology of  Education Ministry;
Institute of Nuclear Science and Technology, Sichuan University, Chengdu 610064, People's Republic of China
}

\begin{abstract}
The $I=\frac{3}{2}$ $\pi K$  $s$-wave scattering phase shift is computed by lattice quantum chromodynamics with
$N_f=3$ flavors of Asqtad-improved staggered fermions.
The energy-eigenvalues of $\pi K$ systems at one center of mass frame and six moving frames
using moving wall source technique are used to get phase shifts by L\"uscher's formula and its extensions.
The calculations are good enough to acquire effective range expansion parameters:
scattering length $a$, effective range $r$, and shape parameter $P$, which are in good agreement
with our explicit analytical  predictions in three-flavor chiral perturbation theory at next-to-leading order.
All results are fairly consistent with experimental measurements, phenomenological studies, and lattice estimations.
Numerical computations are implemented at a fine ($a\approx0.082$~fm, $L^3 T = 40^3 96$) lattice ensemble
with physical quark masses.
\end{abstract}

\date{\today}
\maketitle

\section{Introduction}
Pion-kaon ($\pi K$) scattering at low energies is simplest reaction involving strangeness,
and allows for an explicit detection of three-flavor structure of hadronic interaction,
which is not directly probed in $\pi\pi$ scattering~\cite{Weinberg:1966kf,Gasser:1983yg,Bijnens:1997vq,Colangelo:2001df}.
Thus, the measurement of $\pi K$ scattering phase shift is one of the cleanest processes and a decisive test for the comprehension of the chiral $\rm SU(3)$ symmetry breaking of quantum chromodynamics (QCD).

The use of this symmetry and effective lagrangian is known as chiral perturbation theory ($\chi$PT),
which was introduced by Weinberg~\cite{Weinberg:1978kz},
and systematized by Gasser and Leutwyler~\cite{Gasser:1983yg} for light $u/d$ quark case,
along with its extension to the expansion in strange quark mass~\cite{Gasser:1984gg}.

Earliest work on $\pi K$ scattering is accomplished by Griffith~\cite{Griffith:1968jaz},
and the establishment of $\chi$PT gave rise to predict $\pi K$ scattering amplitude
at leading order (LO) and next-to-leading order (NLO)~\cite{Bernard:1990kw,Bernard:1990kx},
which involve the computable non-analytical contributions and analytical terms with low-energy constants (LEC's)~\cite{Bernard:1990kx,Bernard:1990kw,Dobado:1996ps,SaBorges:1994yy,Roessl:1999iu,Amoros:1999dp,Kubis:2001bx,Nehme:2001wa,
Ananthanarayan:2000cp,GomezNicola:2001as,Nehme:2001wf,Buettiker:2003pp,Bijnens:2004bu,Zhou:2006wm,
Descotes-Genon:2007sqh,Bijnens:2014lea,Pelaez:2016tgi,Pelaez:2020gnd,Cao:2025hqm}.

Experimentally, $\pi K$ scattering length is gained by the scattering phases through Roy-Stainer equations~\cite{Dumbrajs:1983jd,Lang:1976ze,Johannesson:1974ma,Matison:1974sm,Karabarbounis:1980bk,DIRAC:2017hmz}.
It is worth emphasizing that the DIRAC experiment at CERN has
produced the constraints on $\pi K$ scattering lengths using $\pi K$ atoms~\cite{DIRAC:2017hmz}.

The $\pi K$ scattering has been investigated by different lattice groups~\cite{Beane:2006gj,Chen:2006wf,Nagata:2008wk,Sasaki:2010zz,Fu:2011wc,Fu:2011xw,Lang:2012sv,Prelovsek:2013ela,Sasaki:2013vxa,Dudek:2014qha,Wilson:2014cna,Janowski:2014uda,Shepherd:2016dni,Brett:2018jqw,Helmes:2018nug,Wilson:2019wfr, Rendon:2020rtw}.
The first fully-dynamical lattice calculation~\cite{Beane:2006gj}
was carried out  with Asqtad-improved staggered  quark~\cite{Golterman:1985dz,Kaplan:1992bt,FermilabLattice:2010rur}.
Nagata {\it et al.} found that scattering amplitudes can be written as
the combinations of only three diagrams in the isospin limit~\cite{Nagata:2008wk}.

At present, lattice QCD can calculate LEC's values in $I=\frac{3}{2}$ $\pi K$ scattering with robust statistics.
Using LEC's obtained at the nonphysical points,
threshold parameters and effective range expansion (ERE) parameters
at the physical point can be obtained in $\chi$PT~\cite{NPLQCD:2011htk,Fu:2017apw}.

Since $I=\frac{3}{2}$ $\pi K$ scattering is
technically the simplest lattice study of reactions involving strangeness
due to the absence of $\kappa$ resonance, its phase shifts can be precisely gotten,
and it is most-affordable-option 
to look into the validity and scope of relevant NLO expressions.

This work is definitely inspired by Roessl's enlightening work~\cite{Roessl:1999iu},
where $\pi K$ scattering is analysed in the context of $SU(2)$ $\chi$PT to
provide the explicit analytical expression of three threshold parameters ($a$, $b$, $c$),
which, in principle, can be used to guide lattice calculations.
Note that threshold parameters are also offered by
novel analysis from Roy and Steiner type equations~\cite{Buettiker:2003pp}.
Note that the chiral expansion of $SU(2)$ $\chi$PT is expected to converge more rapidly than
in the $SU(3)$ theory~\cite{Roessl:1999iu}.
According to the discussions in Ref.~\cite{Bijnens:2004bu},
our concerned threshold parameters in this work display indication of reasonable convergence, and
acceptable agreement with the dispersive result from Ref.~\cite{Buettiker:2003pp}.

Since $\pi K$ scattering is a reaction with strangeness,
it is highly desirable for one to study in $SU(3)$ $\chi$PT,
where strange quark mass is treated as an expansion parameter.
On the other hand, it is fixed and contained in low-energy constants in $\chi SU(2)$ case~\cite{Roessl:1999iu}.
Hence, analytical expressions
for threshold parameters in $\chi SU(3)$ case are more voluminous than
those in $\chi SU(2)$~\cite{Roessl:1999iu}, as shown later.
As another result, the relevant results for threshold parameters are mainly given numerically,
only scattering length is expressed in analytic formulae~\cite{Kubis:2001bx},
which, along with its variants~\cite{Chen:2006wf,Fu:2011wc,Helmes:2018nug}
were widely used to tackle the relevant lattice data~\cite{Beane:2006gj,Chen:2006wf,Nagata:2008wk,Sasaki:2010zz,Fu:2011wc,Fu:2011xw,Lang:2012sv,Prelovsek:2013ela,Sasaki:2013vxa,Dudek:2014qha,Wilson:2014cna,Janowski:2014uda,Shepherd:2016dni,Brett:2018jqw,Helmes:2018nug,Wilson:2019wfr, Rendon:2020rtw}.
Unfortunately, to the best of our knowledge,
no alike analytical $\chi SU(3)$  expressions are described for two threshold parameters ($b$ and $c$).

\newpage

Note that ERE parameters ($a$, $r$, $P$) are related to range,
depth and detailed shape of nuclear potential~\cite{Blatt:1949zz}.
Moreover, the effective range $r$ depends not only on the range
but also on the depth of nuclear potential well~\cite{Blatt:1949zz}.
The shape term  $Pk^4$ is first parameter in expansion of phase shift $k\cot\delta$,
which implies a shape of potential~\cite{Blatt:1949zz}.
If scattering experiments or lattice calculations could be made accurate enough to
deduce shape parameter $P$, the shape of nuclear potential can be delimited.
Hence, it is highly desirable for one to extend the work in Ref.~\cite{Kubis:2001bx}
to offer concrete analytical $\chi SU(3)$ expressions for $r$ and $P$,
which are helpful to handle relevant lattice data.

For this aim, we exploringly pick up slope parameters $b$ and $c$ at NLO
on the basis of $\pi K$ scattering amplitude derived from $\rm SU(3)$ $\chi$PT in Ref.~\cite{GomezNicola:2001as},
and rearrange it with NPLQCD's skill~\cite{NPLQCD:2011htk}.
Accordingly, three LEC's are needed to deal with $I=\frac{3}{2} \, \pi K$ scattering at NLO in $\chi$PT.
Note that lattice or experimental data of $\pi K$ phase shifts directly
connect to ERE parameters ($a$, $r$, $P$). In view of this,
useful tools are initiatively provided to directly
connect threshold parameters ($a$, $b$, $c$) with $r$ and $P$ respectively.
It's best to understand that these methods are effective
in the elastic region~\cite{Adhikari:1983ii,Beane:2003da}.

In principle, one can opt any published $\pi K$ scattering amplitude
to do this quite laborious and meaningful task.
Admittedly, our derivation is definitely relied on one loop $\pi K$
amplitude in Ref.~\cite{GomezNicola:2001as}, since it meets the exact perturbative unitarity
when expressed in terms of physical constants~\cite{GomezNicola:2001as}.
Most of all, it was delivered just in terms of pion decay constant $f_\pi$, which is important for
lattice study due to expensive computation of kaon decay constant $f_K$.
As a double-check, our relevant results are in astonishingly good agreement
with those in Ref.~\cite{Bijnens:2004bu}, since its amplitude also obeys the relations
from the crossing symmetry and unitarity~\cite{Bijnens:2004bu}.

Note that $s$-wave $\pi K$ scattering lengths ($a^+, a^-$) are offered just in terms of
$f_\pi$~\cite{Kubis:2001bx} based on their scattering amplitudes in Ref.~\cite{Bernard:1990kx}.
As shown later, our expression for  $I=\frac{3}{2}$ $s$-wave $\pi K$ scattering length
is certainly in well agreement with it.
Moreover, in addition to $b$ and $c$, analytical expressions for $I=\frac{3}{2}$ $p$-wave and $d$-wave
$\pi K$ scattering length are also supplied.

From our phenomenological prediction in $\rm SU(3)$ $\chi$PT,
and using mesonic low-energy constants $L_i(i=1,\cdots,8)$ from BE14 in Ref.~\cite{Bijnens:2014lea},
ERE parameters ($a$, $r$ and $P$) at the physical point yield
\begin{eqnarray}
m_\pi a &=& -0.0595(62),  \nonumber \\[1.5mm]
m_\pi r &=&  16.92(4.01), \nonumber \\[1.5mm]
P       &=& -8.76(1.91),  \nonumber
\end{eqnarray}

\noindent which are in fair agreement with other phenomenological determinations in Refs.~\cite{Bernard:1990kx,Dobado:1996ps,Roessl:1999iu,Buettiker:2003pp,Pelaez:2016tgi,Pelaez:2020gnd}.

To numerically verify relevant derived expressions, one MILC fine ($a\approx0.082$~fm, $L^3 T = 40^3 96$)
lattice ensemble with $N_f=3$ flavors of Asqtad-improved staggered
dynamical quarks~\cite{Golterman:1985dz,Kaplan:1992bt,FermilabLattice:2010rur}
is used to compute $I=\frac{3}{2}$ $\pi K$ scattering,
where L\"uscher's technique and its extensions~\cite{Luscher:1986pf,Luscher:1990ux,Luscher:1990ck,Rummukainen:1995vs,Davoudi:2011md,Doring:2011vk,Gockeler:2012yj,Kim:2005gf,Christ:2005gi,Doring:2012eu,
Fu:2011xz,Leskovec:2012gb} are employed to get scattering phases with lattice-calculated energy eigenstates
for total momenta $\mathbf{P}=[0,0,0]$, $[0,0,1]$, $[0,1,1]$, $[1,1,1]$, $[0,0,2]$, $[0,0,3]$, and $[0,0,4]$.
The moving wall source technique is used to
calculate relevant quark-line diagrams~\cite{Sharpe:1992pp,Kuramashi:1993ka}.

Usually, one only uses a point at center of mass frame to evaluate scattering length,
and truncation of the effective range $r$ and higher terms
is typically considered as an important source of systematic error~\cite{Beane:2006gj,Chen:2006wf,Nagata:2008wk,Sasaki:2010zz,Fu:2011wc,Fu:2011xw,Lang:2012sv,Prelovsek:2013ela,Sasaki:2013vxa,Dudek:2014qha,Wilson:2014cna,Janowski:2014uda,Shepherd:2016dni,Brett:2018jqw,Helmes:2018nug,Wilson:2019wfr, Rendon:2020rtw}.
PACS-CS finds that $I=\frac{3}{2} \, \pi K$ scattering length from computation at near physical point
is slight deviation from its previous work~\cite{Sasaki:2010zz},
and reason is not quite clear~\cite{Sasaki:2013vxa}.
Actually, according to works in Refs.~\cite{NPLQCD:2011htk,Fu:2017apw},
the effective range $r$ is supposed to be larger and larger when approaching the physical point.
Thus, PACS-CS's inspirational work~\cite{Sasaki:2013vxa} not only urges us to
derive the relevant NLO expressions to interpret lattice results,
but also motivates us to study $\pi K$  scattering directly at the physical point,
which is already exploringly investigated by RBC-UKQCD
to avoid a potential error due to chiral extrapolation~\cite{Janowski:2014uda}.

The $\pi K$  correlators are performed with physical quark masses,
i.e., light valence $u$ quark mass $am_u=0.0009004$,
and valence strange quark mass $am_s=0.02468$~\cite{DeTar:2018uko}.
From discussions~\cite{Lepage:1989hd,Fu:2016itp},
if one uses fine gauge configurations, employs lattice ensembles with relatively large spatial dimensions $L$,
and sums correlators over all $96$ time slices,
the signals are anticipated to be significantly improved~\cite{Fu:2016itp}.
It allows us to not only measure scattering length,
but also preliminarily examine the effective range and shape parameter.
A three-parameter fit of the phase shifts gives
\begin{eqnarray}
m_\pi a &=& -0.0588(28),  \nonumber \\[1.5mm]
m_\pi r &=&  21.54(6.90), \nonumber \\[1.5mm]
P       &=& -6.98(3.80), \nonumber
\end{eqnarray}

\noindent which are in reasonable accordance with recent experimental and theoretical
determinations~\cite{Bernard:1990kx,Dobado:1996ps,Roessl:1999iu,Buettiker:2003pp,Pelaez:2016tgi,Pelaez:2020gnd} as well as corresponding lattice calculations available in the literature
~\cite{Beane:2006gj,Chen:2006wf,Nagata:2008wk,Sasaki:2010zz,Fu:2011wc,Fu:2011xw,Lang:2012sv,Prelovsek:2013ela,Sasaki:2013vxa,Dudek:2014qha,Wilson:2014cna,Janowski:2014uda,Shepherd:2016dni,Brett:2018jqw,Helmes:2018nug,Wilson:2019wfr, Rendon:2020rtw}.

This paper is organized as follows.
L\"uscher's method and lattice scheme are discussed in Sec.~\ref{sec:Methods}.
Lattice results are given in Sec.~\ref{sec:piKscattering},
along with relational fits, which are used to gain threshold parameters.
A derivation of the relevant $\chi$PT formulas at NLO
and its numerical analysis are presented in Sec.~\ref{sec:chiPT_PK}.
A simple summary and discussion are shown in Sec.~\ref{sec:conclude}.
The derivation of the scattering amplitude at NLO in $\chi$PT is courteously dedicated to the Appendix~\ref{app:ChPT PK_NLO}.
Near threshold behavior of the $s$-wave amplitude is discussed in Appendix~\ref{app:ChPT PK_PWA},
and that of $p$-wave and $d$-wave amplitudes are treated in Appendix~\ref{app:ChPT PK_PD_AB}.

\newpage
\section{Method of measurement}
\label{sec:Methods}
In the current study, we will measure the $s$-wave $\pi K$ system with the isospin representation of $(I, I_z)=\left(\frac{3}{2},\frac{3}{2}\right)$.
The calculations are carried out for total
momenta $\mathbf{P}=[0,0,0]$, $[0,0,1]$, $[0,1,1]$, $[1,1,1]$, $[0,0,2]$, $[0,0,3]$, and $[0,0,4]$,
which are written in units of $2\pi/L$.

\subsection{Center of mass frame}
In the center-of-mass (CoM) frame, the energy eigenvalues of
non-interacting $\pi K$ system read
$$
E = \sqrt{m_\pi^2+ |{\mathbf p}|^2} + \sqrt{m_K^2 + |{\mathbf p}|^2}  ,
$$
where $m_\pi$ is pion mass, $m_K$ is kaon mass, and ${\mathbf p}=\frac{2\pi}{L}{\mathbf n}$,
${\mathbf n}\in \mathbb{Z}^3$.
The lowest energy for ${\mathbf n} \ne 0$
is beyond $t$-channel cut, which approximately starts at ${k^2}={m_\pi^2}$~\cite{Raposo:2023nex}.

The energy eigenstates of $\pi K$ system are displaced
by the hadronic interaction from $E$ to $\overline{E}$,
$$
\overline{E} = \sqrt{m_\pi^2 + k^2} + \sqrt{m_K^2 + k^2} ,
\quad k=\frac{2\pi}{L}q ,
$$
where the dimensionless scattering momentum $q \in \mathbb{R}$.
Solving it for the scattering momentum $k$, one gets
\begin{equation}
\label{eq:MF_k_e}
k^2  = \frac{1}{4}\left( \overline{E} + \frac{m_\pi^2 - m_K^2}{\overline{E}} \right)^2 - m_\pi^2  ,
\end{equation}
which is handily used to calculate $k$~\cite{Fu:2011xw}.

The $s$-wave $\pi K$ scattering phase $\delta_0$ is linked to the energy $\overline{E}$
by the L\"uscher formula~\cite{Luscher:1990ux,Luscher:1990ck,Luscher:1986pf},
\begin{equation}
\label{eq:CMF}
k \cot\delta(k)=\frac{2}{L\sqrt{\pi}} {\mathcal{Z}_{00}(1;q^2)} ,
\end{equation}
where the zeta function is formally defined by
\begin{equation}
\label{eq:Zeta00_CM}
\mathcal{Z}_{00}(s;q^2)=\frac{1}{\sqrt{4\pi}}
\sum_{{\mathbf n}\in\mathbb{Z}^3} \frac{1}{\left(|{\mathbf n}|^2-q^2\right)^s} ,
\end{equation}
which can be usually evaluated by the way described in Ref.~\cite{Yamazaki:2004qb}.
As a consistency check, we exploit the L\"uscher method~\cite{Luscher:1990ux,Luscher:1990ck,Luscher:1986pf}
to get scattering length~\cite{Beane:2006gj,Nagata:2008wk}.
Both methods are found to arrive at the consistent results.

\subsection{Moving frame}
Using a moving frame with non-zero total momentum ${\mathbf P}=\frac{2\pi}{L}{\mathbf d}$,
${\mathbf d}\in\mathbb{Z}^3$,
energy level of free pion and kaon is
$$
E_{MF} = \sqrt{m_\pi^2+|{\mathbf p}_1|^2} + \sqrt{m_K^2+ |{\mathbf p}_2|^2} ,
$$
where ${\mathbf p}_1$ and ${\mathbf p}_2$ are three-momenta of pion and kaon, respectively,
which meet periodic boundary condition,
${\mathbf p}_1=\frac{2\pi}{L}{\mathbf n}_1$, ${\mathbf p}_2=\frac{2\pi}{L}{\mathbf n}_2$,
${\mathbf n}_1,{\mathbf n}_2\in \mathbb{Z}^3$,
and total momentum ${\mathbf P}$ satisfies
$
{\mathbf P} = {\mathbf p}_1 + {\mathbf p}_2 .
$

In the existence of hadronic interaction between pion and kaon,
$E_{MF}$ is shifted to
$$
E_{CM} = \sqrt{m_\pi^2 + k^2} + \sqrt{m_K^2 + k^2},
\quad k = \frac{2\pi}{L} q  ,
$$
where $q\in\mathbb{R},\,k=|{\mathbf p}|$,  CoM momentum ${\mathbf p}$ is quantized to the values
$
{\mathbf p} =\frac{2\pi}{L}{\mathbf r},\,{\mathbf r} \in P_{\mathbf d} ,
$
and the set $P_{\mathbf d}$  is~\cite{Fu:2011xw}
\begin{equation}
\label{eq:set_Pd_MF}
P_{\mathbf d} = \left\{ {\mathbf r} \left|  {\mathbf r} = \vec{\gamma}^{-1}
\left[{\mathbf n}+\frac{{\mathbf d}}{2}
\left(1+\frac{m_K^2\hspace{-0.05cm}-\hspace{-0.05cm}m_\pi^2}{E_{CM}^2}\right)
\right], \right.  {\mathbf n}\in\mathbb{Z}^3 \right\} , \nonumber
\end{equation}
where $\vec{\gamma}^{-1}$ is inverse Lorentz transformation
operating in direction of  CoM velocity ${\mathbf v}$,
$
\vec{\gamma}^{-1}{\mathbf p} =
\gamma^{-1}{\mathbf p}_{\parallel}+{\mathbf p}_{\perp} ,
$
where ${\mathbf p}_{\parallel}$ and ${\mathbf p}_{\perp}$ are
the components of ${\mathbf p}$ parallel
and perpendicular to ${\mathbf v}$, respectively~\cite{Rummukainen:1995vs}.
The energy $E_{CM}$ is linked to $E_{MF}$  via $E_{CM}^2 = E_{MF}^2-{\mathbf P}^2.$

To get more eigenenergies, we implement six moving frame~${\rm MF}i(i=1-6)$,
i.e., MF1 is taken with ${\mathbf d}={\mathbf e}_3$,
MF2 with ${\mathbf d}={\mathbf e}_2+{\mathbf e}_3$,
MF3 with ${\mathbf d}={\mathbf e}_1+{\mathbf e}_2 + {\mathbf e}_3$,
MF4 with ${\mathbf d}=2{\mathbf e}_3$,
MF5 with ${\mathbf d}=3{\mathbf e}_3$, and
MF6 with ${\mathbf d}=4{\mathbf e}_3$.
For MF1 and MF4, first excited state is also considered.
The scattering phase shifts can be gained from eigen-energies of $\pi K$ system enveloped in a cubic torus
by L\"uscher technique~\cite{Luscher:1986pf,Luscher:1990ux,Luscher:1990ck}
and its extensions~\cite{Rummukainen:1995vs,Davoudi:2011md,Fu:2011xz,Leskovec:2012gb,Gockeler:2012yj}
\begin{equation}
k \cot\delta(k)  =  \frac{2}{\gamma L \sqrt{\pi}} Z_{00}^{\bf d}(1; q^{2})\,,
\label{eqn:RumGott}
\end{equation}
where boost factor $\gamma=E_{MF}/E_{CM}$.
The evaluation of zeta functions $\mathcal{Z}_{00}^{{\mathbf d}} (1; q^2)$
is discoursed in Ref.~\cite{Fu:2011xw}.

\subsection{ $\pi K$ correlator function}
\label{SubSec:pK4pFunc}
The isospin-$\frac{3}{2}$ $\pi K$ state with momentum ${\mathbf P} = {\mathbf p}_1 + {\mathbf p}_2$
is built with following interpolating operator~\cite{Beane:2006gj,Nagata:2008wk}.
$$
{\cal O}_{\pi K}^{I=\frac{3}{2}} (\mathbf{P},t) = \pi^{+}(\mathbf{p}_1, t) K^{+}(\mathbf{p}_2, t+1)
$$
with interpolating pion and kaon operators denoted by
\begin{eqnarray}
{\pi^+}(t) = -\sum_{\bf{x}} \bar{d}({\bf{x}}, t)\gamma_5 u({\bf{x}},t) \nonumber \\[0.6mm]
{K^+}(t) = \sum_{\bf{x}} \bar{s}({\bf{x}}, t)\gamma_5 u({\bf{x}},t)\,.  \nonumber
\end{eqnarray}

\begin{table*}[th!]
\caption{ \label{tab:MILC_configs}
Simulation parameters of a MILC lattice ensemble.
Lattice dimensions are written in lattice units with spatial ($L$) and temporal ($T$) size.
Column $3$ indicates gauge coupling $\beta$.
The fourth block gives the sea-quark masses of light and strange quark $a m_l^\prime/a m_s^\prime$
compared with physical light and strange quark masses $a m_l/a m_s$.
Columns $6$ and $7$ render pion mass and kaon mass in MeV, respectively.
The ratio $m_\pi/f_\pi$ is given in Column $8$.
The number of gauge configurations and the number of time slices calculated quark propagators are shown in last Column.
}
\begin{ruledtabular}
\begin{tabular}{llllcccll}
$\approx a [\rm fm]$ & $L^3\times T$ &$\beta$ & $a m_l^\prime/a m_s^\prime$ & $a m_l$/$a m_s$ &  $m_\pi({\rm MeV})$ & $m_K({\rm MeV})$ & $m_\pi/f_\pi$  &$N_{\rm cfg} \times N_{\rm slice}^{\pi K}$   \\
\hline
$0.082$  & $40^3\times96$   & $7.08$    & $0.0031/0.031$  & $0.0009004$/$0.02468$ & $139$  & $496$  & $1.07(1)$ & $492\times 96$
\end{tabular}
\end{ruledtabular}
\end{table*}

Using operators $O_\pi(x_1), O_\pi(x_3)$ for pions at points $x_1, x_3$,
and operators $O_K(x_2), O_K(x_4)$ for kaons at points $x_2, x_4$, respectively,
then $\pi K$ four-point functions with total momentum ${\mathbf P}$ can be expressed as
\begin{eqnarray}
\label{EQ:4point_pK_mom}
C_{\pi K}({\mathbf P}, t_4,t_3,t_2,t_1) &=&
\sum_{\mathbf{x}_1,\mathbf{x}_3}\,\,\sum_{\mathbf{x}_2,\mathbf{x}_4}
e^{ i{{\mathbf p}_1} \cdot ({\mathbf{x}}_3 -{\mathbf{x}}_1) } e^{ i{{\mathbf p}_2} \cdot ({\mathbf{x}}_4 -{\mathbf{x}}_2) }
\nonumber \\[0.8mm]
&&\times\langle
{\cal O}_{\pi}({\bf{x}}_4, t_4) {\cal O}_{\pi}({\bf{x}}_3, t_3) \nonumber \\[0.8mm]
&&\times {\cal O}_{\pi}^{\dag}({\bf{x}}_2, t_2) {\cal O}_{\pi}^{\dag}({\bf{x}}_1, t_1) \rangle ,
\end{eqnarray}
where ${\mathbf P} = {\mathbf p}_1 + {\mathbf p}_2$,
and one typically selects $t_1=0$, $t_2=1$, $t_3=t$, and $t_4=t+1$ to hinder color Fierz rearrangement
of quark lines~\cite{Kuramashi:1993ka},
and $t$ stands for time difference, namely, $t\equiv t_3 - t_1$.

In the isospin limit, two quark-line diagrams contribute to
$I=\frac{3}{2}$ $\pi K $ scattering amplitude~\cite{Nagata:2008wk}.
The quark line diagrams contributing to $\pi K$ four-point function
denoted in Eq.~(\ref{EQ:4point_pK_mom})
are displayed in Fig.~1 of the previous work~\cite{Fu:2011wc,Fu:2011xw},
labeling them as direct $(D)$, and cross $(C)$, respectively.
Note that Nagata et al. call $D$ and $C$ as $A$ and $H$, respectively~\cite{Nagata:2008wk}.

The moving wall source technique is usually exploited
to compute quark-line diagrams~\cite{Kuramashi:1993ka}.
In our previous works~\cite{Fu:2011wc,Fu:2011xw},
a detailed procedure is devoted to describe
these diagrams in CoM frame~\cite{Fu:2011wc}
with the light quark propagator $G$~\cite{Kuramashi:1993ka},
and the relevant  representations in the moving frame are analogically offered in Ref.~\cite{Fu:2011xw}.
To deliver them in the generic frame,
we employ an up quark source with $1$, along with
anti-up quark source with $e^{i{\mathbf p}_1 \cdot {\mathbf{x}} }$ and
anti-strange quark source with $e^{i{\mathbf p}_2 \cdot {\mathbf{x}} }$
on each site for  pion and kaon creation operator, respectively, then relevant expressions
can be expressed in terms of the quark propagator $G$
as~\cite{Kuramashi:1993ka,Fu:2011wc,Fu:2011xw}.
\begin{eqnarray}
\label{eq:dcr}
C^D_{\pi K}({\mathbf P},t_4,t_3,t_2,t_1) &=&
\sum_{ \mathbf{x}_3}  e^{ i{{\mathbf p}_1} \cdot {\mathbf{x}}_3 } \sum_{ \mathbf{x}_4} e^{ i{{\mathbf p}_2} \cdot {\mathbf{x}}_4 }
\nonumber \\[0.6mm]
&&\times \langle \mbox{Tr}[G_{t_1}^{\dag}({\mathbf{x}}_3,t_3) G_{t_1}({\mathbf{x}}_3,t_3)] \nonumber \\[0.6mm]
&&\times \mbox{Tr}[G_{t_2}^{\dag}({\mathbf{x}}_4,t_4)G_{t_2}({\mathbf{x}}_4,t_4)] \rangle, \nonumber \\[2.0mm]
C^C_{\pi K}({\mathbf P},t_4,t_3,t_2,t_1) &=&
\sum_{ \mathbf{x}_3} \sum_{ \mathbf{x}_4}
e^{ i{{\mathbf p}_1} \cdot {\mathbf{x}}_3 } e^{ i{{\mathbf p}_2} \cdot {\mathbf{x}}_4 } \nonumber \\[0.6mm]
&&\times \langle \mbox{Tr}
[G_{t_1}^{\dag}({\mathbf{x}}_3,t_3)G_{t_2}({\mathbf{x}}_3,t_3) \nonumber \\[0.6mm]
&&\times G_{t_2}^{\dag}({\mathbf{x}}_4,t_4)G_{t_1}({\mathbf{x}}_4,t_4)] \rangle .
\end{eqnarray}
where ${\mathbf P} = {\mathbf p}_1 + {\mathbf p}_2$.
The combinations of the quark propagators $G$ for quark-line diagrams
that are adopted for $\pi K$ correlation functions
are illustrated in Fig.~3 of Ref.~\cite{Nagata:2008wk} and Fig.~1 in Refs.~\cite{Fu:2011wc,Fu:2011xw}.

The corresponding $I=\frac{3}{2}$  $\pi K$ scattering correlation functions
can be delivered in terms of  just two quark-line diagrams~\cite{Nagata:2008wk},
\begin{eqnarray}
\label{EQ:phy_I12_32}
C_{\pi K }^{I=\frac{3}{2}}(t) &=& D - N_f C ,
\end{eqnarray}
where $N_f$ is the staggered-flavor factor due to the number of tastes
intrinsic to Kogut-Susskind formulation~\cite{Sharpe:1992pp}.

\subsection{Lattice Calculation}
We use the gauge configurations with three Asqtad-improved
staggered sea quarks~\cite{Kaplan:1992bt,Golterman:1985dz,FermilabLattice:2010rur},
and analyze $\pi K$ correlation function on a $0.082$ fm MILC lattice ensemble.
The simulation parameters are listed in Table~\ref{tab:MILC_configs}.~\footnote{
We are indebted to professor Carleton DeTar for kindly providing some MILC lattice ensembles
for our initial lattice study. More lattice gauge configurations can be obtained
by updating the present gauge configurations to create some fresh trajectories or
directly generate new ones with the aid of MILC codes~\cite{MILC:DeTar}.
Most of all, more than twenty years ago, it is Carleton DeTar who ignites the passion
for us to learn $\pi K$ scattering.
This work specifically answers his assignment,
and we here admire his sharp insight of physical essence.
}
Lattice ensemble is gauge fixed to the Coulomb gauge before computing quark propagators.

The moving wall source technique is well established to determine the relevant correlators
with high quality~\cite{Kuramashi:1993ka},
and this approach is extensively exploited to two-particle system
with the arbitrary momenta~\cite{Fu:2011xw,Fu:2012gf,Fu:2011wc,Fu:2017apw}.

The correlators are measured on all $T$ time slices,
namely, the correlator $C_{\pi K}(t)$ is computed by
\begin{eqnarray}
 C_{\pi K}(t) &=&
\frac{1}{T}\sum_{t_s=0}^{T-1} \left\langle
\left(\pi K\right)(t+t_s)\left(\pi K\right)^\dag(t_s)\right\rangle .
\nonumber
\end{eqnarray}
After averaging the correlators over all the $T$ slices,
the statistics are supposed to be substantially improved~\footnote{
For each configuration, we measure $3T=288$ $u/s$ quark propagators,
and each inversion for $u$ quark takes about $7000$ iterations.
Part of calculations are actually done in our previous work~\cite{Fu:2017apw}.
However, computations were steadily carried out about three years.
In practice, all quark propagators are saved, and loaded when needed.
So, for each time-slice estimation,
it averagely expends only a quark propagator computation for each color.
From this perspective, it is actually ``cost-effective''~\cite{Fu:2017apw}.
}.

According to discussions~\cite{Lepage:1989hd,Fu:2016itp},
noise-to-signal ratio of $\pi K$ correlator is enhanced roughly $\propto 1/\sqrt{N_{\rm slice} L^3}$,
where $L$ is lattice spatial dimension, and $N_{\rm slice}$ is
number of time slices calculated propagators.
In this work, we adopt lattice ensembles with relatively large $L=40$.
Hence, the relevant signals are turned out to be indeed good enough.
Admittedly, most efficient approach is to adopt anisotropic gauge configurations~\cite{NPLQCD:2011htk},
where the signals are exponentially improved.
Keep in mind that our signals are semi-empirically anticipated to be
significantly enhanced~\cite{Fu:2016itp}, but algebraically.

We compute two-point pion and kaon correlators with the zero and none-zero
momenta ($\mathbf{0}$ and $\mathbf{p}$) as well,
\begin{eqnarray}
C_\pi({\mathbf 0}, t) &=& \frac{1}{T}\sum_{t_s=0}^{T-1}
\langle 0|\pi^\dag ({\mathbf 0}, t+t_s) W_\pi({\mathbf 0}, t_s) |0\rangle, \label{eq:pi_cor_PW_k000}\\[0.2mm]
C_\pi({\mathbf p}, t) &=& \frac{1}{T}\sum_{t_s=0}^{T-1}
\langle 0|\pi^\dag ({\mathbf p}, t+t_s) W_\pi({\mathbf p}, t_s) |0\rangle, \label{eq:pi_cor_PW_k100}\\[0.2mm]
C_K({\mathbf 0}, t) &=& \frac{1}{T}\sum_{t_s=0}^{T-1}
\langle 0|K^\dag ({\mathbf 0}, t+t_s) W_K({\mathbf 0}, t_s) |0\rangle,     \label{eq:pi_cor_KW_k000}\\[0.2mm]
C_K({\mathbf p}, t) &=& \frac{1}{T}\sum_{t_s=0}^{T-1}
\langle 0|K^\dag ({\mathbf p}, t+t_s) W_K({\mathbf p}, t_s) |0\rangle,     \label{eq:pi_cor_KW_k100}
\end{eqnarray}
where $\pi$, $K$, $W_\pi$ and  $W_K$ are pion and kaon point-source and wall-source operator,
respectively~\cite{FermilabLattice:2010rur},
and summation over lattice space point in sink is not written here for simple notation.

The pion mass $m_\pi$, energy $E_\pi({\mathbf p})$, and kaon mass $m_K$, energy $E_K({\mathbf p})$
can be robustly obtained at large $t$ from Eqs.~(\ref{eq:pi_cor_PW_k000})-(\ref{eq:pi_cor_KW_k100}), respectively~\cite{FermilabLattice:2010rur},
\begin{eqnarray}
\label{eq:pi_fit_PW_k000} \hspace{-0.6cm} C_\pi({\mathbf 0}, t) &=&
A_\pi(\mathbf{0}) \left[e^{-m_\pi t}+e^{-m_\pi(T-t)}\right] +\cdots, \\[0.5mm]
\label{eq:pi_fit_PW_k100}\hspace{-0.6cm} C_\pi({\mathbf p}, t) &=&
A_\pi(\mathbf{p})\left[e^{-E_\pi({\mathbf p}) t}+e^{-E_\pi({\mathbf p})(T-t)}\right] + \cdots,\\[0.5mm]
\label{eq:pi_fit_KW_k000}\hspace{-0.6cm} C_K({\mathbf 0}, t) &=&
A_K(\mathbf{0}) \left[e^{-m_K t}+e^{-m_K(T-t)}\right] +\cdots, \\[0.5mm]
\label{eq:pi_fit_KW_k100} \hspace{-0.6cm} C_K({\mathbf p}, t) &=&
 A_K(\mathbf{p}) \left[e^{-E_K({\mathbf p}) t}+e^{-E_K({\mathbf p})(T-t)}\right] + \cdots,
\end{eqnarray}
where ellipses hint oscillating parity partners, and overlapping amplitudes
$A_\pi(\mathbf{0})$, $A_K(\mathbf{0})$, $A_\pi(\mathbf{p})$ and $A_K(\mathbf{p})$
are used to estimate wrap-around contributions~\cite{Gupta:1993rn,Umeda:2007hy,Nagata:2008wk}.

For moving frame just with ground state, the energy $E_{\pi K}$  is extracted  from  $\pi K$ four-point function~\cite{Golterman:1985dz,DeTar:2014gla}
\begin{eqnarray}
\label{eq:E_pionpion}
\hspace{-0.6cm} C_{\pi K}(t)  &=&
Z_{\pi K}\cosh\left[E_{\pi K}\left(t - \frac{T}{2}\right)\right] \nonumber \\[0.5mm]
&& + (-1)^t Z_{\pi K}^{\prime}\cosh \left[E_{\pi K}^{\prime} \left(t-\frac{T}{2}\right)\right] + \cdots
\end{eqnarray}
for a large $t$ to suppress excited states.
In our practice, the pollution by the ``wraparound'' effects~\cite{Nagata:2008wk,Gupta:1993rn,Umeda:2007hy}
should be considered~\cite{Fu:2012gf}.
In practice, pion decay constants $f_\pi$ can be efficiently estimated by the approach in Ref.~\cite{Beane:2005rj},
which are listed in Table~\ref{tab:MILC_configs} in $m_\pi/f_\pi$.

For MF1 and MF4, ground and first excited states can be separated
with variational method~\cite{Luscher:1990ck}
by calculating a $2 \times 2 $ correlation matrix
$C(t)$ whose components can be estimated by Eq.~(\ref{eq:dcr}).
For this goal, we build a ratio of the correlation function matrices as~\cite{Fu:2011wc,Fu:2011xw}
\begin{equation}
M(t,t_R) = C(t)  \, C^{-1}(t_R)  ,
\label{eq:M_def}
\end{equation}
\noindent with some reference time $t_R$~\cite{Luscher:1990ck}
to extract two lowest energy eigenvalues $\overline{E}_n$ ($n=1,2$),
which can be obtained by a fit to two eigenvalues
$\lambda_n (t,t_R)$ ($n=1,2$) of the correlation matrix $M(t,t_R)$~\cite{Gupta:1993rn,Golterman:1985dz,DeTar:2014gla,Fu:2011xw,Fu:2016itp,Umeda:2007hy},
\begin{eqnarray}
\label{Eq:asy}
\lambda_n (t, t_R) &=&
       A_n \cosh\left[-E_n\left(t-\frac{T}{2}\right)\right] \nonumber \\[0.2mm]
       &&+
(-1)^t B_n \cosh\left[-E_n^{\prime}\left(t-\frac{T}{2}\right)\right] ,
\end{eqnarray}
for a large $t$ to suppress the excited states.

\section{FITTING ANALYSES}
\label{sec:piKscattering}
\subsection{Lattice Phase Shift}
The $\pi K$ four-point functions are calculated with valence quarks at its physical value,
namely, $am_u = 0.0009004$ and $am_s=0.02468$~\cite{DeTar:2018uko}.
Using two-point pion and kaon correlators, we can precisely extract the pion mass
and kaon mass, which are summarized in Table~\ref{tab:m_pi_K}.
\begin{table}[b!]	
\caption{\label{tab:m_pi_K}
Summary of pion mass and kaon mass.
The third and fourth blocks show pion mass and kaon mass in lattice units, respectively.
The first and second blocks give pion mass and kaon mass in GeV.
}
\begin{ruledtabular}
\begin{tabular}{cccc}
$m_\pi({\rm GeV})$ & $m_K({\rm GeV})$ & $a m_\pi$      & $a m_K $   \\[0.5mm]
\hline
\\[-3.5mm]
 $0.139(1)$        & $0.496(4)$        & $0.05786(5)$  & $0.2101(1)$ \\
\end{tabular}
\end{ruledtabular}
\end{table}

In order to intuitively demonstrate lattice calculations,
one ordinarily calculates the ratios~\cite{Kuramashi:1993ka},
\begin{eqnarray}
\label{EQ:ratio}
R^X(t) = \frac{ C_{\pi K}^X(0,1,t,t+1) }{ C_\pi (0,t) C_K(1,t+1) }, \quad  X = D, C,\, {\rm and} \,\,C^W , \nonumber
\end{eqnarray}
\noindent where $C_\pi$ and $C_K$ are pion and kaon correlators with a designated momentum, respectively,
and $C^W$ suggests the ratio of cross amplitude obtained by properly eliminating
relevant wraparound pollution~\cite{Gupta:1993rn,Umeda:2007hy,Nagata:2008wk,Fu:2017apw}.

\begin{figure}[b!]
\includegraphics[width=8.5cm,clip]{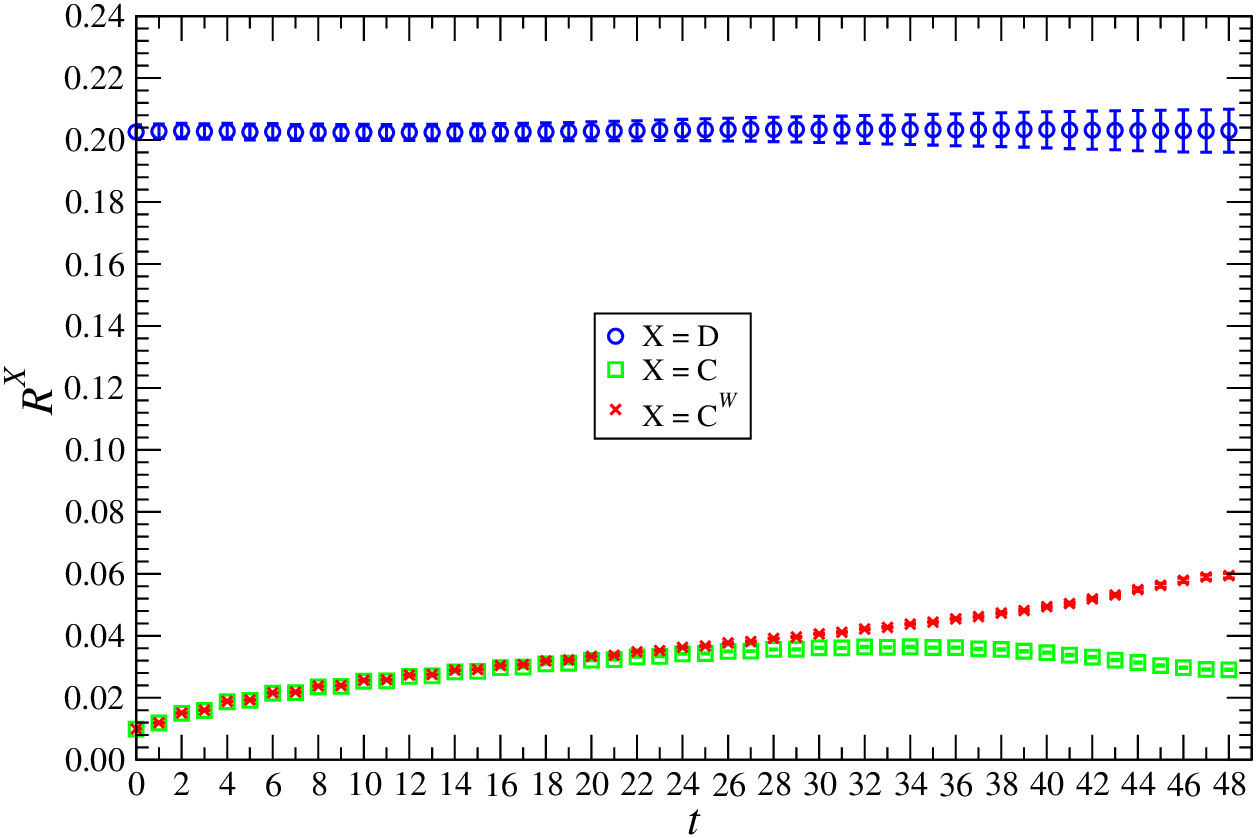}
\caption{\label{fig:rcd_r_000_I2}
Individual amplitude ratios $R^X(t)$ of $\pi K$ correlations calculated by moving wall source technique
at $\mathbf{P}=[0,0,0]$:
direct diagram (blue circle) displaced by $0.8$,
crossed diagram (green octagon), and crossed diagram (red cross)
with consideration of the removal of wraparound pollution.
}
\end{figure}

Two contributions to $\pi K$ correlators at $\mathbf{P}=[0,0,0]$ as the functions of $t$ are displayed
in Fig.~\ref{fig:rcd_r_000_I2}, which are displayed  as individual ratios $R^X, \, X=D,\, C,\, {\rm and} \,\,C^W $.
The ratio values of direct amplitude $R^D$ are quite close to unity,
indicating a pretty weak interaction~\cite{Kuramashi:1993ka,Fu:2012gf}.
Actually, $I=\frac{3}{2}$ $\pi K$  scattering is perturbative
at low momenta and small quark masses, as mandated by $\chi$PT.
Consequently, pion-kaon energies deviate slightly from the noninteracting energies~\cite{Kuramashi:1993ka,Fu:2012gf}.
\begin{table*}[th!]	
\caption{\label{tab:pp_Io_kcotk}
Summaries of lattice results for fitted energies $E_{\pi K}$ of $I=3/2$ $\pi K$ system.
For each total momentum $\mathbf{P}$, energy level, two single-hadron are denoted by $({\mathbf p}_1, {\mathbf p}_2)$
for three-momenta of $\pi$ and $K$, respectively.
The fourth block shows the fitted energies $E_{\pi K}$ in lattice units.
Column five gives the fitting range,
and Column six indicates the number of degrees of freedom (d.o.f) for the fit.
The seven block is the center-of-mass scattering momentum $k^2$ in terms of $m_\pi^2$,
and Column eight gives the values of $k \cot\delta/m_\pi$, which are calculated
by L\"uscher formulae~(\ref{eq:CMF}) or (\ref{eqn:RumGott}).
}
\begin{ruledtabular}
\begin{tabular}{cccccccc}
\\[-4mm]
$\mathbf{P}$   & {\rm Level} & $({\mathbf p}_1, {\mathbf p}_2)$ & $a E_{\pi K}$  &  {\rm Range} & $\chi^2/{\rm dof}$
& $k^2/m_\pi^2$   & $k\cot\delta/m_\pi $   \\[0.3mm]
\hline
\\[-3.5mm]
$[0,0,0]$ &$n=0$ & $([0,0,0],[0,0,0])$ & $0.27108(12)$  & $19-29$ & $8.3/7$   & $0.0675(38)$   & $-16.22(84)$   \\[0.3mm]
$[0,0,1]$ &$n=0$ & $([0,0,0],[0,0,1])$ & $0.32439(12)$  & $16-29$ & $18.9/10$ & $0.3958(44)$   & $-13.90(83)$   \\[0.3mm]
$       $ &$n=1$ & $([0,0,1],[0,0,0])$ & $0.38250(60)$  & $16-29$ & $4.3/10$  & $2.917(26)$    & $-6.85(1.08)$  \\[0.3mm]
$[0,1,1]$ &$n=0$ & $([0,0,0],[0,1,1])$ & $0.36722(19)$  & $14-36$ & $32.1/19$ & $0.6714(67)$   & $-14.30(1.53)$ \\[0.3mm]
$[1,1,1]$ &$n=0$ & $([0,0,0],[1,1,1])$ & $0.40529(29)$  & $14-25$ & $13.2/8$  & $0.9452(111)$  & $-11.61(1.82)$ \\[0.3mm]
$[0,0,2]$ &$n=0$ & $([0,0,1],[0,0,1])$ & $0.43219(29)$  & $16-28$ & $9.2/9$   & $0.8206(112)$  & $-13.71(5.80)$ \\[0.3mm]
          &$n=1$ & $([0,0,0],[0,0,2])$ & $0.43873(40)$  & $11-29$ & $34.6/15$ & $1.1557(163)$  & $-16.60(5.16)$
\\[0.3mm]
$[0,0,3]$ &$n=0$ & $([0,0,1],[0,0,2])$ & $0.54720(61)$  & $11-24$ & $13.8/10$ & $0.2241(263)$  & $-12.76(4.86)$ \\[0.3mm]
$[0,0,4]$ &$n=0$ & $([0,0,1],[0,0,3])$ & $0.68466(116)$ & $10-25$ & $10.4/12$ & $0.0483(592)$  & $-14.38(8.87)$ \\[0.3mm]
\end{tabular}
\end{ruledtabular}
\end{table*}

Lattice-computed correlators are precise enough to enable us to erase wraparound pollution,
which can be nicely estimated from the procedure provided in Ref.~\cite{Nagata:2008wk}.
From Fig.~\ref{fig:rcd_r_000_I2}, the contribution from
finite-$T$ effects is obviously observed as $t$ approaches to $T/2=48$.
Besides, it is pretty impressive to watch that the ratios of  $C^W$ are nearly in a straight line up to $t=48$.
Furthermore, it is interesting to note that the ratio of $C$ is approximately
half to that of $C^W$ as $t$ closes to $T/2$,
as expected from analytical arguments in Ref.~\cite{Nagata:2008wk}.
Above all, wraparound pollution is viewed from Fig.~\ref{fig:rcd_r_000_I2}
about starting as early as at $t\sim20$,
as emphasized in Ref.~\cite{Janowski:2014uda} that the physical point analysis has large ``around-the-world'' effects.

\begin{figure}[b!]
\includegraphics[width=8.5cm,clip]{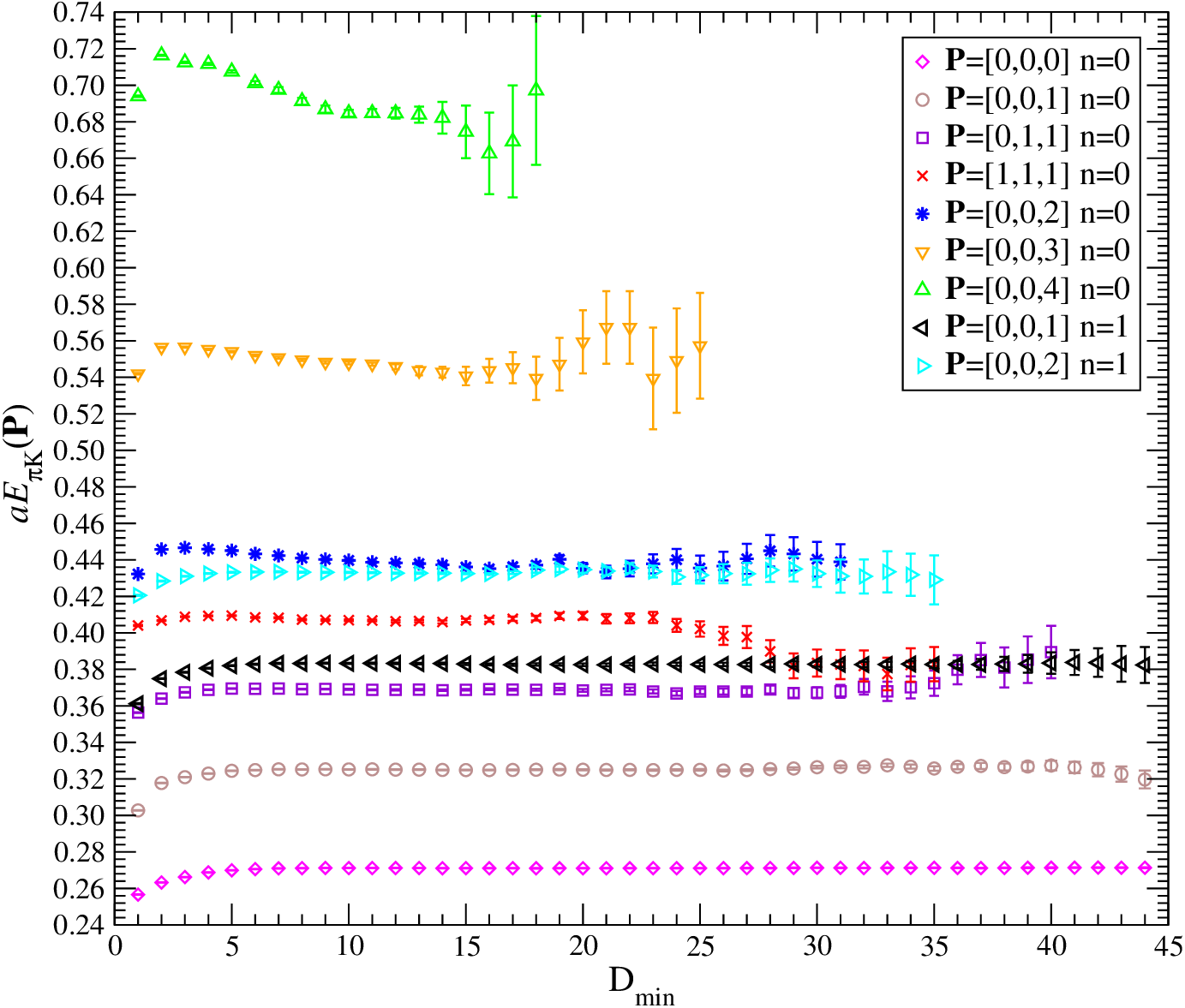}
\caption{\label{fig:eff_eng_I0}
Effective energy $E_{\pi K}$ plots as the functions of $\rm D_{min}$
for $I=\frac{3}{2}$  $\pi K$ scattering in lattice units.
}
\end{figure}

A persuasive way to handle lattice data is
the use of ``effective energy'' plot~\cite{FermilabLattice:2010rur,Fu:2012gf}.
In practice, $I=\frac{3}{2}$ $\pi K$ four-point correlators were fit by
modifying minimum fitting distances $\rm D_{min}$,
and setting maximum distance $\rm D_{max}$ either at $T/2$
or where fractional statistical uncertainties exceed about $20\%$
for two successive time slices~\cite{FermilabLattice:2010rur}.
In our default fits we use Bayesian priors centered around the noninteracting energies
with a prior uncertainty~\cite{DeTar:2018uko}.
The ``effective energy'' plot as the function of $\rm D_{min}$ is shown in Fig.~\ref{fig:eff_eng_I0}.
For $\mathbf{P}=[0,0,0]$, it is pretty delightful that plateau is
obviously watched from $\rm D_{min} = 5 \sim 42$.

The energies $a E_{\pi K}$  are opted from  ``effective energy'' plots,
and selected by comprehensive consideration of a ``plateau'' in the energy $E_{\pi K}$
as the function of  $\rm D_{min}$, a good confidence level,
and $\rm D_{min}$ large enough to effectively restrain excited states.
The fitted values of $a E_{\pi K}$, fit range and quality ($\chi^2/{\rm dof}$)
are tabulated in Table~\ref{tab:pp_Io_kcotk}.
Now it is straightforward to insert $a E_{\pi K}$ into L\"uscher formula~(\ref{eq:CMF})~or~(\ref{eqn:RumGott})
to secure phase shifts of $k \cot \delta/m_\pi$, which are listed in Table~\ref{tab:pp_Io_kcotk},
where statistical errors of $k^2$ are estimated from
statistical errors of $E_{\pi K}$, $m_\pi$ and $m_K$.
It is worth noting that the scattering momenta $k$ are cited in units of
$m_\pi$ in order to enable following analysis
to be independent of scale setting~\cite{Beane:2006gj,NPLQCD:2011htk}.

\newpage
\subsection{The effective range approximation parameters}
\label{se:erap}
As explained latter, the effective range expansion 
is an expansion of real part
of inverse partial wave scattering amplitude in powers of
the magnitude of the CoM three-momentum of pion or kaon,
namely~\cite{KalmahalliGuruswamy:2020uxi,Landau:1991wop,Adhikari:1983ii,Beane:2003da},
\begin{equation}
k\,\cot{\delta} = \frac{1}{a} + \frac{1}{2} r k^2  + P k^4 + {\cal{O}}(k^6) ,
\label{eq:effrange_s}
\end{equation}
where $a$ is scattering length, $r$ represents effective range of hadronic interaction,
and $P$ is shape parameter.
It is worth noting that $P$ is actually corresponding to quantity $T$ denoted in ``nuclear physics"~\cite{Beane:2006gj}
for neutron-proton scattering~\cite{Blatt:1949zz},
which can be written in the form of an integral involving the effective range $r$.
Since the coefficient $T$ has the dimensions of a volume, and determines the curvature of $ k\,\cot{\delta}$,
it is related to the non-dimensional coefficients $P$ as~\cite{Blatt:1949zz},
$$
T = P r^3,
$$
as practised in Ref.~\cite{NPLQCD:2011htk}.
Note that  ``particle physics" definition of $T$ is opposite in sign~\cite{Beane:2006gj}.

As aforementioned, the scattering momenta $k$ are, in practice,
cited in units of $m_\pi$ in order to allow subsequent analysis
to be independent of scale setting~\cite{Beane:2006gj,NPLQCD:2011htk}.
Hence, Eq.~(\ref{eq:effrange_s}) can be recast as
\begin{eqnarray}
\frac{k\cot{\delta}}{m_\pi} &=& \frac{1}{m_\pi a} + \frac{m_\pi r}{2}
\left(\frac{k^2}{m_\pi^2}\right)
+ m_\pi^3 P\left(\frac{k^2}{m_\pi^2}\right)^2
+ \cdots , \nonumber\\
\label{eq:effrange}
\end{eqnarray}
where, in what follows, shape parameter $P$ is scaled by $m_\pi^3$,
for convenience's sake, sometimes $P \equiv m_\pi^3 P $~\cite{Fu:2017apw}.
For simple notation, $a \equiv a_0^\mathrm{3/2}$, similar for $r$, {\it etc.}

\begin{figure}[t!]
\includegraphics[width=8.0cm,clip]{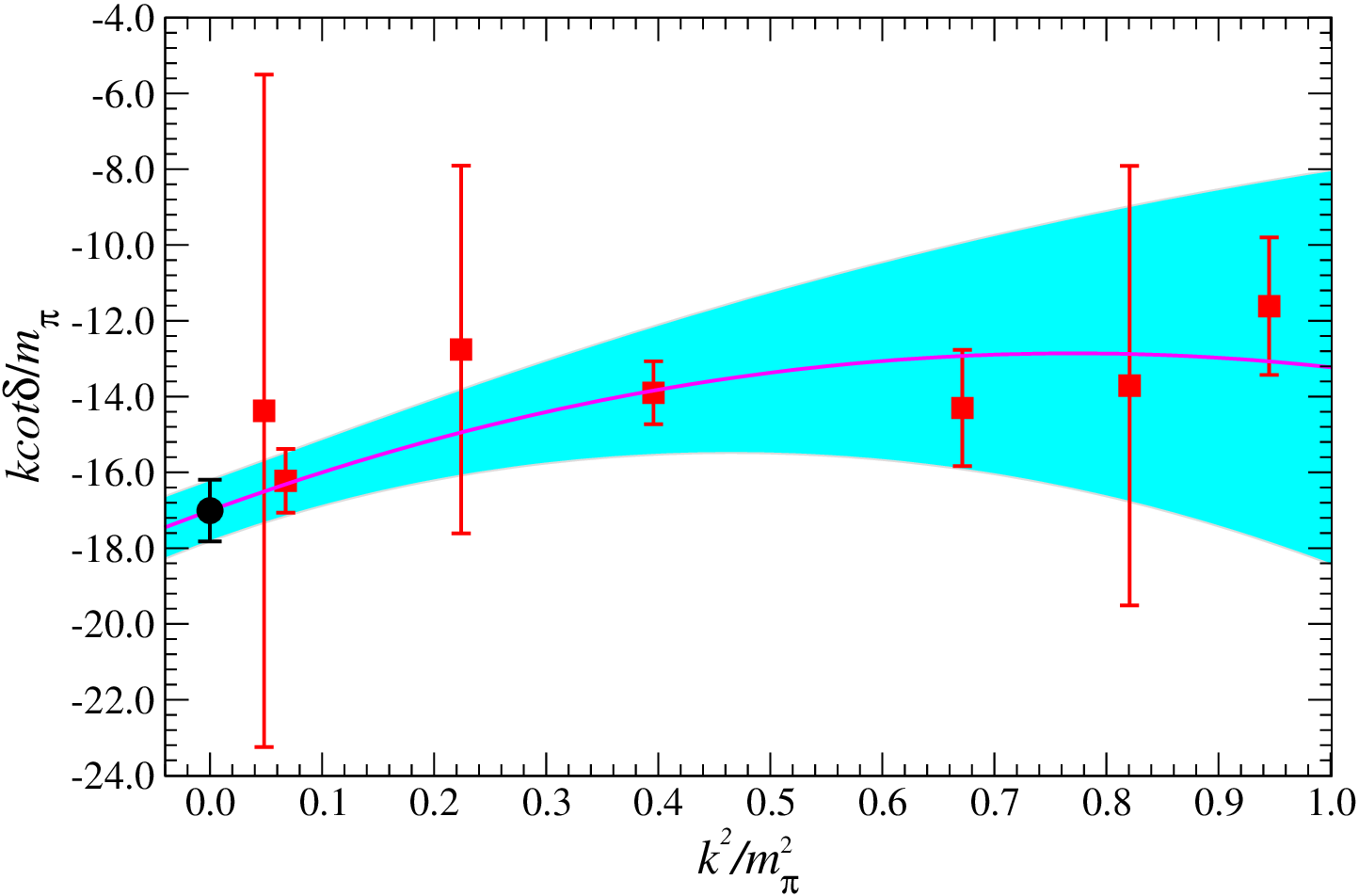}
\caption{
A three-parameter fit to $k\cot\delta/m_\pi$ over region $k^2/m_\pi^2<1.0$.
The shaded cyan band shows statistical error, and
solid magenta curve is central values.
The black circle in this figure indicates fit value of $1/(m_\pi a)$.
}
\label{fig:pipiEREfitA}
\end{figure}

Owing to lack of funding and limited computational resources,
only nine lattice data are at disposal.
Besides, lack of lattice ensembles with different $L$
for a given pion mass is another critical reason~\cite{NPLQCD:2011htk}.
Lattice-determined values of $k \cot{\delta}/m_\pi$ within $t$-channel cut ${k^2}={m_\pi^2}$
are summarized in Table~\ref{tab:pp_Io_kcotk}, and exhibited in Fig.~\ref{fig:pipiEREfitA}

It is interesting to observe that the values of $k\cot\delta/m_\pi$ are
not roughly linear in $k^2$  in the region $k^2/m_\pi^2<1.0$,
which reflects the fact that the shape parameter $P$
indeed has a impact on the curvature~\cite{NPLQCD:2011htk}.
As a matter of fact, according to the quantitatively analytical discussions in Sec.~\ref{sec:chiPT_PK},
the second term and third term in Eq.~(\ref{eq:effrange})
both contribute significantly for values of $k\cot\delta/m_\pi$.

\begin{table}[b!]
\caption{Summaries of the effective range expansion parameters evaluated
from lattice determinations of $k\cot\delta/m_\pi$.
}
\label{tab:fitstoERT}
\begin{ruledtabular}
\begin{tabular}{c  l l}
Quantity  &     Fit   & FitS \\[0.5mm]
\hline
\\[-3.0mm]
$m_\pi a$          & $-0.0588(28)$ & $-0.0589(30)$  \\[0.5mm]
$m_\pi r$          & $21.54(6.90)$ & $25.98(9.69)$  \\[0.5mm]
$P$                & $-6.98(3.80)$ & $-13.08(7.15)$ \\[1.0mm]
$\chi^2/{\rm dof}$ & $1.74/4$      & $0.703/3$      \\[0.5mm]
$(m_\pi a)_{000}$  & $-0.0616(32)$ &                \\[0.5mm]
$R_a$              & $4.85$        &                \\[0.5mm]
\end{tabular}
\end{ruledtabular}
\end{table}

Since lattice measurements indicate that
curvatures have quadratic (and higher) dependence on $k^2$ during the region $k^2/m_\pi^2<1$,
in this work, three leading ERE parameters in Eq.~(\ref{eq:effrange})
are fit to lattice evaluations of $k\cot\delta/m_\pi$.
Fitted values of $m_\pi a$, $m_\pi r$ and $P$ are given in the second column of Table~\ref{tab:fitstoERT}.
A fit of lattice computations is also illustrated in Fig.~\ref{fig:pipiEREfitA},
where the shaded cyan band corresponds to statistical error,
the solid magenta curve is the central values,
and the black circle manifests the corresponding fit value of $1/(m_\pi a)$.

Admittedly, fitted values of effective range $m_\pi r$ and shape parameter $P$ listed in Table~\ref{tab:fitstoERT}
contain roughly reasonable statistical errors,
it is tolerable to use these data to subsequently get threshold parameters ($b$ and $c$).
For more sophisticated lattice study in the future,
it is highly desirable for one to produce more reliable data in the region $k^2/m_\pi^2<1.0$.

The truncation of the effective range $r$ and higher terms
is normally considered as an important source of systematic error~\cite{Beane:2006gj,Chen:2006wf,Nagata:2008wk,Sasaki:2010zz,Fu:2011wc,Fu:2011xw,Lang:2012sv,Prelovsek:2013ela,Sasaki:2013vxa,Dudek:2014qha,Wilson:2014cna,Janowski:2014uda,Shepherd:2016dni,Brett:2018jqw,Helmes:2018nug,Wilson:2019wfr, Rendon:2020rtw}, and as in our former work~\cite{Fu:2011wc},
only one data from center of mass frame ($\mathbf{P}=[0,0,0]$) is used to secure scattering length $(m_\pi a)_{000}$,
which is listed in Table~\ref{tab:fitstoERT},
and in fair agreement with RBC-UKQCD's result~\cite{Janowski:2014uda}.
To make this difference more intuitive, in Table~\ref{tab:fitstoERT},
we also list the relative error~\cite{Fu:2017apw},
$$
R_{a} = \left|\frac{m_\pi a-(m_\pi a)_{000}}{m_\pi a}\right|\times100\%\,.
$$
We find that $R_{a}$ is nontrivial,
which means that this systematic error can not be neglected when approaching physical point.
It should be recognized that this error is well under control when slightly away from physical point
(please consult Sec.~\ref{sec:chiPT_PK}).
We view this as one of most important results in this work.

Fitting all data and then ``pruning'' the highest-$k^2/m_\pi^2$ point
and refitting provides a useful measure of the stability of the extracted
parameters with respect to the upper bound of the fitted momentum range.
Fitted values of $m_\pi a$, $m_\pi r$ and $P$ are given in last column of Table~\ref{tab:fitstoERT}.
The shape parameter $P$ is found to be sensitive to it.
To enhance its stability, it is highly desirable to produce more data points
in the region $0.5 <k^2/m_\pi^2<1.0$.

\section{$\pi K$ scattering near the threshold in chiral $SU(3)$ perturbation theory}
\label{sec:chiPT_PK}
The low-energy theorems imposed by chiral symmetry indicate that
each scattering parameter can be related to relevant LEC's in $\chi$PT,
as illustrated by NPLQCD~\cite{NPLQCD:2011htk}.
Here we offer NLO $\chi$PT expressions of $I=\frac{3}{2}$ $\pi K$ scattering
on the basis of scattering amplitude in Ref.~\cite{GomezNicola:2001as}.

\subsection{Threshold parameters in $\chi$PT}
\label{sec:TPCHIPT}
In the elastic region, isospin-$\frac{3}{2}$  $\pi K$ $s$-wave scattering amplitude
$t(s)\equiv t_{\ell=0}^{I=\frac{3}{2}}(s)$ can be written in terms of the phase-shift $\delta$ as~\cite{Bernard:1990kw,Kubis:2001bx}
\begin{eqnarray}
t(s)= \frac{\sqrt{s}}{2k}\frac{1}{2i} \left[ e^{2i\delta(s)}-1\right],
\label{eq:unitary}
\end{eqnarray}
where $s=\left(\sqrt{m_\pi^2 + k^2} + \sqrt{m_K^2 + k^2}\right)^2$
is Mandelstam variable in the $s$-channel,
and $k$ is the magnitude of the center-of-mass three-momentum.

In Appendix~\ref{app:ChPT PK_NLO}, the detailed procedures are courteously provided to expand all terms of
the NLO isospin-$\frac{3}{2}$ $\pi K$ scattering amplitude in powers of $k^2$,
as a consequence, the Taylor series of the $\pi K$ scattering amplitude in $k^2$ has a pretty concise form,
\begin{widetext}
\vspace{-0.5cm}
\begin{eqnarray}
t(k)&=&
-\frac{m_K m_\pi}{8\pi f_\pi^2}-\frac{\left(m_K+m_\pi\right)^2}{16\pi m_K m_\pi f_\pi^2}k^2
+ \frac{\left(m_K^2-m_\pi^2\right)^2}{64\pi  m_K^3 m_\pi^3 f_\pi^2}k^4 \nonumber\\[1.5mm]
&&+\frac{m_K^3m_\pi^3}{8\pi\left(m_K+m_\pi\right)^2f_\pi^4}
\left[32\frac{\left(m_K + m_\pi\right)^2 }{m_K m_\pi}\left(2L_1 + 2L_2 + L_3 - 2 L_4 - \frac{L_5}{2} + 2L_6 + L_8\right)
- 16\frac{\left(m_K + m_\pi\right)^2 }{m_K^2}L_5\right]  \nonumber\\[1.5mm]
&&+\frac{m_K m_\pi}{16\pi f_\pi^4}k^2\left[64\frac{m_K^2+m_\pi^2}{m_K m_\pi}\left(2 L_1 + 2L_2 + L_3 - L_4\right)
+ 128 L_2 - 16\frac{\left(m_K + m_\pi\right)^2 }{m_K^2} L_5\right]\nonumber\\[1.5mm]
&&+\frac{\left(m_K - m_\pi\right)^2}{64\pi m_K m_\pi f_\pi^4}k^4
\left[\frac{1024}{3}\frac{m_K m_\pi}{\left(m_K - m_\pi\right)^2}(2L_1 + L_2 + L_3)
+ 256\frac{\left(m_K + m_\pi\right)^2}
{\left(m_K - m_\pi\right)^2}L_2 + 16\frac{\left(m_K + m_\pi\right)^2 }{m_K^2} L_5\right]\nonumber\\[1.5mm]
&&+\frac{m_K^3 m_\pi^3}{128\pi^3\left(m_K+m_\pi\right)^2 f_\pi^4}\left[A_0 + A_\pi^0\ln\frac{m_\pi^2}{\mu^2}
 +A_K^0\ln\frac{m_K^2}{\mu^2} + A_\eta^0\ln\frac{m_\eta^2}{\mu^2} + A_T^0\, t(m_\pi,m_K)\right]\nonumber\\[1.5mm]
&&+\frac{m_K m_\pi}{256  \pi^3 f_\pi^4}k^2\left[ B^0 + B_\pi^0\ln\frac{m_\pi^2}{\mu^2}
 + B_K^0\ln\frac{m_K^2}{\mu^2} + B_\eta^0\ln\frac{m_\eta^2}{\mu^2}+ B_T^0\, t(m_\pi,m_K)\right] \nonumber\\[1.5mm]
&&+\frac{\left(m_K-m_\pi\right)^2}{1024 \pi^3 m_K m_\pi f_\pi^4}k^4\left[
C^0 + C_\pi^0 \ln\frac{m_\pi^2}{\mu^2} + C_K^0\ln\frac{m_K^2}{\mu^2} + C_\eta^0 \ln\frac{m_\eta^2}{\mu^2}
+ C_T^0\, t(m_\pi,m_K)\right],
\label{eq:t0_3_2}
\end{eqnarray}
where the low-energy constants $L_i(i=1,\cdots,8)$ are denoted in $\chi$PT of chiral order $4$
by Gasser and Leutwyler~\cite{Gasser:1984gg}, $\mu$ is the chiral scale,
the function $t(m_\pi, m_K)$ is denoted as~\cite{Sasaki:2013vxa}
\begin{eqnarray}
t(m_\pi, m_K) &=& \frac{\sqrt{2m_K^2 + m_\pi m_K - m_\pi^2}}{m_K - m_\pi}
\arctan\hspace{-0.1cm}\left[\frac{2(m_K-m_\pi)}{m_K+2m_\pi}\sqrt{\frac{m_K + m_\pi}{2m_K-m_\pi}}\right],
\label{eq:t_tan}
\end{eqnarray}
and the constants $A^0$, $A_\pi^0$, $A_K^0$, $A_\eta^0$, $A_T^0$; $B^0$, $B_\pi^0$, $B_K^0$, $B_\eta^0$, $B_T^0$,
and $C^0$, $C_\pi^0$, $C_K^0$, $C_\eta^0$, $C_T^0$ are coefficients denoted as
\begin{eqnarray}
A^0     &=& \frac{86\left(m_K+m_\pi\right)^2}{9m_K m_\pi} \nonumber \\[1.5mm]
A_\pi^0 &=& \frac{\left(m_\pi+m_K\right)\left(8m_K^2+11m_Km_\pi-5m_\pi^2\right)}
            {2m_K^2\left(m_K-m_\pi\right)}\nonumber\\[1.5mm]
A_K^0   &=&-\frac{\left(m_\pi+m_K\right)\left(67 m_K^2 + 23 m_K m_\pi - 8 m_\pi^2\right)}
             {9m_\pi m_K\left(m_K-m_\pi\right)}\nonumber\\[1.5mm]
A_\eta^0&=& \frac{\left(m_\pi+m_K\right)^2\left(24 m_K^2 + 4 m_K m_\pi - 9 m_\pi^2\right)}
             {18m_\pi m_K^2\left(m_K-m_\pi\right)} \nonumber\\[1.5mm]
A_T^0   &=&-\frac{16\left(m_K+m_\pi\right)^2}{9m_K m_\pi} \nonumber\\[1.5mm]
B^0      &=& \frac{2\left(336 m_K^6 - 52 m_K^5 m_\pi + 582 m_K^4 m_\pi^2 + 149 m_K^3 m_\pi^3 +172 m_K^2 m_\pi^4 - 35 m_K m_\pi^5 - 84 m_\pi^6\right)}{9m_K m_\pi\left(m_K-m_\pi\right)^2\left(4m_K^2-m_\pi^2\right)}\nonumber\\[1.5mm]
B_{\pi}^0&=&\frac{64 m_K^6+154 m_K^5m_\pi+75 m_K^4 m_\pi^2 + 92 m_K^3 m_\pi^3 + 8 m_K^2 m_\pi^4 + 78 m_K m_\pi^5 - 15 m_\pi^6}{6 m_K^2\left(m_K + m_\pi\right)\left(m_K - m_\pi\right)^3 }\nonumber\\[1.5mm]
B_K^0&=&-\frac{402 m_K^6 - 11 m_K^5 m_\pi + 592 m_K^4 m_\pi^2 + 516 m_K^3 m_\pi^3 + 682 m_K^2 m_\pi^4 + 259 m_K m_\pi^5 - 48 m_\pi^6}{27m_K m_\pi\left(m_K + m_\pi\right)\left(m_K - m_\pi\right)^3} \nonumber\\[1.5mm]
B_{\eta}^0 &=& \frac{48 m_K^6+380 m_K^5 m_\pi-42 m_K^4 m_\pi^2-87 m_K^3 m_\pi^3+83 m_K^2 m_\pi^4 - 15 m_K m_\pi^5 - 27 m_\pi^6}
{54m_\pi m_K^2\left(m_K -m_\pi\right)^3 } \nonumber\\[1.5mm]
B_T^0 &=&-\frac{4\left(24 m_K^5 - 17 m_K^4 m_\pi - 18 m_K^3 m_\pi^2 +39 m_K^2 m_\pi^3
+ 22 m_K m_\pi^4 + 24 m_\pi^5\right)}
{27m_K m_\pi\left(m_K + m_\pi\right)\left(m_K - m_\pi\right)^2}\nonumber\\[1.5mm]
C^0 &=& \frac{2}{135\left(m_K - m_\pi\right)^6\left(m_K + m_\pi\right)}
\bigg[ 6174 m_K^7 + 13603 m_K^6 m_\pi + 49840 m_K^5 m_\pi^2 + 41752 m_K^4 m_\pi^3
     + 41062 m_K^3 m_\pi^4  \nonumber\\[0.6mm]
&&+ 53383 m_K^2 m_\pi^5 + 18604 m_K m_\pi^6 + 7662 m_\pi^7 \bigg]
+\frac{16\left(56 m_K^2 - 15 m_\pi^2\right)m_K m_\pi^3}
{45\left(m_K - m_\pi\right)^2\left(4 m_K^2 - m_\pi^2\right)^2} \nonumber\\[1.5mm]
%
C_{\pi}^0 &=&\frac{1}{18m_K^2\left(m_K-m_\pi\right)^7}
\bigg[168 m_K^9 + 1336 m_K^8 m_\pi + 5017 m_K^7 m_\pi^2 - 3413 m_K^6 m_\pi^3 + 10517 m_K^5 m_\pi^4 \nonumber\\[0.6mm]
&&- 433 m_K^4 m_\pi^5 + 971 m_K^3 m_\pi^6 + 993 m_K^2 m_\pi^7 - 225 m_K m_\pi^8 + 45 m_\pi^9\bigg] \nonumber\\[1.5mm]
C_K^0&=&-\frac{4137 m_K^7  \hspace{-0.05cm}+\hspace{-0.05cm} 6403 m_K^6 m_\pi
   \hspace{-0.05cm}+\hspace{-0.05cm} 2653 m_K^5 m_\pi^2  \hspace{-0.05cm}+\hspace{-0.05cm} 46111 m_K^4 m_\pi^3
   \hspace{-0.05cm}-\hspace{-0.05cm} 13685 m_K^3 m_\pi^4  \hspace{-0.05cm}+\hspace{-0.05cm} 23353 m_K^2 m_\pi^5
   \hspace{-0.05cm}+\hspace{-0.05cm} 4783 m_K m_\pi^6
   \hspace{-0.05cm}+\hspace{-0.05cm} 357 m_\pi^7}{81\left(m_K-m_\pi\right)^7}\nonumber\\[1.5mm]
C_{\eta}^0 &=&\frac{1}{162m_K^2\left(m_K-m_\pi\right)^7}
\bigg[2388 m_K^9 + 4508 m_K^8 m_\pi + 4217 m_K^7 m_\pi^2 - 181 m_K^6 m_\pi^3 + 7901 m_K^5 m_\pi^4 - 4153 m_K^4 m_\pi^5  \nonumber\\[1.0mm]
&&- 469 m_K^3 m_\pi^6 -447 m_K^2 m_\pi^7 - 405 m_K m_\pi^8 + 81 m_\pi^9\bigg] \nonumber\\[1.5mm]
C_T^0 &=& -\frac{1}{81\left(m_K-m_\pi\right)^6\left(m_K+m_\pi\right)^2\left(2m_K-m_\pi\right)}
\bigg[696 m_K^9 - 2122 m_K^8 m_\pi - 135 m_K^7 m_\pi^2 + 7184 m_K^6 m_\pi^3  \nonumber\\[0.6mm]
&&+13154 m_K^5 m_\pi^4 + 4838 m_K^4 m_\pi^5 + 2513 m_K^3 m_\pi^6
+ 4044 m_K^2 m_\pi^7 - 2164 m_K m_\pi^8 - 1176 m_\pi^9\bigg] .
\label{eq:coeff_mpk}
\end{eqnarray}
\end{widetext}

It is worth noting that for loop functions up to ${\cal O}(p^4)$ involving eta mass $m_\eta$,
Gell-Mann-Okubo mass relation $m_\eta^2=\tfrac{4}{3}m_K^2-\tfrac{1}{3}m_\pi^2$
considerably reduces relvant expressions,
as explained and practised in Refs.~\cite{Kubis:2001bx,Sasaki:2013vxa,Amoros:1999dp}.
As a convention, it is convenient to define
$
L_{\pi K}\equiv 2L_1 + 2L_2 + L_3 - 2 L_4 - L_5/2 + 2L_6 + L_8
$~\cite{Gasser:1984gg}.

Note that the constants $A^0$, $A_\pi^0$, $A_K^0$, $A_\eta^0$ and $A_T^0$ are related with scattering length $a$.
The $B^0$, $B_\pi^0$, $B_K^0$, $B_\eta^0$, $B_T^0$, and $C^0$, $C_\pi^0$, $C_K^0$, $C_\eta^0$, $C_T^0$
are linked with slope parameters $b$ and $c$, respectively.
For the notational compactness, the superscript $0$, $1$, $2$ in relevant physical quantities or constants
just indicate them for the $s$-wave, $p$-wave and $d$-wave, respectively.
The subscript $\pi$, $K$, $\eta$ suggest the corresponding chiral logarithm terms, respectively,
and $T$ hints $\arctan$ term.

It is important and interesting to note that all coefficients denoted in Eq.~(\ref{eq:coeff_mpk}) are dimensionless,
and just rely on  pion mass $m_\pi$ and kaon mass $m_K$ due to the usage of Gell-Mann-Okubo relation,
and $A_\eta^0$ is relevant expression of $\kappa_\eta$ in Eq.~(34) of Ref.~\cite{Fu:2011wc}
or Eq.~(A21) of Ref.~\cite{Helmes:2018nug}
owing to the equality
\begin{eqnarray}
\left(m_K + m_\pi\right)\left(24 m_K^2 + 4 m_K m_\pi - 9 m_\pi^2\right) &=&
\nonumber \\[0.6mm]
&&\hspace{-4.35cm}24 m_K^3 - 5 m_K m_\pi^2 + 28 m_K^2m_\pi - 9 m_\pi^3. \nonumber
\end{eqnarray}

For $\pi K$ scattering, it is traditional that near threshold behavior of 
$t(k)$ can be expanded in a Taylor series
\begin{eqnarray}
\mbox{Re}\; t(k) = \frac{\sqrt{s}}{2}\left(a +k^2 b + k^4 c + {\cal O}(k^6)\right),
\label{eq:threshexp_s}
\end{eqnarray}
where $a$ is usually referred to as the scattering length, $b$ and $c$
are often called as slope parameters, respectively.

Matching threshold expansion in Eq.~(\ref{eq:threshexp_s}) to ERE in Eq.~(\ref{eq:effrange_s}),
the effective range $r$ and shape parameter $P$ can be neatly described
just in terms of threshold parameters:
\begin{eqnarray}
r  &=&  - \frac{2b}{a^2} - 2a  \label{eq:m_pi_r_s} \\[1.2mm]
P  &=&  - b - a^3  + \frac{b^2}{a^3} - \frac{c}{a^2} .
\label{eq:m_pi_P_s}
\end{eqnarray}
It is worth emphasizing that Eq.~(\ref{eq:m_pi_r_s}) and Eq.~(\ref{eq:m_pi_P_s})
are friend and clean, which are not relevant to pion and kaon mass, as a consequence, it is convenient to scale them to any physical quantities according to specific needs.
They can be easily used to inversely secure the
slope parameters $b$ and $c$ from lattice-measured ERE parameters.
\begin{eqnarray}
b &=& -\frac{1}{2} a^2 r  - a^3   \label{eq:m_pi_b} \\[1.2mm]
c &=& \frac{1}{4} a^3 r^2 +   \frac{3}{2} a^4 r   - a^2 P + a^5 \label{eq:m_pi_cs}
\end{eqnarray}

It is interesting and noteworthy that, for $\pi\pi$ scattering,
the compact form of the effective range $r$ has  three terms
and that of shape parameter $P$ contains seven separate components,
respectively~\cite{NPLQCD:2011htk,Fu:2017apw}.

For $\pi \pi$ scattering, one usually uses~\cite{NPLQCD:2011htk}
\begin{equation}
\mbox{Re}\; t(k) = m_\pi a + k^2 b + k^4 c + {\cal O}(k^6).
\label{eq:threshexp_NPLQCD_pp}
\end{equation}
Considering $\sqrt{s}/2 = (m_K+m_\pi)/2 + {\cal O}(k^2)$,
near threshold behavior of $t(k)$ for $\pi K$ scattering
can also be expressed as a power-series expansion in $k^2$ as
\begin{equation}
\mbox{Re}\; t(k) = \frac{m_K+m_\pi}{2}a + k^2 b + k^4 c + {\cal O}(k^6),
\label{eq:threshexp_NPLQCD}
\end{equation}
which is usually in favor of NPLQCD's notations~\cite{NPLQCD:2011htk},
and some physical quantities of $\pi K$ scattering can elegantly reduce to
the relevant $\pi\pi$ case~\cite{NPLQCD:2011htk}, as demonstrated later,
which can, on the other hand, be also used to double-check our relevant derivations.

Since near threshold partial $s$-wave amplitude is parameterized
in terms of the scattering length $a$ and slope parameter $b$
by Nehme~\cite{Nehme:2001wa,Nehme:2001wf} as
$\mbox{Re}\; t(k) =  a + k^2 b + {\cal O}(k^4)$,
it is interesting to note that the value of
scattering length $a$ is dimensionless~\cite{Nehme:2001wa,Nehme:2001wf}.

If one uses the threshold expansion in Eq.~(\ref{eq:threshexp_NPLQCD}),
it is pretty straightforward to acquire the relevant expressions for three threshold parameters ($a^\prime$, $b^\prime$ and $c^\prime$), which are courteously dedicated to the Appendix~\ref{app:ChPT PK_PWA}.
Note that the superscript prime in the relevant physical quantities are here just for difference.
It is easy to verify that three threshold parameters ($a^\prime$, $b^\prime$ and $c^\prime$)
are associated with those with definition in Eq.~(\ref{eq:threshexp_s}) through
\begin{eqnarray}
a &=& a^\prime   \\[1.5mm] \label{eq:a_2_a}
b &=& \frac{2}{m_K + m_\pi} b^\prime - \frac{1}{2 m_\pi m_K} a^\prime   \\[1.2mm] \label{eq:b_2_b}
c &=& \frac{2}{m_\pi+m_K}c^\prime - \frac{1}{m_\pi m_K(m_K + m_\pi)} b^\prime \nonumber \\[1.2mm]
&&+ \frac{m_K^2 + m_K m_\pi + m_\pi^2}{8m_\pi^3 m_K^3} a^\prime \label{eq:c_2_c},
\end{eqnarray}
which can be exploited to compare corresponding results.
It is worth emphasizing that both definitions in Eq.~(\ref{eq:threshexp_s}) and Eq.~(\ref{eq:threshexp_NPLQCD})
lead to same forms of scattering length $a$.

Since it is traditional to match threshold expansion to ERE in Eq.~(\ref{eq:effrange_s}),
it is a natural aftermath that the final expressions for the effective range $r$ and
shape parameter $P$ from two definitions are definitely equivalent,
as shown later in Appendix~\ref{app:ChPT PK_PWA}.
Note that lattice data directly connect to the scattering length $a$, effective range $r$,
and shape parameter $P$.
In practice, one can opt any one of them to fit lattice data.
However, it is more convenient to adopt first one to compare the
results of $b$ and $c$ (or $r$ and $P$) to all relevant results accessible in the literature.

For the sake of simplicity, it is handy to follow notation in Ref.~\cite{NPLQCD:2011htk}
and enlightening work in Ref.~\cite{Beane:2006gj} to denote $z\equiv \mu_{\pi K}^2/f_\pi^2$,
where $\mu_{\pi K}=m_\pi m_K/(m_\pi + m_K)$ is the reduced mass of $\pi K$ system.
From the Taylor series of $t(k)$ in powers of $k^2$ in Eq.~(\ref{eq:t0_3_2}) and its
near threshold behavior in Eq.~(\ref{eq:threshexp_s}),
it is pretty straightforward to acquire NLO $\chi$PT expressions for three threshold parameters:
\begin{widetext}
\vspace{-0.3cm}
\begin{eqnarray}
\mu_{\pi K}a    &=& - \frac{z}{4\pi} +
\frac{z^2}{4\pi}\left[C_1 + \frac{1}{16\pi^2}\chi_{a}(\mu)\right]    \label{eq:aThresholds_as} \\[1.2mm]
\mu_{\pi K}^3 b &=& -\frac{m_K^2 + m_K m_\pi + m_\pi^2}{8\pi\left(m_K + m_\pi\right)^2} z +
\frac{ z^2}{8\pi} \left[C_2 + \frac{1}{16\pi^2}\chi_{b}(\mu)\right]  \label{eq:aThresholds_bs}\\[1.2mm]
\mu_{\pi K}^5 c &=& \frac{m_K^4 + m_K^3 m_\pi + m_K^2 m_\pi^2 + m_K m_\pi^3 + m_\pi^4}
{32 \pi\left(m_K + m_\pi\right)^4} z
+ \frac{z^2}{32\pi} \left[C_3 + \frac{1}{16\pi^2}\chi_{c}(\mu)\right]  \label{eq:aThresholds_cs}
\end{eqnarray}
where the constants $C_1$, $C_2$ and $C_3$ are denoted as
\begin{eqnarray}
C_1 &=& 32\frac{\left(m_K + m_\pi\right)^2 }{m_K m_\pi}
\left[2L_1 + 2L_2 + L_3 - 2 L_4 - \frac{L_5}{2} + 2L_6 + L_8\right]
- 16\frac{\left(m_K + m_\pi\right)^2 }{m_K^2}L_5 \label{eq:C123_1} \\[1.2mm]
C_2 &=&64\left(2 L_1 + 2 L_2 + L_3 - L_4\right)\frac{m_K^2 + m_\pi^2}{ m_K m_\pi}
-32\left(2 L_1 - 2 L_2 + L_3 - 2 L_4 + 2 L_6 + L_8\right)
- \frac{16m_\pi(m_\pi + m_K)}{m_K^2}L_5 \label{eq:c23_p2} \\[1.2mm]
C_3 &=& -32\left(6 L_1 - 2 L_2 + 3 L_3 - 2 L_4 - 2 L_6 - L_8\right)
\frac{m_K^2 + m_\pi^2}{\left(m_K + m_\pi\right)^2} \nonumber\\[0.6mm]
&&+\frac{32}{3}\left(70 L_1 +62 L_2 + 35 L_3 - 6 L_4 + 6 L_6 + 3 L_8\right)\frac{m_\pi m_K}{\left(m_K + m_\pi\right)^2}
  +\frac{16 m_\pi^3}{m_K^2\left(m_K + m_\pi\right)}L_5 \label{eq:c23_p3}
\end{eqnarray}
and three known functions
\begin{eqnarray}
\chi_{a}(\mu) &=&A^0 + A_\pi^0\ln\frac{m_\pi^2}{\mu^2}
+A_K^0\ln\frac{m_K^2}{\mu^2} + A_\eta^0\ln\frac{m_\eta^2}{\mu^2} + A_T^0\,t(m_\pi,m_K) \label{eq:chi_a}\\[1.2mm]
\chi_{b}(\mu) &=&b^0 + b_\pi^0\ln\frac{m_\pi^2}{\mu^2}
+ b_K^0\ln\frac{m_K^2}{\mu^2} + b_\eta^0\ln\frac{m_\eta^2}{\mu^2}+ b_T^0\, t(m_\pi,m_K) \label{eq:chi_b}\\[1.2mm]
\chi_{b}(\mu) &=&
c^0 + c_\pi^0 \ln\frac{m_\pi^2}{\mu^2} + c_K^0\ln\frac{m_K^2}{\mu^2} + c_\eta^0 \ln\frac{m_\eta^2}{\mu^2}
+ c_T^0\, t(m_\pi,m_K) \label{eq:chi_c}
\end{eqnarray}
are clearly dependent upon the chiral scale $\mu$ with chiral logarithm terms.
It is evident that the corresponding results of the scattering length $a$ at NLO in Refs.~\cite{Kubis:2001bx,Helmes:2018nug,Fu:2011wc}
are neatly reproduced in Eq.~(\ref{eq:aThresholds_as}), as expected.
The explicit formula of the slope parameters $b$ and $c$
are originally rendered in Eq.~(\ref{eq:aThresholds_bs}) and Eq.~(\ref{eq:aThresholds_cs}).
The constant $A^0$, $A_\pi^0$, $A_K^0$, $A_\eta^0$, $A_T^0$ are the coefficients denoted in Eq.(~\ref{eq:coeff_mpk}),
and constants $b^0$, $b_\pi^0$, $b_K^0$, $b_\eta^0$, $b_T^0$,
and $c^0$, $c_\pi^0$, $c_K^0$, $c_\eta^0$, $c_T^0$ are the dimensionless coefficients denoted as
\begin{eqnarray}
b^0 &=& \frac{4\left(168 m_K^6 - 112 m_K^5 m_\pi + 463 m_K^4 m_\pi^2 + 10 m_K^3 m_\pi^3
+43 m_K^2 m_\pi^4 + 4 m_K m_\pi^5 - 42 m_\pi^6\right)}
{9 m_\pi m_K\left(m_K - m_\pi\right)^2\left(4 m_K^2 - m_\pi^2\right)} \nonumber\\[1.2mm]
%
b_\pi^0 &=&\frac{64 m_K^6 + 130 m_K^5 m_\pi + 90 m_K^4 m_\pi^2 + 149 m_K^3 m_\pi^3 -
   55 m_K^2 m_\pi^4 + 93 m_K m_\pi^5 - 15 m_\pi^6}
   {6 m_K^2\left(m_K - m_\pi\right)^3\left(m_K + m_\pi\right)} \nonumber\\[1.2mm]
%
b_K^0&=&-\frac{402 m_K^6 - 212 m_K^5 m_\pi + 925 m_K^4 m_\pi^2 + 477 m_K^3 m_\pi^3 + 565 m_K^2 m_\pi^4 + 283 m_K m_\pi^5 - 48 m_\pi^6}{27 m_K m_\pi\left(m_K - m_\pi\right)^3\left(m_K + m_\pi\right)} \nonumber\\[1.2mm]
%
b_\eta^0&=&\frac{48 m_K^6 + 308 m_K^5 m_\pi + 90 m_K^4 m_\pi^2 - 108 m_K^3 m_\pi^3 + 17 m_K^2 m_\pi^4 + 12 m_K m_\pi^5 - 27 m_\pi^6}
{54  m_\pi m_K^2\left(m_K - m_\pi\right)^3} \nonumber\\[1.2mm]
%
b_T^0 &=&-\frac{4(24 m_K^5 - 29 m_K^4 m_\pi - 6 m_K^3 m_\pi^2 + 51 m_K^2 m_\pi^3 +
    10 m_K m_\pi^4 + 24 m_\pi^5)}
{27 m_\pi m_K\left(m_K - m_\pi\right)^2\left(m_K + m_\pi\right)} \nonumber\\[1.2mm]
%
c_0 &=&\frac{2}{135\left(m_K - m_\pi\right)^4\left(m_K + m_\pi\right)^3}
\bigg[4299 m_K^7 + 15223 m_K^6 m_\pi + 46990 m_K^5 m_\pi^2 + 43417 m_K^4 m_\pi^3 + 45607 m_K^3 m_\pi^4 \nonumber\\[0.6mm]
&&+ 51973 m_K^2 m_\pi^5 + 18784 m_K m_\pi^6 + 5787 m_\pi^7 \bigg]
-\frac{8m_\pi^2 m_K\left(20 m_K^3 -112 m_K^2 m_\pi - 5 m_K m_\pi^2 + 30 m_\pi^3\right)}
{45\left(m_K + m_\pi\right)^2\left(4 m_K^2 - m_\pi^2\right)^2} \nonumber\\[1.2mm]
%
c_\pi^0 &=&\frac{1}{18 m_K^2\left(m_K - m_\pi\right)^5\left(m_K + m_\pi\right)^3}
\bigg[168 m_K^{10} + 1192 m_K^9 m_\pi + 6080 m_K^8 m_\pi^2 + 2492 m_K^7 m_\pi^3 + 6816 m_K^6 m_\pi^4 \nonumber\\[0.6mm]
&& + 10573 m_K^5 m_\pi^5 - 215 m_K^4 m_\pi^6 + 2582 m_K^3 m_\pi^7
   + 354 m_K^2 m_\pi^8 - 135 m_K m_\pi^9 + 45 m_\pi^{10}\bigg] \nonumber\\[1.2mm]
%
c_K^0 &=& -\frac{1}{81\left(m_K - m_\pi\right)^5\left(m_K + m_\pi\right)^3}
\bigg[2328 m_K^8 + 13828 m_K^7 m_\pi + 4076 m_K^6 m_\pi^2 + 52469 m_K^5 m_\pi^3 + 32153 m_K^4 m_\pi^4 \nonumber\\[0.6mm]
&&  + 12158 m_K^3 m_\pi^5 + 27206 m_K^2 m_\pi^6 + 3433 m_K m_\pi^7 + 573 m_\pi^8\bigg] \nonumber\\[1.2mm]
%
c_\eta^0 &=&\frac{1}{162 m_K^2\left(m_K - m_\pi\right)^5\left(m_K + m_\pi\right)^2)}
 \bigg[2316 m_K^9 + 2192 m_K^8 m_\pi + 9200 m_K^7 m_\pi^2 - 2524 m_K^6 m_\pi^3 +
 6944 m_K^5 m_\pi^4 \nonumber\\[0.6mm]
&&- 2923 m_K^4 m_\pi^5 - 1120 m_K^3 m_\pi^6 - 402 m_K^2 m_\pi^7 - 324 m_K m_\pi^8 + 81 m_\pi^9\bigg]\nonumber\\[1.2mm]
%
c_T^0&=&\frac{1}{81\left(m_K^2 - m_\pi^2\right)^4\left(2 m_K - m_\pi\right)}
 \bigg[168 m_K^9 + 10 m_K^8 m_\pi + 351 m_K^7 m_\pi^2 - 2456 m_K^6 m_\pi^3 - 16706 m_K^5 m_\pi^4 \nonumber\\[0.6mm]
&& - 7094 m_K^4 m_\pi^5 - 809 m_K^3 m_\pi^6 - 3972 m_K^2 m_\pi^7  + 2932 m_K m_\pi^8 + 744 m_\pi^9\bigg]\, .
\label{eq:coeff_mpk_s}
\end{eqnarray}
\end{widetext}
It is worth while to note that all coefficients denoted in Eq.~(\ref{eq:coeff_mpk_s})
are just dependent on pion mass $m_\pi$ and kaon mass $m_K$ due to the usage of Gell-Mann-Okubo relation,
and the analytical expressions for threshold parameters are
definitely much more voluminous than those in chiral $SU(2)$ perturbation theory,
as expected in Ref.~\cite{Roessl:1999iu}.

From Eqs.~(\ref{eq:m_pi_r_s}) and (\ref{eq:m_pi_P_s}), it is obvious that the second term in $r$
dominates its behaviors for large $z$-values,
and last two terms in $P$ controls its features for small $z$-values.
Substituting Eqs.~(\ref{eq:aThresholds_as})-(\ref{eq:aThresholds_cs}) into
Eqs.~(\ref{eq:m_pi_r_s})-(\ref{eq:m_pi_P_s}), after some algebraic operations,
it is straightforward to achieve NLO $\chi$PT descriptions
for ERE parameters as
\begin{widetext}
\vspace{-0.3cm}
\begin{eqnarray}
\mu_{\pi K} r &=&
\frac{m_K^2 + m_K m_\pi + m_\pi^2}{\left(m_K + m_\pi\right)^2}\frac{4\pi}{z} +  C_4 + \chi_{r}(\mu) \label{eq:m_kpi_r_s} \\[1.5mm]
%
\mu_{\pi K}^3 P&=&
-\frac{3 m_K^4+5 m_K^3 m_\pi+7 m_K^2 m_\pi^2+5 m_K m_\pi^3+3 m_\pi^4} {2\left(m_K + m_\pi\right)^4}\frac{\pi}{z} + C_5
+ \chi_P(\mu), \label{eq:m_kpi_P_s}
\end{eqnarray}
where the constants $C_i(i=4-5)$ are solely denoted in terms of the constants $C_i(i=1-3)$ via
\begin{eqnarray}
C_4 &=&
4\pi\left[ 2\frac{m_K^2 + m_K m_\pi + m_\pi^2}{\left(m_K + m_\pi\right)^2}C_1 -  C_2\right]  \label{eq:r_C4_s} \\[1.5mm]
%
C_5 &=&
\frac{\pi}{2}\left[-2\frac{4 m_K^4 + 7 m_K^3 m_\pi + 10 m_K^2 m_\pi^2 + 7 m_K m_\pi^3 + 4 m_\pi^4}
{\left(m_K +m_\pi\right)^4} C_1
+\frac{4\left(m_K^2 + m_K m_\pi + m_\pi^2\right)}{\left(m_K + m_\pi\right)^2} C_2 - C_3 \right] \label{eq:P_C5_s}
\end{eqnarray}
and the chiral logarithm terms for the effective range $r$ and shape parameter $P$
\begin{eqnarray}
\hspace{-0.5cm}\chi_{r}(\mu)&=&
4\pi\left[ 2\frac{m_K^2 + m_K m_\pi + m_\pi^2}{\left(m_K + m_\pi\right)^2}\chi_{a}(\mu) - \chi_{b}(\mu) \right]  \\[1.5mm]
%
\hspace{-0.5cm}\chi_P(\mu)&=&
\frac{\pi}{2}\left[-2\frac{4 m_K^4 + 7 m_K^3 m_\pi + 10 m_K^2 m_\pi^2 + 7 m_K m_\pi^3
+ 4 m_\pi^4}{\left(m_K +m_\pi\right)^4}  \chi_a(\mu)
+\frac{4\left(m_K^2 + m_K m_\pi + m_\pi^2\right)}{\left(m_K + m_\pi\right)^2} \chi_b(\mu)
- \chi_c(\mu) \right]
\end{eqnarray}
are liner combinations of $\chi_a(\mu)$, $\chi_b(\mu)$,
and $\chi_c(\mu)$ denoted in Eqs.~(\ref{eq:chi_a})-(\ref{eq:chi_c}). 
Meanwhile, from Eqs.~(\ref{eq:C123_1}), (\ref{eq:c23_p2}), (\ref{eq:c23_p3}), (\ref{eq:r_C4_s}) and (\ref{eq:P_C5_s}),
the constants $C_i (i=4,5)$ can be recast in terms of low-energy constants $L_i$ as
\begin{eqnarray}
C_4 &=& 128\pi\left(6 L_1 + 2 L_2 + 3 L_3 - 6 L_4 + 6 L_6 + 3L_8\right)
- 256 \pi\left(L_4 - 2 L_6 - L_8\right) \frac{m_K^2 + m_\pi^2}{m_K m_\pi} \nonumber\\[0.5mm]
&&-64\pi\frac{\left(m_K + m_\pi\right)\left(2 m_K^2 + 2 m_K m_\pi + m_\pi^2\right)}
{ m_K^2 m_\pi} L_5 \label{eq:r_C4_sd} \\[1.5mm]
%
C_5 &=&  \frac{16 \pi m_K m_\pi}{3\left(m_K + m_\pi\right)^2}\bigg[
    24\left(L_4 - 2 L_6 - L_8\right)\frac{m_K^4 + m_\pi^4}{m_K^2 m_\pi^2} -
     3\left(14 L_1 + 6 L_2 + 7 L_3 - 26 L_4 + 38 L_6 + 19 L_8\right)\frac{m_K^2 + m_\pi^2}{ m_K m_\pi } \nonumber\\[0.5mm]
&&-  \left(118 L_1 + 62 L_2 + 59 L_3 - 102 L_4 + 150 L_6 + 75 L_8\right)\bigg]  \nonumber\\[0.5mm]
&&+\frac{ 8 \pi (8 m_K^4 + 14 m_K^3 m_\pi + 16 m_K^2 m_\pi^2 + 10 m_K m_\pi^3 + 3 m_\pi^4)}
{m_K^2 m_\pi (m_K + m_\pi)}L_5 \,.
\label{eq:P_C5_sd}
\end{eqnarray}
\end{widetext}

It is nice to note that constant $C_{4}$ is solely related to the effective range $r$,
and $C_{5}$ the shape parameter $P$, respectively~\cite{NPLQCD:2011htk}.
In practice, if lattice-measured effective range expansion parameters ($a$, $r$, and
$P$) are at hand, one can use Equations~(\ref{eq:aThresholds_as}), (\ref{eq:m_kpi_r_s}), and~(\ref{eq:m_kpi_P_s})
to acquire three constants: $C_1$, $C_4$ and $C_{5}$,
which can be employed to estimate constants $C_2$ and $C_3$,
then gain the desirable slope parameters $b$ and $c$, respectively.
This option clearly has more actual physical meanings and is definitely more convenient for one
to fit relevant lattice data since the phase shift $k\cot\delta$ directly
connect to the effective range expansion parameters ($a$, $r$, and $P$).

It is useful to note that, from Eq.~(\ref{eq:r_C4_s}) and  Eq.~(\ref{eq:P_C5_s}),
 the constants $C_2$ and $C_3$ can be delivered as
\begin{eqnarray}
C_2 &=& 2\frac{m_K^2 + m_K m_\pi + m_\pi^2}{\left(m_K + m_\pi\right)^2}C_1
                       - \frac{1}{4\pi}C_4 \\[2.0mm]
C_3 &=& -\frac{4 m_K^4 + 9 m_K^3 m_\pi + 14 m_K^2 m_\pi^2 + 9 m_K m_\pi^3 + 4 m_\pi^4}
                   {\left(m_K +m_\pi\right)^4} C_1 \nonumber\\[0.6mm]
&&+\frac{m_K^2 + m_K m_\pi + m_\pi^2}{\pi\left(m_K + m_\pi\right)^2} C_4  + \frac{2}{\pi}C_5\,.
\label{eq:newConst23_s}
\end{eqnarray}

We should remark at this point that it is also straightforward to
secure the desirable slope parameters $b$ and $c$
using Eqs.~(\ref{eq:m_pi_r_s})-(\ref{eq:m_pi_P_s}).
As a matter of fact, both methods arrive at the consistent results.

It has to be mentioned that one can use Eqs.~(\ref{eq:r_C4_s}) and (\ref{eq:P_C5_s})
or Eqs.~(\ref{eq:r_C4_sd}) and (\ref{eq:P_C5_sd}) to
equally estimate the effective range $r$ and shape parameter $P$
with given low-energy constants $L_i$.
Since there are correlations between various $L_i$,
relevant uncertainties from both methods are usually not identical, and the latter is often small.

\subsection{Numerical analysis}
It is important to note that relevant phenomenological predictions of threshold parameters
and ERE parameters contain the computable non-analytical contribution
which just rely on the pion mass $m_\pi$, kaon mass $m_K$ and the chiral scale $\mu$,
and the analytical terms which are dependent on seven coupling constants: $L_1$, $L_2$, $L_3$,
$L_4$, $L_5$, $L_6$, and $L_8$~\cite{Gasser:1984gg}.

However, with future more precise measurements of $L_i$,
it is useful to present its first part for prospective reasonable comparison~\cite{Bijnens:2004bu}.
Fortunately, Bijnens {\it et al.} kindly listed this corresponding numerical results in Table~2 of Ref.~\cite{Bijnens:2004bu},
which are calculated with the standard three-flavor $\chi$PT expression for $\pi K$ scattering~\cite{Bijnens:2004bu}.

In the present study, in order to numerically confirm our relevant derivations,
we indeed require a fair and convincing comparison with other corresponding published data.
For $I=\frac{3}{2}$ $\pi K$ scattering, it leads to the central value contributions to
the various concerned threshold parameters at order $p^2$ and $p^4$
with all $L_i$  set equal to zero, which are given in Table~\ref{tb:spd_abc}.

It is pretty nice and pleasure to note that our relevant predictions are astonishingly in good agreement
with  the corresponding numerical results in Table~2 of Ref.~\cite{Bijnens:2004bu},
which,  on the other hand, partially indicates the reliability of our results in this work.
Note that we adopt the latest PDG values of pion mass, kaon mass, and pion decay constant~\cite{ParticleDataGroup:2024cfk}.

\begin{table}[t!]
\caption{\label{tb:spd_abc}
The contributions at order $p^2$ and $p^4$ for $I=\frac{3}{2}$ $\pi K$ scattering length $a$,
slope parameter $b$ and slope parameter $c$ with LECs set equal to zero,
where the chiral scale $\mu$ is taken as the physical $\rho$ mass.
Please consult Appendix~\ref{app:ChPT PK_PD_AB} for near threshold behavior of
$p$-wave and $d$-wave amplitudes.
}
\begin{ruledtabular}
\begin{tabular}{lcc}
           & $p^2 $   &  $p^4 $      \\
\hline
$m_\pi   a_0$ & $-0.0713$   & $0.0148$      \\[0.8mm]
$m_\pi^3 a_1$ &             & $0.0000279$   \\[0.8mm]
$m_\pi^5 a_2$ &             & $0.0000942$   \\[0.8mm]
$m_\pi^3 b_0$ & $-0.0486$   & $0.0190$      \\[0.8mm]
$m_\pi^5 b_1$ &             & $-0.000209$   \\[0.8mm]
$m_\pi^5 c_0$ & $0.0124$    & $0.00141$     \\
\end{tabular}
\end{ruledtabular}
\end{table}

At the physical point, it is interesting and important to
note that all constants $C_1$, $C_2$, $C_3$, $C_4$ and $C_5$
are solely dependent on seven coupling constants: $L_1$, $L_2$, $L_3$, $L_4$, $L_5$
$L_6$, and $L_8$~\cite{Gasser:1984gg}, in addition to the chiral scale $\mu$.

In the present study, for the mesonic low-energy constants $L_i(i=1,\cdots,8)$
at the renormalization scale $\mu = m_\rho = 775.26 {\rm MeV}$~\cite{ParticleDataGroup:2024cfk},
we adopt the values of preferred fit (BE14)
given in Table~3 of Ref.~\cite{Bijnens:2014lea}:
\begin{eqnarray}
10^3L_1 &=& 0.53(6)  \nonumber \\[0.6mm]
10^3L_2 &=& 0.81(4)   \nonumber\\[0.6mm]
10^3L_3 &=& -3.07(20) \nonumber\\[0.6mm]
10^3L_4 &=& 0.3       \nonumber\\[0.6mm]
10^3L_5 &=& 1.01(6)   \nonumber\\[0.6mm]
10^3L_6 &=& 0.14(5)   \nonumber\\[0.6mm]
10^3L_8 &=& 0.47(10).
\nonumber
\end{eqnarray}

For our calculations we quote:
$m_\pi = 139.5706~{\rm MeV}$,  $m_K = 493.677~{\rm MeV}$, $f_\pi = 130.2~{\rm MeV}$,
extracted from PDG~\cite{ParticleDataGroup:2024cfk}.
It is worth emphasizing that, in this work, the physical eta mass $m_\eta$  is not extracted from PDG~\cite{ParticleDataGroup:2024cfk},
however, $m_\eta$  is estimated by the Gell-Mann-Okubo mass relation
since this relation is used to derive the relevant expressions.
The values of constant $C_i(i=1-5)$ are estimated at $m_{\rho}$ as
\begin{eqnarray}
C_1 &=& -0.064767 \pm 0.123783  \nonumber\\[0.6mm]
C_2 &=& -0.014457 \pm 0.097284  \nonumber\\[0.6mm]
C_3 &=& 0.14487   \pm  0.01973  \nonumber\\[0.6mm]
C_4 &=& -1.1664   \pm  3.8972   \nonumber\\[0.6mm]
C_5 &=& 0.22047   \pm 1.12131 ,
\label{eq:C123_CGL}
\end{eqnarray}
where the statistical errors are added in quadrature of the LECs's uncertainties.
\begin{table}[t!]
\caption{\label{tb:rP_bc}
Summaries of two ERE parameters ($r$ and $P$) compared to other results accessible in the literature.
The effective range $m_\pi r$ and shape parameter $P$ are estimated
with the help of Eq.~(\ref{eq:m_pi_r_s}) and Eq.~(\ref{eq:m_pi_P_s}).
Our NLO prediction is taken with the chiral scale $\mu$  at the physical $\rho$ mass.
Last block is a lattice determination at the physical point.
}
\begin{ruledtabular}
\begin{tabular}{lcl}
{\rm Reference}                         & $m_\pi r$           &  $m_\pi^3 P$     \\
\hline
Bernard1990~\cite{Bernard:1990kx}       & $8.90  \pm 8.10$    & \\[0.3mm]
Dobado1996~\cite{Dobado:1996ps}         & $21.76 \pm 4.33 $   & \\[0.3mm]
Pelaez2022~\cite{Pelaez:2020gnd}        & $37.42 \pm 10.74$   & \\[0.3mm]
Roessl1999~\cite{Roessl:1999iu}         & $10.74 \pm  9.42$   & $-10.32\pm 4.30$ \\[0.3mm]
B\"{u}ttiker2004~\cite{Buettiker:2003pp}& $36.96 \pm 13.02$   & $-24.16\pm 6.63$ \\[0.3mm]
This work                               & $16.92 \pm 4.01$    & $-8.76 \pm 1.91$ \\[2mm]
This work(Lat)                          & $21.54 \pm 6.90$    & $-6.98 \pm 3.80$ \\
\end{tabular}
\end{ruledtabular}
\end{table}

With the evaluated values of constant $C_i(i=1-5)$ in Eq.~(\ref{eq:C123_CGL}),
and using Eqs~(\ref{eq:aThresholds_as}), (\ref{eq:aThresholds_bs}),
(\ref{eq:aThresholds_cs}), (\ref{eq:m_pi_r_s}) and (\ref{eq:m_pi_P_s}),
the predictions of threshold parameters and ERE parameters at the physical point are
\begin{eqnarray}
m_\pi a   &=& -0.0595(62)    \nonumber\\[1.2mm]
m_\pi^3 b &=& -0.0297(40)    \nonumber\\[1.2mm]
m_\pi^5 c &=& 0.0162(35)     \nonumber\\[1.2mm]
m_\pi r   &=& 16.92 \pm 4.01 \nonumber\\[1.2mm]
P         &=& -8.76 \pm 1.91 ,
\label{eq:abcrP_phy}
\end{eqnarray}
which are presented in Table~\ref{tb:rP_bc} and Table~\ref{tab:Comp_arp}
for the comparisons with other relevant results.
It is important to note that if using Eqs.~(\ref{eq:m_kpi_r_s}) and (\ref{eq:m_kpi_P_s}),
the identical values of $m_\pi r$ and $P$ can be obtained within statistical uncertainties.

From Eq.~(\ref{eq:abcrP_phy}), the ratio of the effective range $m_\pi r$ to
shape parameter $P$ at the physical point can be computed as $-1.93(62)$,
which indicates that the second term and third term in Eq.~(\ref{eq:effrange}) both contribute significantly
for the lattice-measured values of the phase shift $k\cot\delta$.
Note that our lattice data in Table~\ref{tab:fitstoERT} indicate that this ratio is $-3.08(1.95)$.

\begin{table*}[ht!]
\centering
\caption{
A compilation of the various theoretical (or phenomenological), empirical values(Expt.), and lattice determinations of
$I=\frac{3}{2}$ $\pi K$ scattering length $a$, slope parameter $b$ and $c$ at the physical point.
Together with every reference, for an easier comparison, the first author name or the collaborations are given.
}
\begin{ruledtabular}
\begin{tabular}{lllll}
{\rm Reference}                    & $m_\pi a$             & $m_\pi^3 b$     & $m_\pi^5 c$   &  {\rm Remarks}  \\[0.3mm]
\hline
NPLQCD2006~\cite{Beane:2006gj}     & $-0.0574(16)$         &                 &     &{\rm lattice, Domain-wall valence} \\[0.3mm]
Nagata2009~\cite{Nagata:2008wk}    & $-0.0837^{+0.0506}_{-0.0640}$&          &     &{\rm lattice, improved Iwasaki} \\[0.3mm]
Sasaki2010~\cite{Sasaki:2010zz}    & $-0.0500(68)$         &                 &     &{\rm lattice, improved Wilson}\\[0.3mm]
Sasaki2014~\cite{Sasaki:2013vxa}   & $-0.0602(31)(26)$     &                 &     &{\rm lattice, improved Wilson}\\[0.3mm]
RBC-UKQCD2014~\cite{Janowski:2014uda}& $-0.0674(33)$       &                 &     &{\rm lattice, domain wall valence} \\[0.3mm]
ETM2018~\cite{Helmes:2018nug}      & $-0.059(2)$           &                 &     &{\rm lattice, twisted mass} \\[0.3mm]
This work(Lat)                     & $-0.0588(28)$         & $-0.0370(124)$  & $0.00097(2043)$ &{\rm lattice, staggered, moving wall source}\\[2.3mm]
Griffith1968~\cite{Griffith:1968jaz}& $-0.11$              &                &    &  {\rm Tree-level}\\[0.3mm]
Bernard1990~\cite{Bernard:1990kx}  & $-0.05(2)$            & $-0.011(5)$    &    &  {\rm One-loop chiral perturbation theory}\\[0.3mm]
Kubis2002~\cite{Kubis:2001bx}      & $-0.0543(170)$            &        &    &  {\rm  Isospin breaking, chiral perturbation theory}\\[0.3mm]

Dobado1996~\cite{Dobado:1996ps}    & $-0.049(4)$           & $-0.026(3)$    &          & {\rm Inverse amplitude method }\\[0.3mm]
Roessl1999~\cite{Roessl:1999iu}    & $-0.05(1)$            & $-0.0133(105)$ &$0.0223(17)$&{\rm chiral $SU(2)$ perturbation theory}\\
Nehme2002(Set1)~\cite{Nehme:2001wa}& $-0.0557(166)$        &                &        &{\rm Isospin breaking, chiral perturbation theory} \\[0.3mm]
Buettiker2004~\cite{Buettiker:2003pp}& $-0.0448(77)$       & $-0.037(3)$    & $0.018(2)$ & {\rm Dispersive Roy-Steiner} \\[0.3mm]
Bijnens2004~\cite{Bijnens:2004bu}    & $-0.047$            & $-0.027$       &           & {\rm Chiral $SU(3)$ perturbation theory at NNLO} \\[0.3mm]
Zhou2006~\cite{Zhou:2006wm}          & $-0.042(2)$         &                &           & {\rm Fit to the LASS data}\\[0.3mm]
Bijnens-Ecker2014~\cite{Bijnens:2014lea} & $-0.048(-0.047)$~\footnote{
Note that main number is from fit BE14, and that in parentheses is from the free fit.
}
&                &          & {\rm ChPT NNLO fit BE14 (free fit)} \\
Pelaez2016~\cite{Pelaez:2016tgi}   & $-0.054^{+0.010}_{-0.014}$&            &          & {\rm Fit constrained with FDR} \\[0.3mm]
Pelaez2022~\cite{Pelaez:2020gnd}   & $-0.0480(67)$         & $-0.043(3)$    &     & {\rm Sum rules with CFD input}\\[0.3mm]
This work                          & $-0.0595(62)$         & $-0.0297(40)$  & $0.0162(35)$ & {\rm Chiral $SU(3)$ perturbation theory} \\[2.3mm]

Empirical values(Expt.)~\cite{Bernard:1990kx}          & $-0.13 \cdots -0.05$   &                  &    &Mean experimental values\\[0.3mm]
\end{tabular}
\label{tab:Comp_arp}
\end{ruledtabular}
\end{table*}

\subsection{Comparisons with the relevant results}
\label{sec:Results}

At present, there are many investigations about $I=\frac{3}{2}$ $\pi K$ scattering length $m_\pi a$.
In Table~\ref{tab:Comp_arp}, we compare our lattice data in Table~\ref{tab:fitstoERT}
to these relevant results accessible in the literature.
Our lattice result of $m_\pi a$ is reasonable agreement with the newer experimental and
theoretical determinations as well as lattice calculations.
Note that the empirical value (Expt.) given in Ref.~\cite{Bernard:1990kx}
is just the mean experimental values for $m_\pi a $  listed in
Refs.~\cite{Dumbrajs:1983jd}-\cite{Karabarbounis:1980bk}.

\begin{figure}[t!]
\includegraphics[width=8.5cm,clip]{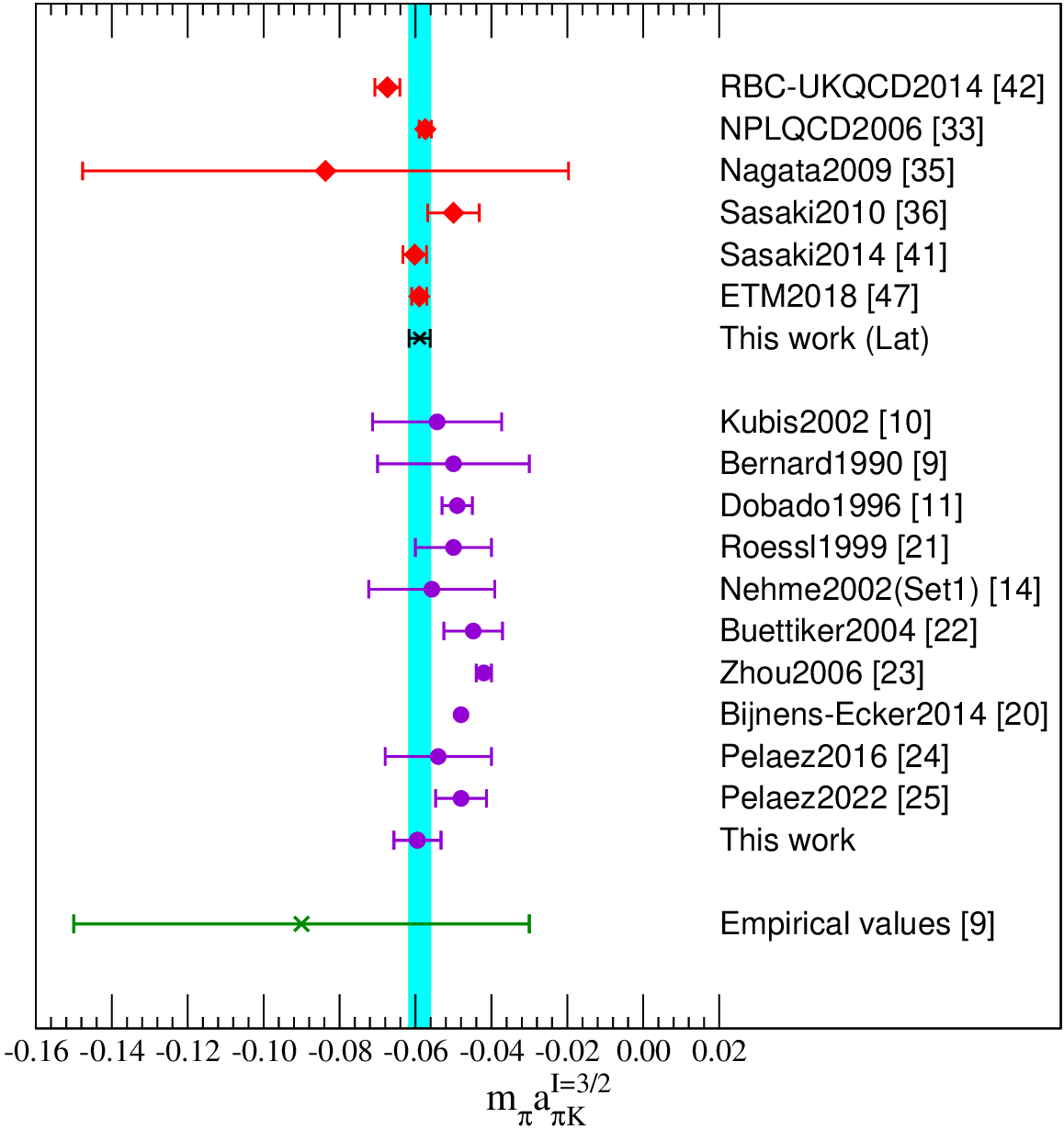}
\caption{\label{fig:compare_a}
A collection of  various lattice QCD, theoretical (or phenomenological) and
empirical (Expt.) results of $m_\pi a$ listed in Table~\ref{tab:Comp_arp}.
The red diamonds are lattice determinations,
purple circles are theoretical (or phenomenological) studies,
and empirical value (Expt.) is represented by a green cross.
Our lattice result is shown by a black cross, and our phenomenological prediction is also indicated by a purple circle.
For an easier comparison, the cyan  strip corresponds to the statistical error.
}
\end{figure}

To make our report of these results more intuitive,
these results are demonstrated graphically in Fig.~\ref{fig:compare_a} as well,
where the various results of $m_\pi a$ are compatible with each other within errors
except that of Zhou~\cite{Zhou:2006wm}.
In the figure, we do not plot the results of Griffith~\cite{Griffith:1968jaz} and Bijnens~\cite{Bijnens:2004bu}
due to the absence of statistical error in the papers.
We should remark at this point that our result is obtained at the physical kinematics,
which certainly rules out a likely error due to chiral extrapolation~\cite{Janowski:2014uda}, moreover,
our novel three-parameter fit to $k\cot\delta/m_\pi$ over region $k^2/m_\pi^2<1.0$ is
turned out to nicely avoid a nontrivial systematic error.

To the best of our knowledge, only a few numerical values are for the slope parameter
$b$~\cite{Bernard:1990kx,Dobado:1996ps,Roessl:1999iu,Buettiker:2003pp,Pelaez:2016tgi,Pelaez:2020gnd},
and just two for the shape parameter $c$ up to now~\cite{Roessl:1999iu,Buettiker:2003pp}.
Using Eqs.~(\ref{eq:m_pi_r_s})-(\ref{eq:m_pi_P_s}), the corresponding effective range $m_\pi r$,
and shape parameter $P$ can be readily obtained,
which are specifically are tabulated in Table~\ref{tb:rP_bc},
in addition to our lattice determinations which are listed in Table~\ref{tab:fitstoERT}.

From our novel three-parameter fit to the lattice-measured $k\cot\delta/m_\pi$ over region $k^2/m_\pi^2<1.0$,
we can directly get the scattering length $m_\pi a$, effective range $m_\pi r$ and shape parameter $P$,
which can be used to inversely secure the slope parameters $b$ and $c$ from Eqs.~(\ref{eq:m_pi_b}) and~(\ref{eq:m_pi_cs}).
Actually, from the estimated values in Eq.~(\ref{eq:abcrP_phy}),
it is easily verify that these equations can be reasonably simplified to
\begin{eqnarray}
b &=& -\frac{1}{2} a^2 r ,  \label{eq:m_pi_r_s2} \\[1.2mm]
c &=& \frac{1}{4} a^3 r^2 + \frac{3}{2} a^4 r - a^2 P \label{eq:m_pi_P_s2} .
\end{eqnarray}
\noindent Using our lattice-measured ERE parameters ($a$, $r$, and $P$) listed in Table~\ref{tab:fitstoERT},
and Equations~(\ref{eq:m_pi_b}) and~(\ref{eq:m_pi_cs}) or
Equations~(\ref{eq:m_pi_r_s2}) and~(\ref{eq:m_pi_P_s2}), we can obtain the values of $b$ and $c$,
which are listed in Table~\ref{tab:Comp_arp}.

\begin{figure}[b!]
\includegraphics[width=8.5cm,clip]{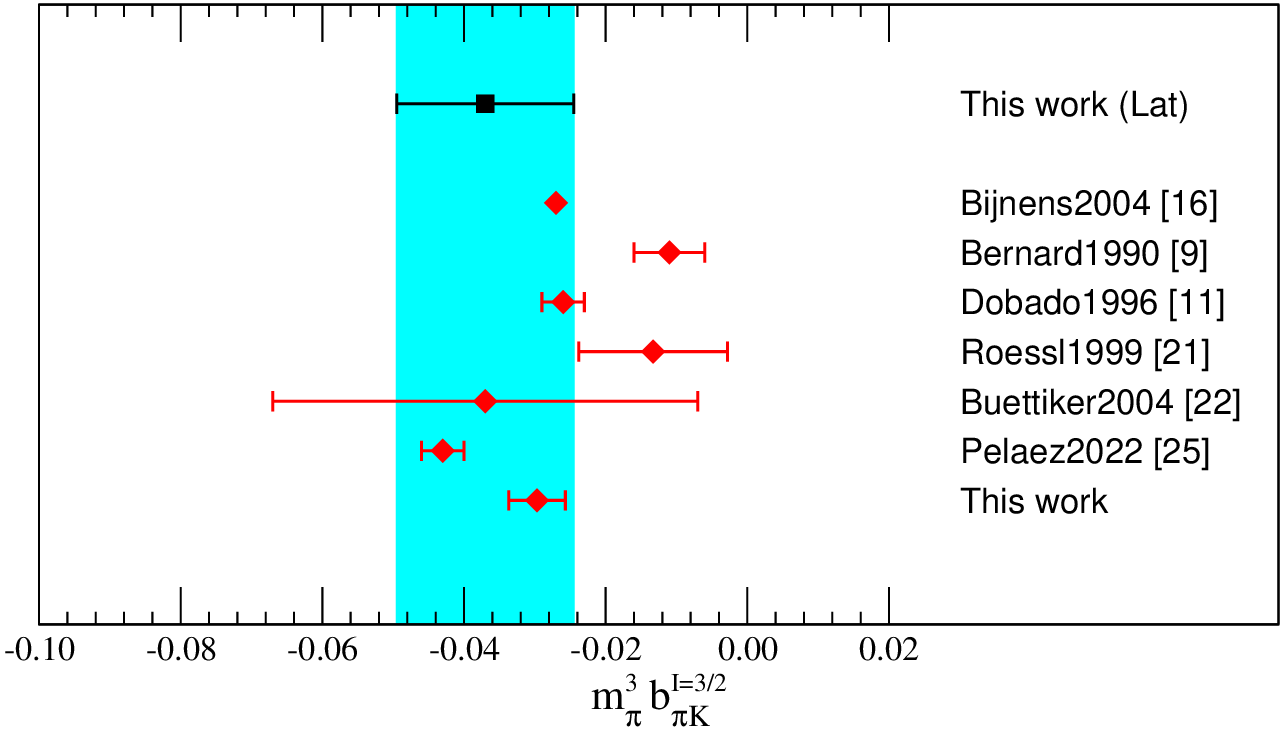}
\caption{\label{fig:compare_I32b}
A collection of  various lattice QCD and theoretical (or phenomenological) of $m_\pi^3 b_{\pi K}^{I=3/2}$
listed in Table~\ref{tb:rP_bc}.
The red diamonds are theoretical (or phenomenological) studies.
Our lattice result is shown by a black square,
and our phenomenological prediction is also indicated by a red diamond.
For an easier comparison, the cyan strip corresponds to the statistical error of our lattice result.
}
\end{figure}

To make our report of these results more intuitive,
these results of slope parameter $b$
are as well offered graphically in Fig.~\ref{fig:compare_I32b},
where the various outcomes of $m_\pi^3 b_{\pi K}^{I=3/2}$$(m_\pi^3 b)$
are reasonably compatible with other results within errors.

It should be worthwhile to stress that our theoretical prediction of slope parameter $c$
is in good accordance with two relevant data listed in Table~\ref{tab:Comp_arp}.
Admittedly, our lattice-derived value of slope parameter $c$
listed in Table~\ref{tab:Comp_arp} contains rather large statistical errors,
as a result, it is not convincing to compare with other data.
This kinds of work should wait for more robust lattice data in the future.

\section{Summary and outlook}
\label{sec:conclude}

With tremendously improved computing capabilities,
lattice simulation on $I=\frac{3}{2}$ $\pi K$ scattering can calculate its phase shifts with robust statistics,
which can be used to acquire the information on its threshold parameters.
Unfortunately, lattice calculations of $\pi K$ scattering are carried out by various lattice groups~\cite{Beane:2006gj,Chen:2006wf,Nagata:2008wk,Sasaki:2010zz,Fu:2011wc,Fu:2011xw,Lang:2012sv,Prelovsek:2013ela,Sasaki:2013vxa,Janowski:2014uda,Dudek:2014qha,Wilson:2014cna,Shepherd:2016dni,Brett:2018jqw,Helmes:2018nug,Wilson:2019wfr, Rendon:2020rtw} to mainly get scattering length
on the strength of analytical $\chi SU(3)$ expressions in Ref.~\cite{Kubis:2001bx}
or its variants~\cite{Chen:2006wf,Fu:2011wc,Sasaki:2013vxa,Helmes:2018nug}.
This work offers alike analytical $\chi SU(3)$ expressions for $r$ and $P$, 
which are helpful to handle lattice data.

To this end, we exploringly pick up  slope parameters $b$ and $c$
on the basis of one-loop $\pi K$ scattering amplitude derived from $\rm SU(3)$ $\chi$PT in Ref.~\cite{GomezNicola:2001as}.
Consequently, just three LEC's are needed to analyze $I=\frac{3}{2}$ $\pi K$ scattering at NLO $\chi$PT~\cite{NPLQCD:2011htk},
which are believed to be a helpful tools in the study of $\pi K$ scattering
since it builds a convenient bridge between phase shift $k\cot\delta$ and ERE parameters ($r$, $P$ or $b$, $c$)
with a clean analytical $\chi SU(3)$ expressions at NLO
like that for the scattering length $a$ in Ref.~\cite{Kubis:2001bx}.
In addition, for $\pi K$ scattering, we deduce two relation equations connecting $a$, $b$ and $c$ with
the effective range $r$ and  shape parameter $P$ respectively.

From the  phenomenological prediction in $\rm SU(3)$ $\chi$PT, and using authors's preferred fit (BE14) in Ref.~\cite{Bijnens:2014lea},
ERE parameters ($a$, $r$ and $P$) at the physical point yield
\begin{eqnarray}
m_\pi a &=& -0.0595(62)  \nonumber \\[1.2mm]
m_\pi r &=&  16.92(4.01) \nonumber \\[1.2mm]
P       &=& -8.76(1.91),  \nonumber
\end{eqnarray}
\noindent which are in fair agreement with other phenomenological determinations in Refs.~\cite{Bernard:1990kx,Dobado:1996ps,Roessl:1999iu,Buettiker:2003pp,Pelaez:2016tgi,Pelaez:2020gnd}.

To numerically verify our derived expressions,
we use one MILC fine ($a\approx0.082$~fm, $L^3 \times T = 40^3\times 96$)
lattice ensemble with $N_f=2+1$ flavors of Asqtad-improved staggered
dynamical quarks~\cite{Golterman:1985dz,Kaplan:1992bt,FermilabLattice:2010rur}
to compute $s$-wave $I=\frac{3}{2}$ $\pi K$ scattering,
where L\"uscher's technique~\cite{Luscher:1986pf,Luscher:1990ux,Luscher:1990ck} and its extensions~\cite{Rummukainen:1995vs,Davoudi:2011md,Doring:2011vk,Gockeler:2012yj,Kim:2005gf,Christ:2005gi,Doring:2012eu,
Fu:2011xz,Leskovec:2012gb} are employed to get scattering phase shift
with moving wall source technique~\cite{Kuramashi:1993ka}.

To get robust ERE parameters, lattice computations are done
for seven total momenta.
Unfortunately, some of energy levels, especially for one $\mathbf{P}$ = $[0,0,1]$,
either turned out to be beyond the elastic region or the relevant signals were not good enough,
which provides some valuable experience for more sophisticated calculations.

A three-parameter fit of the phase shifts gives
\begin{eqnarray}
m_\pi a &=& -0.0588(28)  \nonumber \\[1.2mm]
m_\pi r &=&  21.54(6.90) \nonumber \\[1.2mm]
P       &=& -6.98(3.80), \nonumber
\end{eqnarray}
which are in reasonable agreement with recent experimental and theoretical
determinations~\cite{Bernard:1990kx,Dobado:1996ps,Roessl:1999iu,Buettiker:2003pp,Pelaez:2016tgi,Pelaez:2020gnd} as well as corresponding lattice calculations available in the literature~\cite{Beane:2006gj,Chen:2006wf,Nagata:2008wk,Sasaki:2010zz,Fu:2011wc,Fu:2011xw,Lang:2012sv,Prelovsek:2013ela,Sasaki:2013vxa,Dudek:2014qha,Wilson:2014cna,Janowski:2014uda,Shepherd:2016dni,Brett:2018jqw,Helmes:2018nug,Wilson:2019wfr, Rendon:2020rtw}.

Admittedly, our fitted values of the shape parameter $P$
listed in Table III contain the rather large statistical errors,
as a result, it is not convincing to use this data to secure the slope parameter $c$.
This kinds of investigation should be waiting for more robust lattice data in the future.
Moreover, the extracted shape parameter $P$ is found to be sensitive to the upper bound of the fitted momentum range.
To improve its stability, it is highly desirable for one to calculate more data points
in the region $0.5 <k^2/m_\pi^2<1.0$.
One thing greatly comforting us is our lattice-derived result of slope parameter $b$
is fair compatible with other results within uncertainty
due to its independent of the value of $P$.

Note that lattice-measured values of $k\cot\delta/m_\pi$ have relatively large errors,
which lead to fairly large statistical errors for extracted quantities ($m_\pi a$, $m_\pi r$ and $P$).
Besides, we just calculate $7$  points, simply due to lack of computational resources,
and robust extraction of shape parameter $P$
definitely needs more lattice data within $k^2/m_\pi^2< 1.0$.
Admittedly, most efficient way to improve the statistical errors of $P$ is
working on lattice ensembles with different size $L$ for a given pion mass,
as is done for $I=2$ $\pi\pi$ scattering in Ref.~\cite{NPLQCD:2011htk}.
Alternatively, it is  high desirable to investigate with large $L$ (i.e., $64$, etc.),
so more lattice points can be acquired within $k^2/m_\pi^2<1.0$.
A systematic study of this issue is absolutely important,
for our present preliminary study, the tentative estimates are sufficient.

It is important to note that the calculation of $\pi K$ scattering amplitude to NNLO
in standard three flavor $\chi$PT is presented in Ref.~\cite{Bijnens:2004bu},
and for $I=\frac{3}{2}$ $s$-wave $\pi K$ scattering,
the importance of the contributions from ${\cal{O}}(p^6)$ is easy to see.
Using the relevant data in Ref.~\cite{Bijnens:2004bu},
we can evaluate NNLO term (including the resonance estimate) contributes
roughly $30\%$ for the effective range.
Therefore, the higher-order sensitivity should be enhanced in the more sophisticated investigation.
Planned ambitious work for the future is in an effort to systematically
learn more knowledge on three flavor $\chi$PT at NNLO and relevant Wilson $\chi$PT
for the better comprehension of lattice research on $\pi K$ scattering~\cite{Sasaki:2013vxa,Janowski:2014uda}.

According to the arguments in Ref.~\cite{Sasaki:2013vxa},
the systematic error of the finite volume could be non-negligible.
The contribution to $k \cot{\delta}$ is estimated to be the order of $e^{-m_\pi L}$~\cite{Sasaki:2013vxa}.
For an lattice ensemble with $m_\pi L = 2.3$ in this work,
it is smaller than $10\%$ of $k \cot{\delta}$~\cite{Sasaki:2013vxa}.
Since it is typically much smaller than our statistical errors,
this systematic error can be reasonably ignored,  as practised in Ref.~\cite{Sasaki:2013vxa}.

The lattice spacing error could be one of an important
source of the uncertainty of the relevant physical quantities.
According to the discussions in Ref.~\cite{Helmes:2018nug}, the lattice
spacing dependence of them is mild for the lattice calculations,
and can be hence ignored in this preliminary study.
For more accurate lattice study in the future, systematic study with different lattice spacings near the
physical point is desirably demanded for quantitative understanding,
as suggested in pioneering works~\cite{Sasaki:2013vxa,Janowski:2014uda}.

\newpage
\section*{Acknowledgments}
This project was strongly and gratuitously supported by the High Performance Computing Center of Sichuan University.
We would like to deliver our earnest gratitude to MILC for allowing us to use some initial gauge configurations
and  GPL open of MILC codes.
We are deeply indebted to Carleton DeTar for his providing initial gauge configurations
and instructing required knowledge and computational skills for generating lattice gauge configurations.
We would like to express our appreciation to many warm-hearted persons's scientific charity
for donating us enough removable hard drives to store light quark propagators,
and contributing us their idle computation quota.
We specially convey our  gratitude to Institute of Nuclear Science and Technology, Sichuan
University, from which computer resources and electricity costs were freely furnished.

\appendix
\newpage
\section{$I=\frac{3}{2}$ NLO $\pi K$ scattering amplitude}
\label{app:ChPT PK_NLO}
One loop $\pi K$ scattering amplitude was initially offered in Ref.~\cite{Bernard:1990kx}.
Nevertheless, it does not meet the exact perturbative unitarity when expressed in terms of physical constants,
as pointed out in Ref.~\cite{GomezNicola:2001as}.
The corrected one was expressed just in terms of $f_\pi$ in Eq.~(B5) of Ref.~\cite{GomezNicola:2001as}.
In this Appendix, $I=\frac{3}{2}$ $\pi K$ partial scattering amplitude at NLO is derived
on the basis of the work in Ref.~\cite{GomezNicola:2001as}.

For elastic $K \pi$ scattering, Mandelstam variables are related to
the CoM three-momentum $k$ and scattering angle $\theta$ as~\cite{Nehme:2001wf}
\begin{eqnarray}
s &=& \left[\sqrt{m_\pi^2 + k^2} + \sqrt{m_K^2 + k^2}\right]^2 \nonumber \\[2.0mm]
u &=& \left[\sqrt{m_\pi^2 + k^2} - \sqrt{m_K^2 + k^2}\right]^2 -2k^2\left(1+\cos\theta\right)  \nonumber \\[2.0mm]
t &=& -2k^2\left(1-\cos\theta\right). \label{eq:sut_kx}
\end{eqnarray}
%
Using notations in Ref.~\cite{GomezNicola:2001as}, it is pretty easy to show that
\begin{widetext}
\vspace{-0.2cm}
\begin{eqnarray}
{\bar J}_{\pi\pi}(t)  &=& -\frac{k^2}{48\pi^2 m_\pi^2} + \frac{k^4}{240\pi^2m_\pi^4} + \frac{\cos\theta k^2}{48\pi^2m_\pi^2} -\frac{\cos\theta k^4}{120\pi^2m_\pi^4} + \frac{\cos^2\hspace{-0.08cm}\theta k^4}{240\pi^2m_\pi^4} + {\cal{O}}(k^6) \label{eq:JPPKP} \\[1.5mm]
{\bar J}_{KK}(t)      &=& -\frac{k^2}{48\pi^2 m_K^2}   + \frac{k^4}{240\pi^2m_K^4} + \frac{\cos\theta k^2}{48\pi^2m_K^2}-\frac{\cos\theta k^4}{120\pi^2m_K^4} + \frac{\cos^2\hspace{-0.08cm}\theta k^4}{240\pi^2m_K^4} + {\cal{O}}(k^6) \label{eq:JKKKP} \\[1.5mm]
{\bar J}_{\eta\eta}(t)&=& -\frac{k^2}{48\pi^2m_\eta^2} + \frac{k^4}{240\pi^2m_\eta^4}
+ \frac{\cos\theta k^2}{48\pi^2m_\eta^2} -\frac{\cos\theta k^4}{120\pi^2m_\eta^4}
+ \frac{\cos^2\hspace{-0.08cm}\theta k^4}{240\pi^2m_\eta^4} + {\cal{O}}(k^6) \label{eq:JEEKP}\\[1.5mm]
{\bar J}_{K\pi}(s) &=&
\frac{1}{16\pi^2}-\frac{m_K m_\pi}{16\pi^2\left(m_K^2-m_\pi^2\right)}\ln\frac{m_\pi^2}{m_K^2} \nonumber\\[1.0mm]
&&-\frac{k^2}{8\pi^2m_K m_\pi}
-\frac{m_K-m_\pi}{32\pi^2\left(m_K+m_\pi\right)m_K m_\pi}k^2\ln\frac{m_\pi^2}{m_K^2}\nonumber\\[1.0mm]
&&+\frac{\left(m_K + m_\pi\right)^2}{48 \pi^2 m_K^3 m_\pi^3}k^4
  +\frac{m_K^2 - m_\pi^2}{128\pi^2 m_K^3 m_\pi^3} k^4\ln\frac{m_\pi^2}{m_K^2}
  + {\cal{O}}(k^6) \label{eq:JKPS}  \\[1.5mm]
%
{\bar J}_{K\pi}(u) &=&
\frac{1}{16\pi^2}+\frac{m_K m_\pi}{16\pi^2\left(m_K^2-m_\pi^2\right)}\ln\frac{m_\pi^2}{m_K^2} \nonumber\\[1.0mm]
&&+\frac{(m_K^2+m_\pi^2)}{8\pi^2\left(m_K-m_\pi\right)^2m_K m_\pi} k^2
+\frac{(m_K+m_\pi)\left(m_K^2 + m_\pi^2\right)}{32\pi^2\left(m_K-m_\pi\right)^3m_K m_\pi} k^2\ln\frac{m_\pi^2}{m_K^2}
\nonumber\\[1.0mm]
%
&&-\frac{m_K^6 - 8 m_K^5 m_\pi - 3 m_K^4 m_\pi^2 - 4 m_K^3 m_\pi^3 - 3 m_K^2 m_\pi^4 - 8 m_K m_\pi^5 + m_\pi^6}
{48\pi^2\left(m_K-m_\pi\right)^4 m_K^3 m_\pi^3}k^4 \nonumber\\[1.0mm]
&&-\frac{\left(m_K + m_\pi\right)\left(m_K^6 - 6 m_K^5 m_\pi - m_K^4 m_\pi^2 - 4 m_K^3 m_\pi^3
 -m_K^2 m_\pi^4 - 6 m_K m_\pi^5 + m_\pi^6\right)}
{128\pi^2\left(m_K - m_\pi\right)^5 m_K^3 m_\pi^3}k^4 \ln\frac{m_\pi^2}{m_K^2}\nonumber\\[1.0mm]
%
&&+\frac{1}{4\pi^2\left(m_K-m_\pi\right)^2}\cos\theta k^2
  +\frac{m_K + m_\pi}{16\pi^2\left(m_K-m_\pi\right)^3} \cos\theta k^2\ln\frac{m_\pi^2}{m_K^2}\nonumber\\[1.0mm]
%
&&+\frac{\left(m_K^2 + m_\pi^2\right)\left(m_K^2+10 m_K m_\pi + m_\pi^2\right)}
{24\pi^2\left(m_K-m_\pi\right)^4 m_K^2 m_\pi^2}\cos\theta k^4+
\frac{\left(m_K+m_\pi\right)\left(m_K^2+m_\pi^2\right)}{8\pi^2\left(m_K-m_\pi\right)^5m_K m_\pi}\cos\theta k^4\ln\frac{m_\pi^2}{m_K^2}\nonumber\\[1.0mm]
%
&&+\frac{m_K^2 + 10 m_K m_\pi + m_\pi^2}{24\pi^2\left(m_K-m_\pi\right)^4 m_K m_\pi}\cos^2\hspace{-0.08cm}\theta k^4
+\frac{m_K+m_\pi}{8\pi^2\left(m_K-m_\pi\right)^5}\cos^2\hspace{-0.08cm}\theta k^4\ln\frac{m_\pi^2}{m_K^2}
+ {\cal{O}}(k^6)  \label{eq:JKPU} \\[1.5mm]
%
\hspace{-1.5cm}{\bar J}_{K\eta}(u) &=&
\frac{1}{16\pi^2} + \frac{(2m_K-m_\pi)(5m_K + 2m_\pi)}
{48\pi^2\left(m_K^2 - m_\pi^2\right)}\ln\frac{m_\eta^2}{m_K^2}-\frac{t(m_\pi,m_K)}{12\pi^2}\nonumber\\[1.0mm]
&&+\frac{(m_K^2+m_\pi^2)k^2}{16\pi^2\left(m_K-m_\pi\right)^2m_K m_\pi}
-\frac{(m_K + m_\pi)(m_K^2+m_\pi^2)}{96\pi^2\left(m_K-m_\pi\right)^3m_K m_\pi}k^2\ln\frac{m_\eta^2}{m_K^2}\nonumber\\[1.0mm]
&&-\frac{\left(5m_K+2m_\pi\right)\left(m_K^2+m_\pi^2\right)}
{96\pi^2\left(m_K-m_\pi\right)^2\left(m_K+m_\pi\right) m_K m_\pi}k^2\, t(m_\pi,m_K)\nonumber\\[1.0mm]
%
&&-\frac{4 m_K^7 - 17 m_K^6 m_\pi - 22 m_K^5 m_\pi^2 - 14 m_K^4 m_\pi^3 - 8 m_K^3 m_\pi^4 - 25 m_K^2 m_\pi^5 - 14 m_K m_\pi^6
+ 4 m_\pi^7}{256\pi^2\left(m_K - m_\pi\right)^4\left(m_K + m_\pi\right)m_K^3 m_\pi^3}k^4\nonumber\\[1.0mm]
%
&&+\frac{\left(m_K + m_\pi\right)\left(m_K^6 - 6 m_K^5 m_\pi - m_K^4 m_\pi^2 - 4 m_K^3 m_\pi^3 - m_K^2 m_\pi^4
- 6 m_K m_\pi^5 + m_\pi^6\right)}
{384\pi^2\left(m_K - m_\pi\right)^5 m_K^3 m_\pi^3}k^4\ln\frac{m_\eta^2}{m_K^2}\nonumber\\[1.0mm]
%
&&+\frac{k^4}{1536\pi^2 m_K^3 m_\pi^3\left(m_K-m_\pi\right)^4\left(2m_K-m_\pi\right)\left(m_K+m_\pi\right)^2}
\bigg[40 m_K^9 - 150 m_K^8 m_\pi - 241 m_K^7 m_\pi^2  \nonumber\\[1.0mm]
&&- 24 m_K^6 m_\pi^3 - 70 m_K^5 m_\pi^4 - 166 m_K^4 m_\pi^5 - 105 m_K^3 m_\pi^6 + 116 m_K^2 m_\pi^7
+ 36 m_K m_\pi^8 - 8m_\pi^9\bigg]t(m_\pi,m_K) \nonumber\\[1.0mm]
%
&&+\frac{\cos\hspace{-0.08cm}\theta k^2}{8\pi^2\left(m_K -m_\pi\right)^2}
-\frac{m_K + m_\pi}{48\pi^2\left(m_K-m_\pi\right)^3} \cos\hspace{-0.08cm}\theta k^2\ln\frac{m_\eta^2}{m_K^2}
-\frac{\left(5m_K+2m_\pi\right)\cos\hspace{-0.08cm}\theta k^2}{48\pi^2\left(m_K-m_\pi\right)^2\left(m_K + m_\pi\right)}
t(m_\pi,m_K) \nonumber\\[1.0mm]
%
&&+\frac{\left(13m_K +10m_\pi\right)\left(m_K^2 + m_\pi^2\right)}
{64\pi^2\left(m_K - m_\pi\right)^4\left(m_K + m_\pi\right)m_K m_\pi}\cos\theta k^4
-\frac{\left(m_K + m_\pi\right)\left(m_K^2 + m_\pi^2\right)}
{24\pi^2\left(m_K - m_\pi\right)^5 m_K m_\pi}\cos\theta k^4\ln\frac{m_\eta^2}{m_K^2}
\nonumber\\[1.0mm]
%
&&-\frac{\left(m_K^2 + m_\pi^2\right)\left(106 m_K^3 + 117 m_K^2 m_\pi - 48 m_K m_\pi^2 - 32 m_\pi^3\right)}
{384\pi^2\left(m_K - m_\pi\right)^4\left(m_K + m_\pi\right)^2\left(2m_K-m_\pi\right)m_K m_\pi}
\cos\theta k^4 t(m_\pi,m_K)\nonumber\\[1.0mm]
%
&&+\frac{13m_K +10m_\pi}{64\pi^2\left(m_K - m_\pi\right)^4\left(m_K + m_\pi\right)}\cos^2\hspace{-0.08cm}\theta k^4
-\frac{m_K + m_\pi}{24\pi^2\left(m_K - m_\pi\right)^5}\cos^2\hspace{-0.08cm}\theta k^4
\ln\frac{m_\eta^2}{m_K^2}\nonumber\\[1.0mm]
%
&&-\frac{106 m_K^3 + 117 m_K^2 m_\pi - 48 m_K m_\pi^2 - 32 m_\pi^3}
{384\pi^2\left(m_K - m_\pi\right)^4\left(m_K + m_\pi\right)^2\left(2m_K-m_\pi\right)} \cos^2\hspace{-0.08cm}\theta k^4
t(m_\pi,m_K) + {\cal{O}}(k^6) \label{eq:JKEU} \\[1.5mm]
{\bar J}_{K\eta}(s)&=&
\frac{1}{16\pi^2} + \frac{\left(5 m_K - 2 m_\pi\right)\left(2 m_K + m_\pi\right)}
{48\pi^2\left(m_K^2 - m_\pi^2\right)}\ln\frac{m_\eta^2}{m_K^2}
-\frac{1}{12\pi^2} \, t_2(m_\pi, m_K) \nonumber\\[1.0mm]
&&- \frac{1}{16\pi^2 m_K m_\pi} k^2 + \frac{m_K - m_\pi}{96\pi^2m_K m_\pi\left(m_K+m_\pi\right)} k^2\ln\frac{m_\eta^2}{m_K^2}
+\frac{5m_K-2m_\pi}{96\pi^2 m_K m_\pi\left(m_K-m_\pi\right)} k^2 t_2(m_\pi, m_K) \nonumber\\[1.0mm]
&&+\frac{4 m_K^3 + m_K^2 m_\pi + 2 m_K m_\pi^2 - 4 m_\pi^3}
{256\pi^2 m_K^3 m_\pi^3\left(m_K - m_\pi\right)} k^4
-\frac{m_K^2 - m_\pi^2}{384\pi^2 m_K^3 m_\pi^3}k^4\ln\frac{m_\eta^2}{m_K^2}\nonumber\\[1.0mm]
&&-\frac{40 m_K^5 - 10 m_K^4 m_\pi - 17 m_K^3 m_\pi^2 - 52 m_K^2 m_\pi^3 + 4 m_K m_\pi^4 + 8 m_\pi^5}
{1536\pi^2 m_K^3 m_\pi^3\left(2 m_K + m_\pi\right)\left(m_K - m_\pi\right)^2}k^4\, t_2(m_\pi, m_K) + {\cal{O}}(k^6)
\label{eq:JPQKP}
\end{eqnarray}
where, in this work, we follow the notation in Ref.~\cite{Sasaki:2013vxa} to denote function $t(m_\pi, m_K)$ in Eq.~(\ref{eq:t_tan}), and
\begin{equation}
t_2(m_\pi, m_K) = \frac{\sqrt{2m_K^2 - m_\pi m_K - m_\pi^2}}{m_K + m_\pi}
\arctan\hspace{-0.05cm}\left[\frac{2(m_K+m_\pi)}{m_K-2m_\pi}\sqrt{\frac{m_K - m_\pi}{2m_K+m_\pi}}\right] .
\end{equation}
Note that $\bar{\bar J}_{K\eta}(u) \equiv {\bar J}_{K\eta}(u) - u{\bar J}'_{K\eta}(0)$
and $\bar{\bar J}_{K\pi}(u)\equiv {\bar J}_{K\pi}(u) - u{\bar J}'_{K\pi}(0)$,
where ${\bar J}'_{K\pi}(0)$ and ${\bar J}'_{K\eta}(0)$ are denoted in Ref.~\cite{GomezNicola:2001as}.
Plugging the Eqs.~(\ref{eq:JPPKP})$-$(\ref{eq:JPQKP})  into  Eq.~(B5) of Ref.~\cite{GomezNicola:2001as},
after some massive algebraic manipulations, and considering the Mandelstam variables in Eq.~(\ref{eq:sut_kx})
and the tree-level Gell-Mann-Okubo relation $m_\eta^2 = \frac{4}{3}m_K^2 - \frac{1}{3}m_\pi^2$,
one can finally get the expressions for $T^{\frac{3}{2}}(s,t,u)$ as
\begin{eqnarray}
T^{\frac{3}{2}}(s,t,u)&=&P_0(\cos\theta)\Bigg\{
-\frac{2m_K m_\pi}{f_\pi^2}
-\frac{\left(m_K + m_\pi\right)^2}{m_K m_\pi f_\pi^2} k^2
+\frac{\left(m_K^2 - m_\pi^2\right)^2}{4 m_K^3 m_\pi^3 f_\pi^2}k^4 \nonumber\\[1.2mm]
%
&&+\frac{2m_K^2 m_\pi^2}{f_\pi^4}\left[32\left(2 L_1 + 2L_2 + L_3 - 2 L_4 - \frac{L_5}{2} + 2L_6 + L_8\right) - 16\frac{m_\pi}{m_K}L_5\right] \nonumber\\[1.2mm]
%
&&+\frac{\left(m_K + m_\pi\right)^2}{f_\pi^4}k^2
\left[\frac{64\left(m_K^2+m_\pi^2\right)}{\left(m_K+m_\pi\right)^2}\left(2 L_1 + 2L_2 + L_3 - L_4\right)+ \frac{128 m_K m_\pi}{\left(m_K+m_\pi\right)^2} L_2 - 16\frac{m_\pi}{m_K} L_5\right] \nonumber\\[1.2mm]
%
&&+\frac{\left(m_K^2 - m_\pi^2\right)^2}{4 m_K^2 m_\pi^2 f_\pi^4}k^4\left[
\frac{1024m_K^2 m_\pi^2}{3\left(m_K^2 - m_\pi^2\right)^2}(2L_1 + L_2 + L_3)
+ \frac{256m_K m_\pi}{\left(m_K - m_\pi\right)^2}L_2 + 16\frac{m_\pi}{m_K}L_5 \right]\nonumber\\[1.2mm]
%
&&+\frac{m_K^3 m_\pi^3}{8\pi^2\left(m_K+ m_\pi\right)^2f_\pi^4}\left[A_0 + A_\pi^0\ln\frac{m_\pi^2}{\mu^2}
 + A_K^0 \ln\frac{m_K^2}{\mu^2} + A_\eta^0\ln\frac{m_\eta^2}{\mu^2}
 + A_T^0\, t(m_\pi,m_K)\right]\nonumber\\[1.2mm]
%
&&+\frac{m_K m_\pi}{16\pi^2 f_\pi^4}k^2\Bigg[B^0 +  B_\pi^0\ln\frac{m_\pi^2}{\mu^2}
+ B_K^0\ln\frac{m_K^2}{\mu^2} + B_\eta^0\ln\frac{m_\eta^2}{\mu^2}+ B_T^0\, t(m_\pi,m_K)\Bigg] \nonumber\\[0.8mm]
%
&&+\frac{\left(m_K - m_\pi\right)^2}{64 \pi^2 m_K m_\pi f_\pi^4}k^4
\left[ C^0 + C_\pi^0 \ln\frac{m_\pi^2}{\mu^2} + C_K^0\ln\frac{m_K^2}{\mu^2} + C_\eta^0 \ln\frac{m_\eta^2}{\mu^2}
+ C_T^0\, t(m_\pi,m_K)\right] \Bigg\}\nonumber\\[2.0mm]
%
&+&P_1(\cos\theta)k^2\Bigg\{\frac{32m_K m_\pi}{f_\pi^4}
\left[ 2(2L_2+L_3)-\frac{m_K^2+m_\pi^2}{m_K m_\pi}\left(4 L_1 + L_3 - 2 L_4\right)\right] \nonumber\\[1.2mm]
&&+\frac{32k^2}{f_\pi^4}\hspace{-0.1cm}\left[\frac{m_K^2+m_\pi^2}{m_K m_\pi}(2 L_2+L_3)-2(4L_1+L_3)\right]
\nonumber\\[1.2mm]
%
&&+\frac{m_K m_\pi}{\pi^2 f_\pi^4}\left[A^1 + A_\pi^1 \ln\frac{m_\pi^2}{\mu^2} + A_K^1\ln\frac{m_K^2}{\mu^2}
+A_\eta^1\ln\frac{m_\eta^2}{\mu^2} + A_T^1 \, t(m_\pi,m_K)\right] \nonumber\\[1.2mm]
&&+\frac{k^2}{\pi^2 f_\pi^4}\left[B^1 + B_\pi^1 \ln\frac{m_\pi^2}{\mu^2} + B_K^1\ln\frac{m_K^2}{\mu^2}
+ B_\eta^1\ln\frac{m_\eta^2}{\mu^2} + B_T^1\, t(m_\pi,m_K)\right]
\Bigg\}\nonumber\\[2.0mm]
%
&+&P_2(\cos\theta) k^4 \Bigg\{
\frac{128}{3f_\pi^4}\left( 2 L_1 + L_2 + L_3\right) \nonumber\\[1.2mm]
&&+ \frac{\left(m_K + m_\pi\right)^2}{\pi^2 f_\pi^4 m_K m_\pi}\left[
A^2 + A_\pi^2\ln\frac{m_\pi^2}{\mu^2}+A_K^2\ln\frac{m_K^2}{\mu^2}+A_\eta^2\ln\frac{m_\eta^2}{\mu^2}
+A_T^2\, t(m_\pi,m_K)\right] \Bigg\}
\label{eq:T0_32}
\end{eqnarray}
where the Legendre polynomials $P_0(\cos\theta)=1$, $P_1(\cos\theta) = \cos\theta$, and
$P_2(\cos\theta) = \frac{3}{2}\cos^2\theta - \frac{1}{2}$,
the constants $A^0$, $A_\pi^0$, $A_K^0$, $A_\eta^0$, $A_T^0$; $B^0$, $B_\pi^0$, $B_K^0$, $B_\eta^0$, $B_T^0$;
and $C^0$, $C_\pi^0$, $C_K^0$, $C_\eta^0$, $C_T^0$  are the aforementioned coefficients denoted in Eq.~(\ref{eq:coeff_mpk}).
The constants $A^1$, $A_\pi^1$, $A_K^1$, $A_\eta^1$, $A_T^1$; $B^1$, $B_\pi^1$, $B_K^1$, $B_\eta^1$, $B_T^1$;
and  $A^2$, $A_\pi^2$, $A_K^2$, $A_\eta^2$, $A_T^2$ are the dimensionless coefficients denoted as
\begin{eqnarray}
A^1      &=& -\frac{80 m_K^6 - 524 m_K^5 m_\pi - 458 m_K^4 m_\pi^2 - 397 m_K^3 m_\pi^3 + 96 m_K^2 m_\pi^4 + 131 m_K m_\pi^5 + 4 m_\pi^6}{72m_K m_\pi\left(m_K - m_\pi\right)^2\left(4 m_K^2 - m_\pi^2\right) } \nonumber\\[1.5mm]
A_\pi^1  &=& \frac{4 m_K^4 + 87 m_K^3 m_\pi - 24 m_K^2 m_\pi^2 + 56 m_K m_\pi^3 -9 m_\pi^4}
{48 m_K\left(m_K-m_\pi\right)^3} \nonumber\\[1.5mm]
A_{K}^1  &=&-\frac{m_K^5 + 279 m_K^4 m_\pi - 125 m_K^3 m_\pi^2 + 478 m_K^2 m_\pi^3 - 27 m_K m_\pi^4 - 8 m_\pi^5}
{216 m_K m_\pi\left(m_K - m_\pi\right)^3}\nonumber\\[1.5mm]
A_\eta^1 &=&\frac{56 m_K^5 + 36 m_K^4 m_\pi + 155 m_K^3 m_\pi^2 - 16 m_K^2 m_\pi^3 - 72 m_K m_\pi^4 + 11 m_\pi^5}
{432 m_K m_\pi\left(m_K - m_\pi\right)^3}\nonumber\\[1.5mm]
A_{T}^1  &=&\frac{8 m_K^5 - 12 m_K^4 m_\pi - 7 m_K^3 m_\pi^2 + 2 m_K^2 m_\pi^3 -
   24 m_K m_\pi^4 - 4 m_\pi^5}{54 m_K m_\pi\left(m_K - m_\pi\right)^2\left(m_K + m_\pi\right)}\nonumber\\[2.0mm]
%
B^1     &=&
\frac{399 m_K^7 + 5483 m_K^6 m_\pi + 6335 m_K^5 m_\pi^2 + 15137 m_K^4 m_\pi^3 + 16007 m_K^3 m_\pi^4
+ 7553 m_K^2 m_\pi^5 + 6359 m_K m_\pi^6 + 747 m_\pi^7}
{720 m_K m_\pi\left(m_K - m_\pi\right)^4\left(m_K + m_\pi\right)} \nonumber\\[1.0mm]
&& +\frac{m_\pi^2\left(-56 m_K^2 + 15 m_\pi^2\right)}{120\left(4m_K^2-m_\pi^2\right)^2}\nonumber\\[1.5mm]
%
B_\eta^1&=& \frac{204 m_K^7 + 596 m_K^6 m_\pi + 377 m_K^5 m_\pi^2 + 905 m_K^4 m_\pi^3 - 238 m_K^3 m_\pi^4
+ 242 m_K^2 m_\pi^5 - 367 m_K m_\pi^6 -39 m_\pi^7}
{864 m_K m_\pi\left(m_K - m_\pi\right)^5} \nonumber\\[1.5mm]
%
B_K^1   &=&-\frac{282 m_K^7 + 262 m_K^6 m_\pi + 3397 m_K^5 m_\pi^2 - 263 m_K^4 m_\pi^3
+ 4768 m_K^3 m_\pi^4 - 212 m_K^2 m_\pi^5 + 1081 m_K m_\pi^6 -51 m_\pi^7}
{432 m_K m_\pi\left(m_K-m_\pi\right)^5} \nonumber\\[1.5mm]
%
B_\pi^1 &=&\frac{4 m_K^7 + 244 m_K^6 m_\pi - 43 m_K^5 m_\pi^2 + 1101 m_K^4 m_\pi^3 - 174 m_K^3 m_\pi^4
+ 682 m_K^2 m_\pi^5 + 29 m_K m_\pi^6 + 29 m_\pi^7}
{96 m_K m_\pi\left(m_K - m_\pi\right)^5 } \nonumber\\[1.5mm]
%
B_T^1   &=&\frac{1}{432 m_K m_\pi\left(m_K-m_\pi\right)^4\left(m_K + m_\pi\right)^2\left(2 m_K-m_\pi\right)}
\bigg[48 m_K^9 + 16 m_K^8 m_\pi + 222 m_K^7 m_\pi^2 - 863 m_K^6 m_\pi^3 \nonumber\\[1.0mm]
&& - 1466 m_K^5 m_\pi^4 - 1223 m_K^4 m_\pi^5 - 476 m_K^3 m_\pi^6 + 192 m_K^2 m_\pi^7 + 76 m_K m_\pi^8
+ 120 m_\pi^9\bigg] \nonumber\\[1.5mm]
%
A^2     &=& -\frac{12 m_K^7 + 49 m_K^6 m_\pi - 5885 m_K^5 m_\pi^2 - 7754 m_K^4 m_\pi^3
- 8174 m_K^3 m_\pi^4 - 6851 m_K^2 m_\pi^5 - 413 m_K m_\pi^6 + 6m_\pi^7}
{1080\left(m_K-m_\pi\right)^4\left(m_K+m_\pi\right)^3} \nonumber\\[1.0mm]
&&+ \frac{m_K m_\pi^3\left(56 m_K^2-15 m_\pi^2\right)}
{360\left(m_K + m_\pi\right)^2\left(4 m_K^2 - m_\pi^2\right)^2}\nonumber\\[1.5mm]
A_\pi^2 &=&\frac{m_K m_\pi\left(-10 m_K^5 + 158 m_K^4 m_\pi - 19 m_K^3 m_\pi^2 + 331 m_K^2 m_\pi^3
- 17 m_K m_\pi^4 + 25 m_\pi^5\right)}
{72\left(m_K - m_\pi\right)^5\left(m_K + m_\pi\right)^2}\nonumber\\[1.5mm]
%
A_K^2 &=&-\frac{m_K m_\pi \left(179 m_K^5 + 47 m_K^4 m_\pi + 1553 m_K^3 m_\pi^2
- 49 m_K^2 m_\pi^3 + 650 m_K m_\pi^4 - 64 m_\pi^5\right)}
{324\left(m_K - m_\pi\right)^5\left(m_K + m_\pi\right)^2}\nonumber\\[1.5mm]
%
A_\eta^2 &=&\frac{m_K m_\pi\left(124 m_K^5 + 292 m_K^4 m_\pi + 37 m_K^3 m_\pi^2 + 163 m_K^2 m_\pi^3
-167 m_K m_\pi^4 - 29 m_\pi^5\right)}
{648\left(m_K - m_\pi\right)^5\left(m_K + m_\pi\right)^2} \nonumber\\[1.5mm]
%
A_T^2 &=&\frac{m_K m_\pi}{648\left(m_K^2-m_\pi^2\right)^4\left(2 m_K-m_\pi\right)}
\bigg[160 m_K^7 + 96 m_K^6 m_\pi - 278 m_K^5 m_\pi^2 - 1151 m_K^4 m_\pi^3
- 800 m_K^3 m_\pi^4 + 64 m_K^2 m_\pi^5  \nonumber\\[1.0mm]
&&+ 120 m_K m_\pi^6 + 112 m_\pi^7\bigg]
\label{app:coeff_m}
\end{eqnarray}
In the $s$-channel, it is traditional to decompose $T^{3/2}(s,u,t)$
into the partial waves via
\begin{equation}
T^{3/2}(s,u,t) = 16\pi\sum_{\ell=0}^{\infty}(2\ell+1)P_\ell(\cos\theta)t_{\ell}^{3/2}(s) .
\end{equation}
where $\ell$ is the angular momentum of $\pi K$ system.
Consequently, we get the partial wave for $s$-wave ($\ell=0$),
which is previously denoted in Eq.~(\ref{eq:t0_3_2}).
Although, in this work, we are concentrated on the partial wave for the $s$-wave ($\ell=0$),
on the same time, we also present the partial wave for the $p$-wave ($\ell=1$)
\begin{eqnarray}
\frac{t^{3/2}_1(k)}{k^2}&=&
-\frac{2m_K m_\pi}{3\pi f_\pi^4}\left[\frac{m_K^2+m_\pi^2}
{m_K m_\pi}\left(4 L_1 + L_3 - 2 L_4\right) - 2(2L_2+L_3)\right]\nonumber\\[1.2mm]
&&+\frac{2k^2}{3\pi f_\pi^4}\left[\frac{m_K^2+m_\pi^2}{m_K m_\pi}(2 L_2+L_3)- 2(4L_1+L_3)\right] \nonumber\\[1.2mm]
&&+\frac{m_K m_\pi}{48\pi^3 f_\pi^4}\left[A^1 + A_\pi^1 \ln\frac{m_\pi^2}{\mu^2} + A_K^1\ln\frac{m_K^2}{\mu^2}
+A_\eta^1\ln\frac{m_\eta^2}{\mu^2} + A_T^1 \,t(m_\pi,m_K)\right] \nonumber\\[1.2mm]
&&+\frac{k^2}{48\pi^3 f_\pi^4}\left[B^1 + B_\pi^1 \ln\frac{m_\pi^2}{\mu^2} + B_K^1\ln\frac{m_K^2}{\mu^2}
+ B_\eta^1\ln\frac{m_\eta^2}{\mu^2} + B_T^1\, t(m_\pi,m_K)\right]
\label{eq:t1_32}
\end{eqnarray}
%
and the partial wave for the $d$-wave ($\ell=2$)
\begin{eqnarray}
\frac{t^{3/2}_2(k)}{k^4}&=& \frac{8}{15\pi f_\pi^4} \left( 2 L_1 + L_2 + L_3 \right) \nonumber\\[0.6mm]
&&+\frac{\left(m_K + m_\pi\right)^2}{80 \pi^3 f_\pi^4 m_K m_\pi}
\left[A^2+ A_\pi^2\ln\frac{m_\pi^2}{\mu^2} + A_K^2\ln\frac{m_K^2}{\mu^2}
+ A_\eta^2\ln\frac{m_\eta^2}{\mu^2}+A_T^2\, t(m_\pi,m_K)\right]
\label{eq:t2_32}
\end{eqnarray}
since the scattering amplitude in Eq.~(\ref{eq:T0_32}) includes both information of them.

\section{Near threshold behavior of the $s$-wave amplitude}
\label{app:ChPT PK_PWA}
As a stringent double-check, near threshold behavior of the partial wave amplitude $t_0^{3/2}$
is also expressed as a power-series expansion in Eq.~(\ref{eq:threshexp_NPLQCD}).
Matching Eq.~(\ref{eq:threshexp_NPLQCD}) to Eq.~(\ref{eq:t0_3_2}), it is ready to yield
\begin{eqnarray}
\mu_{\pi K}a^{\prime} &=&\frac{z}{4\pi}
\left[- 1 +  z C_1^{\prime} + \frac{z}{16\pi^2}\chi_{a}^\prime(\mu)\right]
\label{eq:aThresholds_a} \\[2.0mm]
\mu_{\pi K}^2 b^{\prime} &=& \frac{\left(m_K + m_\pi\right)^2}{16\pi m_K m_\pi}z \left[-1
+ z C_2^{\prime} + \frac{z}{16\pi^2} \chi_{b}^\prime(\mu)\right]
\label{eq:aThresholds_b} \\[2.0mm]
\mu_{\pi K}^4 c^{\prime}&=&\frac{\left(m_K - m_\pi\right)^2}{64\pi m_K m_\pi}z
\left[ 1 + z C_3^{\prime} + \frac{z}{16\pi^2}\chi_{c}^\prime(\mu)\right],
\label{eq:aThresholds_c}
\end{eqnarray}
where the superscript prime in relevant physical quantities
in Eqs.~(\ref{eq:aThresholds_a})$-$(\ref{eq:aThresholds_c}) are just for difference
with corresponding ones in Eqs.~(\ref{eq:aThresholds_as})$-$(\ref{eq:aThresholds_cs}), and
three constant $C_i^{\prime}(i=1-3)$ can be conveniently expressed in terms of   $L_i(\mu)$~\cite{Bijnens:1997vq,Colangelo:2001df},
\begin{eqnarray}
C_1^{\prime} &=& 32\frac{\left(m_K + m_\pi\right)^2 }
{m_K m_\pi}\left[2L_1 + 2L_2 + L_3 - 2 L_4 - \frac{L_5}{2} + 2L_6 + L_8\right]
- 16\frac{\left(m_K + m_\pi\right)^2 }{m_K^2}L_5 \label{app:C123_1} \\[1.5mm]
C_2^{\prime} &=& 64\frac{(m_K^2+m_\pi^2)}{m_K m_\pi}\left(2 L_1 + 2L_2 + L_3 - L_4\right)
+  128 L_2 - 16\frac{\left(m_K + m_\pi\right)^2 }{m_K^2} L_5 \label{app:C123_2} \\[1.5mm]
%
C_3^{\prime} &=& \frac{1024}{3}\frac{m_K m_\pi}{\left(m_K - m_\pi\right)^2}(2L_1 + L_2 + L_3)
+ 256\frac{\left(m_K + m_\pi\right)^2}{\left(m_K - m_\pi\right)^2}L_2 + 16\frac{\left(m_K + m_\pi\right)^2 }{m_K^2} L_5 \label{app:C123_3}
\end{eqnarray}
and three known functions
\begin{eqnarray}
\chi_{a}^\prime(\mu) &=& A_0 + A_\pi^0\ln\frac{m_\pi^2}{\mu^2}
+A_K^0\ln\frac{m_K^2}{\mu^2} + A_\eta^0\ln\frac{m_\eta^2}{\mu^2} + A_T^0\, t(m_\pi,m_K) \\[1.2mm]
\chi_{b}^\prime(\mu) &=&B^0 + B_\pi^0\ln\frac{m_\pi^2}{\mu^2}
+ B_K^0\ln\frac{m_K^2}{\mu^2} + B_\eta^0\ln\frac{m_\eta^2}{\mu^2}+ B_T^0\, t(m_\pi,m_K) \\[1.2mm]
\chi_{c}^\prime(\mu) &=&
C^0 + C_\pi^0 \ln\frac{m_\pi^2}{\mu^2} + C_K^0\ln\frac{m_K^2}{\mu^2} + C_\eta^0 \ln\frac{m_\eta^2}{\mu^2}
+ C_T^0\, t(m_\pi,m_K)
\end{eqnarray}
are clearly dependent on the chiral scale $\mu$ with chiral logarithm terms,
and the constants $A^0$, $A_\pi^0$, $A_K^0$, $A_\eta^0$, $A_T^0$;
$B^0$, $B_\pi^0$, $B_K^0$, $B_\eta^0$, $B_T^0$; and
$C^0$, $C_\pi^0$, $C_K^0$, $C_\eta^0$, $C_T^0$ are  the aforementioned coefficients denoted in Eq.~(\ref{eq:coeff_mpk}).
Note that $C_1^\prime \equiv C_1$ denoted in Eq.~(\ref{eq:C123_1}) and
$\chi_{a}^\prime(\mu) \equiv \chi_{a}(\mu)$ denoted in Eq.~(\ref{eq:chi_a}).

Matching the threshold expansion in Eq.~(\ref{eq:threshexp_NPLQCD}) to
ERE in Eq.~(\ref{eq:effrange}), the effective range $r$ and
shape parameter $P$ can be nicely described
just in terms of three threshold parameters~\cite{Bijnens:1997vq}:
\begin{eqnarray}
r &= & \frac{1}{m_\pi m_K a^\prime} - \frac{4b^\prime}{(m_\pi + m_K)a^{\prime2}} - 2 a^\prime \label{app:r_NP} \\[1.2mm]
P &=& \frac{a^\prime}{2m_\pi m_K} - \frac{2b^\prime}{m_\pi + m_K} - a^{\prime3}
- \frac{m_K^2 - m_K m_\pi + m_\pi^2}{8a^\prime(m_\pi m_K)^3}
+ \frac{4 b^{\prime2}}{ \left(m_\pi + m_K\right)^2 a^{\prime3}}
- \frac{b^\prime}{ m_\pi m_K\left(m_\pi + m_K\right) a^{\prime2}} \nonumber\\[0.6mm]
&&- \frac{2 c^\prime}{\left(m_\pi + m_K\right) a^{\prime2}} , \label{app:P_NP}
\end{eqnarray}
where the compact form of $P$ contains seven separate terms.
Note that when $m_\pi = m_K$, $r$ and $P$ gracefully reduce to the $\pi\pi$ cases~\cite{NPLQCD:2011htk},
as expected from denotation in Eq.~(\ref{eq:threshexp_NPLQCD}).
It is worth mentioning that, from Eq.~(\ref{eq:a_2_a}) and Eq.~(\ref{eq:c_2_c}),
it is ready to verify that Eq.~(\ref{app:r_NP}) and Eq.~(\ref{app:P_NP}) nicely
lead to Eq.~(\ref{eq:m_pi_r_s}) and Eq.~(\ref{eq:m_pi_P_s}), respectively
if one replaces $a^\prime$, $b^\prime$ and $c^\prime$ with $a$, $b$ and $c$.

Substituting the Eqs.~(\ref{eq:aThresholds_a})$-$(\ref{eq:aThresholds_c}) into
the Eqs.~(\ref{app:r_NP})$-$(\ref{app:P_NP}), realigning the relevant results in the order of $z$,
it is straightforward to get the NLO $\chi$PT descriptions for the effective range approximation parameters:
\begin{eqnarray}
\mu_{\pi K} r&=&
\frac{m_K^2 + m_K m_\pi +m_\pi^2}{\left(m_K + m_\pi\right)^2} \frac{4\pi}{z}
+ C_4^\prime  + \chi_{\rm r}^\prime(\mu)
\label{app:m_kpi_r} \\[1.5mm]
%
\mu_{\pi K}^3 P &=&
-\frac{3 m_K^4+5 m_K^3 m_\pi+7 m_K^2 m_\pi^2+5 m_K m_\pi^3+3 m_\pi^4}
{2\left(m_K + m_\pi\right)^4}\frac{\pi}{z}
+ C_5^\prime + \chi_{\rm P}^\prime(\mu), \label{app:m_kpi_P}
\end{eqnarray}
where the constants $C_i^\prime(i=4-5)$ are solely denoted
in terms of the constants $C_i^\prime(i=1-3)$ via
\begin{eqnarray}
C_4^\prime &=&
4\pi\left[\frac{2m_K^2 + 3m_K m_\pi + 2m_\pi^2}
{\left(m_K + m_\pi\right)^2} C_1^\prime - C_2^\prime \right] \label{app:C45_4} \\[1.5mm]
%
C_5^\prime &=& -\frac{8 m_\pi^4 + 19 m_\pi^3 m_K + 25 m_\pi^2 m_K^2 + 19 m_\pi m_K^3 + 8 m_K^4}
{2\left(m_K + m_\pi\right)^4}\pi C_1^\prime
+\frac{2m_K^2 + 3m_K m_\pi + 2m_\pi^2}{\left(m_K + m_\pi\right)^2}\pi C_2^\prime \nonumber \\[0.5mm]
&&-\frac{(m_K - m_\pi)^2}{2\left(m_K + m_\pi\right)^2}\pi C_3^\prime \label{app:C45_5}
\end{eqnarray}
and the chiral logarithm terms for the effective range $r$ and shape parameter $P$
\begin{eqnarray}
\chi_{\rm r}^\prime(\mu)&=& 4\pi\left[ \frac{2m_K^2 + 3m_K m_\pi + 2m_\pi^2}
{\left(m_K + m_\pi\right)^2} \chi_{a}(\mu) -\chi_{b}(\mu) \right]
\\[1.5mm]
%
\chi_{\rm P}^\prime(\mu)&=&
-\frac{8 m_\pi^4 + 19 m_\pi^3 m_K + 25 m_\pi^2 m_K^2 + 19 m_\pi m_K^3 + 8 m_K^4}
{2\left(m_K + m_\pi\right)^4}\pi \chi_a(\mu)
+\frac{2m_K^2 + 3m_K m_\pi + 2m_\pi^2}{\left(m_K + m_\pi\right)^2}\pi \chi_b(\mu) \nonumber\\[0.5mm]
&&-\frac{\left(m_K - m_\pi\right)^2}{2\left(m_K + m_\pi\right)^2}\pi \chi_c(\mu)
\end{eqnarray}
are liner combinations of the  $\chi_a(\mu)$, $\chi_b(\mu)$,
and $\chi_c(\mu)$ denoted in Eqs.~(\ref{eq:chi_a})-(\ref{eq:chi_c}), therefore can be calculated by them.
Meanwhile, using Eqs. (\ref{app:C123_1}), (\ref{app:C123_2}), (\ref{app:C123_3}), (\ref{app:C45_4}) and (\ref{app:C45_5}),
we can recast $C_i^\prime(i=4,5)$ in terms of low-energy constants $L_i(i=1,\cdots,8)$,
which are turned out to be completely identical to those in Eq.~(\ref{eq:r_C4_sd}) and Eq.~(\ref{eq:P_C5_sd}), respectively.

The relevant chiral logarithm terms for the effective range $r$ and shape parameter $P$
can be treated in an analogous way.
It is interesting and important to note that, from Eqs.~(\ref{app:m_kpi_r}),~(\ref{app:m_kpi_P})
and Eqs.~(\ref{eq:m_kpi_r_s}),~(\ref{eq:m_kpi_P_s}), the final expressions for the effective range $r$ and
shape parameter $P$ from two definitions are equivalent.

\section{Near threshold behavior of $p$-wave and $d$-wave amplitudes}
\label{app:ChPT PK_PD_AB}
For $p$-wave in Eq.~(\ref{eq:t1_32}) and $d$-wave in Eq.~(\ref{eq:t2_32}) amplitudes,
we use the traditional expansion
\begin{equation}
\mbox{Re}\; t_\ell^I(k)\ = \frac{\sqrt{s}}{2} k^{2\ell}\left\{a_\ell^I +k^2 b_\ell^I + k^4 c_\ell^I + {\cal O}(k^6)\right\}.
\label{app:threshexp_ss}
\end{equation}
Matching Eq.~(\ref{app:threshexp_ss}) to Eq.~(\ref{eq:t1_32}), it is ready to yield
\begin{eqnarray}
\mu_{\pi K}^3 a_1 &=& \frac{2z^2}{3\pi}
\Bigg\{
4\left(2L_2+L_3\right) - \frac{2\left(m_K^2+m_\pi^2\right)}{ m_\pi m_K}\left(4 L_1 + L_3 - 2 L_4\right)  \nonumber\\[0.8mm]
&&+\frac{1}{16\pi^2}\left[A^1 + A_\pi^1 \ln\frac{m_\pi^2}{\mu^2} + A_K^1\ln\frac{m_K^2}{\mu^2}
+A_\eta^1\ln\frac{m_\eta^2}{\mu^2} + A_T^1\,t(m_\pi,m_K)\right] \Bigg\} \nonumber\\[1.5mm]
\mu_{\pi K}^5 b_1 &=& \frac{z^2}{3\pi}\Bigg\{
2(4 L_1 + 4 L_2 + 3 L_3 - 2 L_4)
-\frac{8 m_\pi m_K}{\left(m_K + m_\pi\right)^2}\left(6 L_1 + 3 L_2 + 3 L_3-L_4\right) \nonumber\\[0.8mm]
&&+\frac{1}{16\pi^2}\left[b^1 + b_\pi^1 \ln\frac{m_\pi^2}{\mu^2} + b_K^1\ln\frac{m_K^2}{\mu^2}
+b_\eta^1\ln\frac{m_\eta^2}{\mu^2} + b_T^1\, t(m_\pi,m_K)\right] \Bigg\},
\end{eqnarray}
where the constants $A^1$, $A_\pi^1$, $A_K^1$, $A_\eta^1$, and $A_T^1$ are denoted in Eq.~(\ref{app:coeff_m}),
and the constants
\begin{eqnarray}
b^1&=& \frac{499 m_K^7 + 4728 m_K^6 m_\pi + 6340 m_K^5 m_\pi^2 + 15787 m_K^4 m_\pi^3 +
 16537 m_K^3 m_\pi^4 + 7678 m_K^2 m_\pi^5 + 5724 m_K m_\pi^6 + 727 m_\pi^7}
 {360\left(m_K - m_\pi\right)^4\left(m_K + m_\pi\right)^3} \nonumber \\[0.8mm]
 &&+
\frac{m_K m_\pi^2(20 m_K^3-168 m_K^2 m_\pi-5 m_K m_\pi^2+45 m_\pi^3)}
{180\left(m_K + m_\pi\right)^2\left(4 m_K^2-m_\pi^2\right)^2}\nonumber \\[1.5mm]
 b_\pi^1 &=& \frac{4 m_K^7 + 240 m_K^6 m_\pi - 122 m_K^5 m_\pi^2 + 1295 m_K^4 m_\pi^3 -
 365 m_K^3 m_\pi^4 + 827 m_K^2 m_\pi^5 - 45 m_K m_\pi^6 + 38 m_\pi^7}
 {48\left(m_K - m_\pi\right)^5\left(m_K + m_\pi\right)^2} \nonumber \\[1.5mm]
 b_K^1&=& -\frac{281 m_K^7 - 15 m_K^6 m_\pi + 4079 m_K^5 m_\pi^2 - 1270 m_K^4 m_\pi^3 +
 5876 m_K^3 m_\pi^4 - 736 m_K^2 m_\pi^5 + 1092 m_K m_\pi^6 - 43 m_\pi^7}
 {216\left(m_K - m_\pi\right)^5\left(m_K + m_\pi\right)^2} \nonumber \\[1.5mm]
 b_\eta^1 &=& \frac{148 m_K^7 + 672 m_K^6 m_\pi + 238 m_K^5 m_\pi^2 + 1195 m_K^4 m_\pi^3 -
 353 m_K^3 m_\pi^4 + 103 m_K^2 m_\pi^5 - 273 m_K m_\pi^6 - 50 m_\pi^7}
 {432\left(m_K - m_\pi\right)^5\left(m_K + m_\pi\right)^2}  \nonumber \\[1.5mm]
b_T^1 &=&  -\frac{1}{216\left(m_K^2 - m_\pi^2\right)^4\left(2 m_K -m_\pi\right)}
\bigg[16 m_K^9 - 208 m_K^8 m_\pi - 166 m_K^7 m_\pi^2 + 1107 m_K^6 m_\pi^3 + 1102 m_K^5 m_\pi^4 \nonumber \\[0.6mm]
&& + 1435 m_K^4 m_\pi^5 + 672 m_K^3 m_\pi^6 -472 m_K^2 m_\pi^7 - 28 m_K m_\pi^8 - 104 m_\pi^9\bigg]
\end{eqnarray}
are dimensionless.
Matching Eq.~(\ref{app:threshexp_ss}) to Eq.~(\ref{eq:t2_32}), it is ready to yield
\begin{eqnarray}
\mu_{\pi K}^5 a_2 &=&
\frac{2 z^2}{5\pi}
\left\{\frac{8 m_\pi m_K}{3\left(m_\pi + m_K\right)^2}\left( 2 L_1 + L_2 + L_3 \right) \right.  \nonumber \\[1.2mm]
&&\left.+\frac{1}{16\pi^2}
\left[A^2 + A_\pi^2\ln\frac{m_\pi^2}{\mu^2}   + A_K^2\ln\frac{m_K^2}{\mu^2}
          + A_\eta^2\ln\frac{m_\eta^2}{\mu^2} + A_T^2\, t(m_\pi,m_K)\right]
\right\},
\end{eqnarray}
where the constants $A^2$, $A_\pi^2$, $A_K^2$, $A_\eta^2$, and $A_T^2$ are denoted in Eq.~(\ref{app:coeff_m}).
\end{widetext}



\begin{thebibliography}{90}
\bibitem{Weinberg:1966kf} S.~Weinberg, Pion scattering lengths,
Phys.\ Rev.\ Lett.\  {\bf 17}, 616 (1966).

\bibitem{Gasser:1983yg} J.~Gasser and H.~Leutwyler, Chiral Perturbation Theory to One Loop,
Annals Phys. \textbf{158}, 142 (1984).

\bibitem{Bijnens:1997vq}  J.~Bijnens, G.~Colangelo, G.~Ecker, J.~Gasser and M.~E.~Sainio,
Pion pion scattering at low energy, Nucl.\ Phys.\ B {\bf 508}, 263 (1997)  [Erratum-ibid.\ B {\bf 517}, 639 (1998)].

\bibitem{Colangelo:2001df} G.~Colangelo, J.~Gasser and H.~Leutwyler,
$\pi\pi$ scattering, Nucl.\ Phys.\ B {\bf 603}, 125 (2001). 

\bibitem{Weinberg:1978kz} S.~Weinberg, Phenomenological Lagrangians,
Physica A \textbf{96}, no.1-2, 327-340 (1979).

\bibitem{Gasser:1984gg} J.~Gasser and H.~Leutwyler,
Chiral Perturbation Theory: Expansions in the Mass of the Strange Quark,
Nucl. Phys. B \textbf{250}, 465-516 (1985).

\bibitem{Griffith:1968jaz} R.~W.~Griffith,
Scalar density terms, $K\pi$ and $K K$ scattering lengths, and a symmetry-breaking parameter,
Phys. Rev. \textbf{176}, 1705-1708 (1968).

\bibitem{Bernard:1990kw} V.~Bernard, N.~Kaiser and U.~G.~Meissner,
\ensuremath{\pi}K scattering in chiral perturbation theory to one loop, Nucl. Phys. B \textbf{357}, 129-152 (1991).

\bibitem{Bernard:1990kx} V.~Bernard, N.~Kaiser and U.~G.~Meissner,
Threshold parameters of $\pi K$ scattering in QCD, Phys. Rev. D \textbf{43}, 2757-2760 (1991).

\bibitem{Kubis:2001bx} B.~Kubis and U.~G.~Meissner, Isospin violation in low-energy charged pion kaon scattering,
Phys. Lett. B \textbf{529}, 69-76 (2002).

\bibitem{Dobado:1996ps} A.~Dobado and J.~R.~Pelaez, The Inverse amplitude method in chiral perturbation theory,
Phys. Rev. D \textbf{56}, 3057-3073 (1997).

\bibitem{SaBorges:1994yy} J.~Sa Borges and F.~R.~A.~Simao,
Unitary corrections to current algebra versus chiral perturbation calculations in kaon - pion scattering,
Phys. Rev. D \textbf{53}, 4806-4810 (1996).

\bibitem{GomezNicola:2001as} A.~Gomez Nicola and J.~R.~Pelaez,
Meson meson scattering within one loop chiral perturbation theory and its unitarization,
Phys. Rev. D \textbf{65}, 054009 (2002).

\bibitem{Nehme:2001wa} A.~Nehme, Isospin breaking in low-energy charged pion and kaon elastic scattering,
Eur. Phys. J. C \textbf{23}, 707-718 (2002).

\bibitem{Nehme:2001wf} A.~Nehme and P.~Talavera, Isospin breaking corrections to low-energy pion kaon scattering,
Phys. Rev. D \textbf{65}, 054023 (2002).

\bibitem{Bijnens:2004bu} J.~Bijnens, P.~Dhonte and P.~Talavera,
$\pi K$ scattering in three flavor ChPT, J. High Energy Phys. \textbf{05}, 036 (2004).

\bibitem{Descotes-Genon:2007sqh} S.~Descotes-Genon,
Low-energy $\pi\pi$ and $\pi K$ scatterings revisited in three-flavour resummed chiral perturbation theory,
Eur. Phys. J. C \textbf{52}, 141-158 (2007).

\bibitem{Amoros:1999dp} G.~Amoros, J.~Bijnens and P.~Talavera,
Two point functions at two loops in three flavor chiral perturbation theory,
Nucl. Phys. B \textbf{568}, 319-363 (2000).

\bibitem{Ananthanarayan:2000cp} B.~Ananthanarayan and P.~Buettiker,
Comparison of pion kaon scattering in SU(3) chiral perturbation theory and dispersion relations,
Eur. Phys. J. C \textbf{19}, 517-528 (2001).


\bibitem{Bijnens:2014lea} J.~Bijnens and G.~Ecker, Mesonic low-energy constants,
Ann.\ Rev.\ Nucl.\ Part.\ Sci.\  {\bf 64}, 149 (2014).

\bibitem{Roessl:1999iu} A.~Roessl, Pion kaon scattering near the threshold in chiral SU(2) perturbation theory,
Nucl. Phys. B \textbf{555}, 507-539 (1999).

\bibitem{Buettiker:2003pp} P.~Buettiker, S.~Descotes-Genon and B.~Moussallam,
A new analysis of $\pi K$ scattering from Roy and Steiner type equations, Eur. Phys. J. C \textbf{33}, 409-432 (2004).

\bibitem{Zhou:2006wm} Z.~Y.~Zhou and H.~Q.~Zheng,
An improved study of the kappa resonance and the non-exotic $s$ wave $\pi K$ scatterings up to $\sqrt{s}=2.1$GeV of LASS data,''
Nucl. Phys. A \textbf{775}, 212-223 (2006).

\bibitem{Pelaez:2016tgi} J.~R.~Pelaez and A.~Rodas,
Pion-kaon scattering amplitude constrained with forward dispersion relations up to 1.6 GeV,
Phys. Rev. D \textbf{93}, no.7, 074025 (2016).

\bibitem{Pelaez:2020gnd} J.~R.~Pel{\'a}ez and A.~Rodas,
Dispersive $ \pi K \rightarrow \pi K$ and $ \pi\pi \rightarrow K\bar{K}$  amplitudes from scattering data,
threshold parameters, and the lightest strange resonance $\kappa$ or $K_0^{*}(700)$,
Phys. Rept. \textbf{969}, 1-126 (2022).

\bibitem{Cao:2025hqm} X.~H.~Cao, F.~K.~Guo, Z.~H.~Guo and Q.~Z.~Li,
Revisiting Roy-Steiner-equation analysis of pion-kaon scattering from lattice QCD data,
Phys. Rev. D \textbf{112}, no.3, 034042 (2025).

\bibitem{Dumbrajs:1983jd} O.~Dumbrajs, R.~Koch, H.~Pilkuhn, G.~c.~Oades, H.~Behrens, J.~j.~De Swart and P.~Kroll,
Compilation of Coupling Constants and Low-Energy Parameters. 1982 Edition,
Nucl. Phys. B \textbf{216}, 277-335 (1983).

\bibitem{Matison:1974sm} M.~J.~Matison, A.~Barbaro-Galtieri, M.~Alston-Garnjost, S.~M.~Flatte, J.~H.~Friedman,
 G.~R.~Lynch, M.~S.~Rabin and F.~T.~Solmitz, Study of $K^+\pi^-$ Scattering in the Reaction
 $K^+ p$ ---\ensuremath{>} $K^+\pi^-\Delta^{++}$ at 12-GeV/c, Phys. Rev. D \textbf{9}, 1872 (1974).

\bibitem{Lang:1976ze} C.~B.~Lang,
Meson Meson Scattering Amplitudes in the Modified K-Matrix Formalism,
Nuovo Cim. A \textbf{41}, 73 (1977).

\bibitem{Johannesson:1974ma} N.~O.~Johannesson and J.~L.~Petersen,
Coupled channel study of the $S$ wave $\pi \pi$ ---\ensuremath{>} k anti-k interaction,
Nucl. Phys. B \textbf{68}, 397-412 (1974).

\bibitem{Karabarbounis:1980bk} A.~Karabarbounis and G.~Shaw, Low-energy $K \pi$ scattering,
J. Phys. G \textbf{6}, 583 (1980).

\bibitem{DIRAC:2017hmz} B.~Adeva \textit{et al.} [DIRAC],
Measurement of the $\pi K$ atom lifetime and the $\pi K$ scattering length,
Phys. Rev. D \textbf{96}, 052002 (2017).

\bibitem{Beane:2006gj} S.~R.~Beane, P.~F.~Bedaque, T.~C.~Luu, K.~Orginos, E.~Pallante, A.~Parreno and M.~J.~Savage,
$\pi K$ scattering in full QCD with domain-wall valence quarks, Phys. Rev. D \textbf{74}, 114503 (2006).

\bibitem{Chen:2006wf} J.~W.~Chen, D.~O'Connell and A.~Walker-Loud, Two Meson Systems with Ginsparg-Wilson Valence Quarks,
Phys. Rev. D \textbf{75}, 054501 (2007).

\bibitem{Nagata:2008wk} J.~Nagata, S.~Muroya and A.~Nakamura, Lattice study of $K \pi$ scattering in $I = 3/2$ and $1/2$,
Phys. Rev. C \textbf{80}, 045203 (2009) [erratum: Phys. Rev. C \textbf{84}, 019904 (2011)].

\bibitem{Sasaki:2010zz}  K.~Sasaki, N.~Ishizuka, T.~Yamazaki and M.~Oka,
$S$-wave $\pi K$ scattering length in $2+1$ flavor lattice QCD,
Prog. Theor. Phys. Suppl. \textbf{186}, 187-192 (2010).

\bibitem{Fu:2011wc} Z.~Fu, Lattice study on $\pi K $ scattering with moving wall source,
Phys.\ Rev.\ D {\bf 85}, 074501 (2012).

\bibitem{Fu:2011xw} Z.~Fu, The preliminary lattice QCD calculation of $\kappa$ meson decay width,
J. High Energy Phys. \textbf{01}, 017 (2012);
Z.~Fu and K.~Fu, Lattice QCD study on $K^\ast(892)$ meson decay width,
Phys. Rev. D \textbf{86}, 094507 (2012).

\bibitem{Lang:2012sv} C.~B.~Lang, L.~Leskovec, D.~Mohler and S.~Prelovsek,
$K \pi$ scattering for isospin $1/2$ and $3/2$ in lattice QCD, Phys. Rev. D \textbf{86}, 054508 (2012)

\bibitem{Prelovsek:2013ela} S.~Prelovsek, L.~Leskovec, C.~B.~Lang and D.~Mohler,
$K \pi$ Scattering and the $K^*$ Decay width from Lattice QCD,
Phys. Rev. D \textbf{88}, no.5, 054508 (2013).

\bibitem{Sasaki:2013vxa} K.~Sasaki \textit{et al.} [PACS-CS], Scattering lengths for two pseudoscalar meson systems,
Phys. Rev. D \textbf{89}, no.5, 054502 (2014) [erratum: Phys. Rev. D \textbf{105}, no.1, 019901 (2022)].

\bibitem{Janowski:2014uda} T.~Janowski, P.~A.~Boyle, A.~J{\"u}ttner and C.~Sachrajda,
K-pi scattering lengths at physical kinematics,
PoS \textbf{LATTICE2014}, 080 (2014).

\bibitem{Dudek:2014qha} J.~J.~Dudek \textit{et al.} [Hadron Spectrum],
Resonances in coupled $\pi K -\eta K$ scattering from quantum chromodynamics,
Phys. Rev. Lett. \textbf{113}, no.18, 182001 (2014).

\bibitem{Wilson:2014cna} D.~J.~Wilson, J.~J.~Dudek, R.~G.~Edwards and C.~E.~Thomas,
Resonances in coupled $\pi K, \eta K$ scattering from lattice QCD,
Phys. Rev. D \textbf{91}, no.5, 054008 (2015).

\bibitem{Shepherd:2016dni} M.~R.~Shepherd, J.~J.~Dudek and R.~E.~Mitchell,
Searching for the rules that govern hadron construction,
Nature \textbf{534}, no.7608, 487-493 (2016).

\bibitem{Brett:2018jqw} R.~Brett, J.~Bulava, J.~Fallica, A.~Hanlon, B.~H\"orz and C.~Morningstar,
Determination of $s$- and $p$-wave $I=1/2$ $K\pi$ scattering amplitudes in $N_{\mathrm{f}}=2+1$ lattice QCD,
Nucl. Phys. B \textbf{932}, 29-51 (2018).

\bibitem{Helmes:2018nug} C.~Helmes \textit{et al.} [ETM],
Hadron-Hadron Interactions from $N_f=2+1+1$ Lattice QCD: $I=3/2$ $\pi K$ Scattering Length,
Phys. Rev. D \textbf{98}, no.11, 114511 (2018).

\bibitem{Wilson:2019wfr} D.~J.~Wilson, R.~A.~Briceno, J.~J.~Dudek, R.~G.~Edwards and C.~E.~Thomas,
The quark-mass dependence of elastic $\pi K$ scattering from QCD,
Phys. Rev. Lett. \textbf{123}, no.4, 042002 (2019).

\bibitem{Rendon:2020rtw} G.~Rendon, L.~Leskovec, S.~Meinel, J.~Negele, S.~Paul, M.~Petschlies, A.~Pochinsky, G.~Silvi
and S.~Syritsyn, $I=1/2$ $S$-wave and $P$-wave $K\pi$ scattering and the $\kappa$ and $K^*$ resonances from lattice QCD,
Phys. Rev. D \textbf{102}, no.11, 114520 (2020).

\bibitem{FermilabLattice:2010rur} C.~Bernard \textit{et al.} [Fermilab Lattice and MILC],
Tuning Fermilab Heavy Quarks in 2+1 Flavor Lattice QCD with Application to Hyperfine Splittings,
Phys. Rev. D \textbf{83}, 034503 (2011);
K.~Orginos \textit{et al.} [MILC], Testing improved actions for dynamical Kogut-Susskind quarks,
Phys. Rev. D \textbf{59}, 014501 (1999);
A.~Bazavov {\it et al.} [MILC Collaboration],
Nonperturbative QCD Simulations with 2+1 Flavors of Improved Staggered Quarks,
Rev.\ Mod.\ Phys.\  {\bf 82}, 1349 (2010);
C. Bernard, T. Burch, K. Orginos, D. Toussaint,
T. A. DeGrand, C. DeTar, S. Datta, S. Gottlieb, U. M. Heller, and R. Sugar,
The QCD spectrum with three quark flavors, Phys.\ Rev.\ D {\bf 64}, 054506 (2001);
C. Aubin, C. Bernard, C. DeTar, J. Osborn, S. Gottlieb, E. B.
Gregory, D. Toussaint, U. M. Heller, J. E. Hetrick, and R. Sugar,
Light hadrons with improved staggered quarks: Approaching the continuum limit,
Phys.\ Rev.\ D {\bf 70}, 094505 (2004).

\bibitem{Golterman:1985dz} M.~F.~L.~Golterman, Staggered Mesons,
Nucl.\ Phys.\ B {\bf 273}, 663 (1986).

\bibitem{Kaplan:1992bt} D.~B.~Kaplan, A Method for simulating chiral fermions on the lattice,
Phys.\ Lett.\ B {\bf 288}, 342 (1992).

\bibitem{NPLQCD:2011htk} S.~R.~Beane, E.~Chang, W.~Detmold, H.W.~Lin, T.~C.~Luu,
K.~Orginos, A.~Parreno, M.~J.~Savage, A.~Torok, and A. Walker-Loud, (NPLQCD),
The I=2 $\pi\pi$ S-wave Scattering Phase Shift from Lattice QCD,
Phys. Rev. D \textbf{85}, 034505 (2012).

\bibitem{Fu:2017apw} Z.~Fu and X.~Chen, $I=0$ $\pi\pi$ $s$-wave scattering length from lattice QCD,
Phys. Rev. D \textbf{98}, no.1, 014514 (2018);
Z.~Fu and J.~Wang,
$I=2$ \ensuremath{\pi}\ensuremath{\pi} s-wave scattering length from lattice QCD,
Phys. Rev. D \textbf{110}, no.7, 074513 (2024).

\bibitem{Blatt:1949zz} J.~M.~Blatt and J.~D.~Jackson,
On the Interpretation of Neutron-Proton Scattering Data by the Schwinger Variational Method,
Phys. Rev. \textbf{76}, 18-37 (1949);
J.~D.~Jackson and J.~M.~Blatt, The Interpretation of Low Energy Proton-Proton Scattering,
Rev. Mod. Phys. \textbf{22}, 77-118 (1950);
H.~A.~Bethe, Theory of the Effective Range in Nuclear Scattering,
Phys. Rev. \textbf{76}, 38-50 (1949).

\bibitem{Beane:2003da} S.~R.~Beane, P.~F.~Bedaque, A.~Parre\~no and M.~J.~Savage,
Two nucleons on a lattice, Phys.\ Lett.\  B {\bf 585}, 106 (2004).

\bibitem{Adhikari:1983ii} S.~K.~Adhikari and J.~R.~A.~Torreao,
Effective Range Expansion For The Pion-Pion System,
Phys.\ Lett.\  {\bf 123B}, 452 (1983).

\bibitem{Luscher:1986pf} M.~L\"uscher,
Volume Dependence of the Energy Spectrum in Massive Quantum Field Theories. 2. Scattering States,
Commun.\ Math.\ Phys.\  {\bf 105} (1986) 153.

\bibitem{Luscher:1990ux} M.~L\"uscher,
Two particle states on a torus and their relation to the scattering matrix,
Nucl.\ Phys.\  {\bf B 354}, 531 (1991).

\bibitem{Luscher:1990ck} M.~L\"uscher, U.~Wolff,
How to Calculate the Elastic Scattering Matrix in Two-dimensional Quantum Field Theories by Numerical Simulation,
Nucl.\ Phys.\ B {\bf 339}, 222 (1990).

\bibitem{Doring:2011vk} M.~Doring, U.~-G.~Meissner, E.~Oset and A.~Rusetsky,
Unitarized Chiral Perturbation Theory in a finite volume: Scalar meson sector,
Eur.\ Phys.\ J.\ A {\bf 47}, 139 (2011). 

\bibitem{Rummukainen:1995vs} K.~Rummukainen and S.~A.~Gottlieb,
Resonance scattering phase shifts on a nonrest frame lattice,
Nucl.\ Phys.\  B {\bf 450}, 397 (1995). 

\bibitem{Davoudi:2011md} Z.~Davoudi and M.~J.~Savage,
Improving the Volume Dependence of Two-Body Binding Energies Calculated with Lattice QCD,
Phys. Rev. D \textbf{84}, 114502 (2011).

\bibitem{Fu:2011xz} Z.~Fu,
Rummukainen-Gottlieb's formula on two-particle system with different mass,
Phys.\ Rev.\ D {\bf 85}, 014506 (2012).

\bibitem{Leskovec:2012gb} L.~Leskovec and S.~Prelovsek,
Scattering phase shifts for two particles of different mass and non-zero total momentum in lattice QCD,
Phys.\ Rev.\ D {\bf 85}, 114507 (2012).

\bibitem{Gockeler:2012yj} M.~Gockeler, R.~Horsley, M.~Lage, U.~G.~Meissner, P.~E.~L.~Rakow, A.~Rusetsky,
G.~Schierholz and J.~M.~Zanotti, Scattering phases for meson and baryon resonances on general moving-frame lattices,
Phys. Rev. D \textbf{86}, 094513 (2012).

\bibitem{Kim:2005gf} C.~h.~Kim, C.~T.~Sachrajda and S.~R.~Sharpe,
Finite-volume effects for two-hadron states in moving frames,
Nucl.\ Phys.\  B {\bf 727}, 218 (2005).

\bibitem{Christ:2005gi} N.~H.~Christ, C.~Kim and T.~Yamazaki,
Finite volume corrections to the two-particle decay of states with non-zero momentum,
Phys.\ Rev.\  D {\bf 72}, 114506 (2005).

\bibitem{Doring:2012eu} M.~Doring, U.~G.~Meissner, E.~Oset and A.~Rusetsky,
Scalar mesons moving in a finite volume and the role of partial wave mixing,
Eur.\ Phys.\ J.\ A {\bf 48}, 114 (2012).

\bibitem{Kuramashi:1993ka} Y.~Kuramashi, M.~Fukugita, H.~Mino, M.~Okawa and A.~Ukawa,
Lattice QCD Calculation of Full Pion Scattering Lengths,
Phys. Rev. Lett.  {\bf 71} (1993) 2387;
M.~Fukugita, Y.~Kuramashi, M.~Okawa, H.~Mino and A.~Ukawa,
Hadron scattering lengths in lattice QCD,
Phys. Rev. D \textbf{52}, 3003-3023 (1995);
M.~Fukugita, Y.~Kuramashi, H.~Mino, M.~Okawa and A.~Ukawa,
An Exploratory study of nucleon-nucleon scattering lengths in lattice QCD,
Phys. Rev. Lett. \textbf{73}, 2176-2179 (1994).

\bibitem{Sharpe:1992pp} S.~R.~Sharpe, R.~Gupta and G.~W.~Kilcup,
Lattice calculation of $I = 2$ pion scattering length,
Nucl. Phys. B \textbf{383}, 309-354 (1992).

\bibitem{DeTar:2018uko} C.~DeTar \textit{et al.} [Fermilab Lattice and MILC],
Splittings of low-lying charmonium masses at the physical point,
Phys. Rev. D \textbf{99}, no.3, 034509 (2019).

\bibitem{Lepage:1989hd} G. P. Lepage,
in Proceedings of TASI'89 Summer School, edited by T. DeGrand and D. Toussaint (World
Scientific, Singapore, 1990), p. 97.
The Analysis Of Algorithms For Lattice Field Theory, CLNS-89-971.

\bibitem{Fu:2016itp} Z.~Fu and L.~Wang,
Studying the $\rho$ resonance parameters with staggered fermions,
Phys.\ Rev.\ D {\bf 94}, no. 3, 034505 (2016).

\bibitem{Raposo:2023nex} A.~B.~Raposo and M.~T.~Hansen,
The L{\"u}scher scattering formalism on the t-channel cut,
PoS \textbf{LATTICE2022} (2023), 051.

\bibitem{Yamazaki:2004qb}  
T.~Yamazaki, S.~Aoki, M. Fukugita, K.-I.~Ishikawa, N.~Ishizuka, Y.~Iwasaki, K.~Kanaya, T.~Kaneko, Y.~Kuramashi, M.~Okawa, A.~Ukawa and T.~Yoshie, 
$I = 2$ $\pi\pi$ scattering phase shift with two flavors of O(a) improved dynamical quarks,
Phys.\ Rev.\ D {\bf 70}, 074513 (2004).

\bibitem{Fu:2012gf} Z.~Fu, Preliminary lattice study of $\sigma$ meson decay width,
J. High Energy Phys. 07 (2012) 142;
Z.~Fu, Lattice QCD study of the s-wave $\pi\pi $ scattering lengths in the $I=0$ and $2$ channels
Phys.\  Rev.\  D 87, {\bf 074501} (2013);
Z.~Fu, Preliminary lattice study of the $I=1$ $K \bar{K}$ scattering length,
Eur.\ Phys.\ J.\ C {\bf 72}, 2159 (2012).

\bibitem{MILC:DeTar}
https://web.physics.utah.edu/$\tilde{}$detar/milc/

\bibitem{Gupta:1993rn} R.~Gupta, A.~Patel and S.~R.~Sharpe, I = 2 pion scattering amplitude with Wilson fermions,''
Phys. Rev. D \textbf{48}, 388-396 (1993).

\bibitem{Umeda:2007hy} T.~Umeda, A Constant contribution in meson correlators at finite temperature,
Phys. Rev. D \textbf{75}, 094502 (2007).

\bibitem{DeTar:2014gla} C.~DeTar and S.~H.~Lee,
Variational method with staggered fermions,
Phys.\ Rev.\ D {\bf 91}, 034504 (2015).

\bibitem{Beane:2005rj} S.~R.~Beane, Paulo~F.~Bedaque, Kostas~Orginos, and Martin~J.~Savage (NPLQCD Collaboration),
$I=2$ $\pi\pi$ scattering from fully-dynamical mixed-action lattice QCD,
Phys. Rev. D \textbf{73}, 054503 (2006).


\bibitem{Landau:1991wop} L.~D. Landau and E.~M. Lifshits,
{\em {Quantum Mechanics}}, vol.~v.3 of {\em Course of Theoretical Physics}.
\newblock Butterworth-Heinemann, Oxford, 1991. \newblock

\bibitem{KalmahalliGuruswamy:2020uxi} C.~Kalmahalli Guruswamy, U.~G.~Mei\ss{}ner and C.~Y.~Seng,
Contraction diagram analysis in pion-kaon scattering,
Nucl. Phys. B \textbf{957}, 115091 (2020).

\bibitem{ParticleDataGroup:2024cfk} S.~Navas \textit{et al.} [Particle Data Group],
Review of particle physics, Phys. Rev. D \textbf{110}, no.3, 030001 (2024).

\end{thebibliography}
\end{document}